\documentclass[aps,prd,10pt,superscriptaddress,nofootinbib,reprint]{revtex4-2}
\usepackage[utf8]{inputenc}
\usepackage{textcomp}
\usepackage[dvipsnames]{xcolor}
\usepackage{amssymb,amsfonts,amsmath}
\usepackage[sort&compress]{natbib}
\usepackage[english]{babel}
\usepackage{braket}
\usepackage{graphicx}
\usepackage{bm}
\usepackage{bbm}
\usepackage{braket}
\usepackage{gensymb}
\usepackage{orcidlink}
\usepackage{enumitem}
\usepackage{dcolumn}
\usepackage{mathtools}
\usepackage{IEEEtrantools}
\usepackage{pstricks,pst-coil,pst-node,pst-circ,pst-plot,pst-3dplot,
pst-solides3d,pst-sigsys,pstricks-add,pst-eucl,pst-grad}
\usepackage[normalem]{ulem}
\allowdisplaybreaks[1]
\def\biggg#1{{\hbox{$\left#1\vbox to25.0pt{}\right.$}}}

\usepackage{dcolumn}
\usepackage{lmodern}
\usepackage{microtype}
\usepackage{hyperref}
\hypersetup{
	breaklinks=true,
    colorlinks=true,
    linkcolor=Blue,      
    urlcolor=Blue,
    citecolor=Blue
}
\usepackage{uri}

\newcommand{\beq}{\begin{eqnarray}}
\newcommand{\eeq}{\end{eqnarray}}
\newcommand{\nn}{\nonumber\\}

\newcommand{\rme}{{\rm e}}
\newcommand{\rmd}{{\rm d}}

\def\p{{\boldsymbol p}}

\begin{document}

\title{Thermodynamics of the parity-doublet model: Asymmetric and neutron matter}

\author{J\"urgen Eser\orcidlink{https://orcid.org/0000-0001-5539-9440}}
\email[]{eser@itp.uni-frankfurt.de}

\affiliation{Institut f\"ur Theoretische Physik, Johann Wolfgang Goethe-Universit\"at, Max-von-Laue-Stra\ss e 1, 60438 Frankfurt am Main, Germany}


\author{Jean-Paul Blaizot\orcidlink{https://orcid.org/0000-0002-9824-2467}}
\email[]{jean-paul.blaizot@ipht.fr}

\affiliation{Universit\'e Paris-Saclay, CNRS, CEA, Institut de physique th\'eorique, 91191, Gif-sur-Yvette, France}

\date{\today}

\begin{abstract}
    We consider isospin-asymmetric matter in the parity-doublet model within an extended mean-field calculation, increasing continuously the neutron excess all the way to  pure neutron matter. We compute the liquid-gas and the chiral phase transitions occurring at zero to moderate temperatures, but put special emphasis on the phase structure of matter at zero temperature and large baryon densities. The calculation of the free energy involves the solution of gap equations. This is achieved by transforming these gap equations into ordinary differential equations that control the flow with increasing baryon density of various physical quantities: the isoscalar condensate, the densities of protons and neutrons, as well as those of their respective chiral partners. In this formulation, the initial conditions for the differential equations determine the entire phase structure. It is further demonstrated that the threshold for the onset of the population of the chiral partners is exclusively determined by the fermionic parameters, most notably by the chiral-invariant mass of the nucleon. We underline the role of a parity symmetry energy in driving the equilibration of the nucleons and their parity partners across the chiral transition.  We provide a detailed analysis of the changes in the matter properties as one varies the neutron excess, including a special discussion of the chiral limit, and we compare systematically  the parity-doublet model to its corresponding singlet model, where the chiral partner of the nucleon is neglected. Finally, we focus on neutron matter and  compute the equation of state and the speed of sound. The results are confronted to those of other calculations as well as to recent Bayesian analyses of neutron-star observations. 
\end{abstract}

\maketitle

\section{Introduction}

This paper extends our previous investigation of the predictions of the parity-doublet model \cite{Detar:1988kn, Jido_2000, jido_chiral_2001} for the equation of state of dense baryonic matter. In the previous paper~\cite{Eser:2023oii}, hereafter referred to (I), a major focus was the dynamics of the transition leading to the restoration of chiral symmetry. Let us recall that in the parity-doublet model, chiral symmetry breaking is responsible only for the mass splitting of the parity doublets, and not for the entire masses of the baryons. Thus, as chiral symmetry is restored, which is characterized by the vanishing of a scalar condensate, the baryon doublets become degenerate with a non vanishing mass. The existence of this residual mass (commonly called $m_0$) in the chirally symmetric phase delays chiral symmetry restoration to large baryon density. This is a robust feature of the parity-doublet model, which has been observed in many calculations, see for instance Refs.~\cite{hatsuda_parity_1989, Zschiesche:2006zj, dexheimer_nuclear_2008, sasaki_thermodynamics_2010, Weyrich_2015, marczenko_chiral_2018, yamazaki_constraint_2019, minamikawa_quark-hadron_2021, Larionov:2021ycq, Marczenko_2022a, Minamikawa:2023eky, Fraga:2023wtd, Marczenko:2024trz}. We also underlined in (I) the role, in the transition towards chirally symmetric matter, of a kind of symmetry energy related to the different populations of the members of the doublets, and that we shall refer to here as ``parity symmetry energy.''

In (I) we restricted ourselves to isospin-symmetric matter. This included in particular normal nuclear matter, whose properties were used to determine the parameters of the model. In this paper, we consider isospin-asymmetric matter, with an arbitrary neutron excess, including pure neutron matter as a special limit. The study of pure neutron matter allows us to confront the predictions of the model with  data coming from neutron-star observations, in particular those related to the speed of sound.  We also provide a thorough description of the matter composition as a function of the neutron excess, covering all cases from symmetric matter to pure neutron matter. This allows us to explore more deeply the dynamics of the model as a function of the baryon density.

In our analysis we use the simplest version of the parity-doublet model, which extends minimally that used in (I) (and the one we recently used to compute $\pi\pi$ scattering lengths, obtaining excellent results \cite{Eser:2021ivo}). In addition to the nucleon doublet, which we choose to be the nucleon $N (939)$ and the $N^*(1535)$, the other degrees of freedom that are taken into account include the same meson fields as in (I), namely the isoscalar and isovector scalar fields $\sigma$ and $\pi$, and the  isoscalar vector field $\omega_\mu$. In addition, the description of asymmetric matter requires fixing the isospin symmetry energy, which is achieved by taking into account the contribution of a vector-isovector field $\rho_\mu$.  As in (I), we use the classical field approximation for the mesons and the mean-field approximation for the fermions, keeping the contribution of the fluctuations of the fermion fields (the vacuum fermion loop) which plays an essential role in the chiral transition. Borrowing the terminology from Ref.~\cite{brandes_fluctuations_2021}, we refer to this approximation for the fermions as an extended mean-field approximation. The parameters of the model are kept at their values determined in (I), except of course the coupling constant of the $\rho$-meson field to the baryons, which is adjusted so as to reproduce the empirical value of the symmetry energy. As we did in (I), we found it instructive to compare the predictions of the parity-doublet model with that of a singlet model involving only the positive-parity nucleons. This also offers us the possibility to compare our results to those obtained with similar models, in particular that of Ref.~\cite{brandes_fluctuations_2021}, or that of Ref.~\cite{Alford_2022}, whose parameters are adjusted on neutron-matter properties.

An important technical development, already introduced in (I) but exploited more systematically here, concerns the use of differential equations as an alternative to solving directly the gap equations. These differential equations control the flow of various  quantities, such as the scalar condensate $\sigma$, or the partial densities of various baryon species,  as a function of the total baryon density. This  provides a convenient, and efficient, way to obtain a number of results. It also gives an interesting perspective on the overall structure of the solutions of the gap equation, revealing how the model works and how its predictions are established.  This technical part of the paper is relegated to an appendix in order not to distract the reader from the main stream of the physics discussion, but we view this development as an important contribution of this paper.

In the present analysis, we restrict ourselves to large baryon density but small temperature. The reason is that we ignore the meson fluctuations. These fluctuations, in particular those of the lightest mesons (i.e.\ the pions) are expected to play an increasingly important role as the baryon density decreases \cite{gerber_hadrons_1989, brandt_chiral_2014}, making the mechanisms of chiral symmetry restoration somewhat different depending on whether one considers high temperature and low baryon density or low temperature and high baryon density.

The outline of the paper is as follows: In Sec.~\ref{sec:model} we recall the basic elements of the parity-doublet model, which is used in the present study. With the aim of getting a coherent description of interaction effects when discussing the phase diagram of baryonic matter, we find it convenient to express various derivatives of physical quantities with respect to densities or chemical potentials in terms of a set of Fermi-liquid parameters. Then, in Sec.~\ref{sec:paramters}, we discuss the adjustment of the new vector-isovector coupling constant on the symmetry energy of symmetric nuclear matter. We also comment on the predictions of the model concerning the so-called slope parameter $L$ of the symmetry energy (as well as the higher order coefficients $K_{\rm sym}$ and $Q_{\mathrm{sym}}$). Section~\ref{sec:asym_matter} presents a comprehensive analysis of the phase diagram of asymmetric matter as a function of the neutron excess and the baryon density. Section~\ref{sec:neutron_matter} presents specific results concerning pure neutron matter, and a brief comparison with equations of states inferred from recent neutron-star observations. Conclusions are summarized in Sec.~\ref{sec:summary}. Appendix~\ref{sec:slope} discusses some details of various determinations of the density dependence of the symmetry energy. Finally, Appendix~\ref{sec:diffeqs} is a more technical development, where we explain the usefulness of using differential equations to solve the gap equations.

\vspace*{0.6cm}

\section{Parity-doublet model}
\label{sec:model}

In this section, we recall the main features of the parity-doublet model, fixing the notation, and following here the general presentation given in our previous papers \cite{Eser:2021ivo, Eser:2023oii}.

The model that we consider consists of a baryon parity doublet, identified with the nucleon $N(939)$ and the $N^{*}(1535)$ (with a mass of $1510\ \mathrm{MeV}$ \cite{ParticleDataGroup:2022pth}), which are coupled to a set of meson fields: isoscalar and isovector scalar fields $\sigma$ and $\bm{\pi}$, as well as isoscalar and isovector vector fields, respectively $\omega_\mu$ and $\bm{\rho}_\mu$. At moderate baryon densities, the scalar fields provide attraction between the baryons, while the vector fields provide repulsion, the competition between both leading eventually to the ground state of nuclear matter as a self-bound system (with zero pressure), as we have verified in detail in (I).

The Lagrangian of the parity-doublet model constitutes a generalised (linear) sigma model. A characteristic feature of the model is to allow the baryon to acquire a mass while respecting chiral symmetry. Such a mass survives chiral symmetry restoration at high temperature or high baryon density, both members of the doublet becoming then degenerate with the same mass $m_0$. We shall also compare the results obtained with the parity-doublet model to those obtained with a similar model involving only the positive-parity nucleons and which we refer to as the singlet model.

\subsection{The model}

The parity-doublet model is built on two Dirac spinors $\psi_a$ and $\psi_b$ of opposite parities, with $\psi_a$ having positive parity and $\psi_b$ having negative parity. These fields are transformed under chiral transformations according to the mirror assignment \cite{Detar:1988kn}.

The Lagrangian of the model is the sum of two contributions, ${\cal L}_{\rm F}$ and ${\cal L}_{\rm B}$, representing  respectively the fermionic and bosonic parts of the total Lagrangian. The fermion Lagrangian takes the form
\begin{widetext}
\beq\label{eq:actionF}
{\cal L}_{\rm F}=\left(\bar{\psi}_{a}\ \bar{\psi}_{b}\right)
\left(\begin{array}{cc}
  \gamma^{\mu}(i\partial_{\mu} -{\cal V}_\mu)
	- y_{a} \left(\sigma + i \gamma^5 \bm{\pi} \cdot \bm{\tau} \right) & -m_0 \gamma^5 \\[0.1cm]
  m_0 \gamma^5& \gamma^{\mu}(i\partial_{\mu}-{\cal V}_{\mu}) 
	- y_{b} \left(\sigma - i \gamma^5 \bm{\pi} \cdot \bm{\tau} \right)
\end{array}\right)
\left( \begin{array}{c}
 {\psi}_{a}\\[0.1cm]
{\psi}_{b}
\end{array}
  \right) ,
\eeq
\end{widetext}
where 
\beq
{\cal V}_\mu=g_v \omega_\mu +g_w \bm{\rho}_\mu \cdot \bm{\tau},
\eeq
and $\bm{\tau} = (\tau_{1}, \tau_{2}, \tau_{3})$ denote the three Pauli matrices. Aside from the kinetic term, and the non-diagonal mass term, this Lagrangian exhibits the coupling of the fermions to the mesonic fields
$\sigma$,  $\bm{\pi}$, $\omega_\mu$, and $\bm{\rho}_\mu$. In this paper, these mesonic fields will be treated in the classical approximation (i.e.\ as classical background fields for the fermions), and their  Lagrangian will be specified below. Let us just note at this point that in the states to be considered, which are assumed to be both rotationally and parity invariant,  only the sigma field $\sigma$ and the zeroth components of the vector fields, denoted $\omega$ and $\bm{\rho}$, acquire a classical value. From now on we shall therefore set $\bm{\pi}=0$.  The choice of  unique coupling strengths $g_{v}$ and $g_w$   of the vector mesons to both 
$\psi_{a}$ and $\psi_{b}$ is convenient and in line with previous works on the subject (see e.g.\ Refs.~\cite{Motohiro:2015taa, Eser:2023oii}).  The  Yukawa couplings $y_a$ and $y_b$  between the baryons and the chiral fields are distinct and their values will be fixed by physical constraints. 

The physical fermion states $\psi_{+}$ (positive parity) and $\psi_{-}$ (negative parity) are obtained as linear combinations of states with the same parity, e.g.\ $\psi_a$ and $\gamma^5 \psi_b$. The coefficients of these linear superpositions are determined by diagonalizing the mass matrix. The masses of $\psi_{+}$ and $\psi_{-}$, respectively $M_+$ and $M_-$, are given by
\begin{equation}	M_{\pm} = \frac{1}{2} \left[\pm \sigma \left(y_{a} - y_{b}\right)
	+ \sqrt{\sigma^{2} (y_{a} + y_{b})^{2} + 4 m_{0}^{2}}\right].
	\label{eq:massphys}
\end{equation}
The masses of the members of the parity doublet are degenerate at $m_{0}$ for a vanishing $\sigma$ field, and the mass $M_{+}$ exhibits a minimum at the value $\sigma_{\rm min}$, with
\beq\label{eq:sigmamin}
\sigma^2_{\rm min}=\frac{m_0^2 (y_a-y_b)^2}{y_a y_b (y_a+y_b)^2}.
\eeq
As we shall see in Sec.~\ref{sec:more_insight} and Appendix~\ref{sec:chiral_limit}, the existence of this minimum has a significant impact on the chiral transition in the parity-doublet model.
The fields $\psi_+$ and $\psi_-$ are the physical states that we associate respectively to the nucleon $N(939)$ and its parity partner $N^*(1535)$, at this level of approximation. The fermions are mass-degenerate isospin doublets,
\begin{equation}
    \psi_{+} = \begin{pmatrix} \mathfrak{p} \\ \mathfrak{n} \end{pmatrix}, \qquad
    \psi_{-} = \begin{pmatrix} \mathfrak{p}^{\ast} \\ \mathfrak{n}^{\ast} \end{pmatrix} ,
\end{equation}
with the proton $\mathfrak{p}$, the neutron $\mathfrak{n}$, and their respective chiral partners $\mathfrak{p}^{\ast}$ and $\mathfrak{n}^{\ast}$.

In this physical basis the  two fields $\psi_+$ and $\psi_-$ formally decouple and  the fermionic Lagrangian can then be written as 
\begin{equation}
    {\cal L}_{\rm F}= {\cal L}_{\rm F}^+ + {\cal L}_{\rm F}^-, 
\end{equation}
with 
 \beq
  {\cal L}_{\rm F}^\pm=\bar{\psi}_{\pm}  \left[i\gamma^{\mu}\partial_{\mu}- \gamma^{0}\left(g_v\omega+g_w \bm{\rho} \cdot \bm{\tau}\right) - M_\pm \right]{\psi}_{\pm} , \quad
  \eeq
  whose spectrum is given by 
  \beq
 ( E^{\tau\pm}_\p-g_v\omega-g_w \tau\rho)^2= \p^2+M_\pm^2 ,
  \eeq
  with $\rho$ the third isospin component of $\bm\rho$, and $\p$ the three-momentum. The variable and superscript $\tau$ in $E^{\tau\pm}_\p$ is $\tau=+1$ for a proton $\mathfrak{p}$ or its chiral partner $\mathfrak{p}^*$, and $\tau=-1$  for the neutron $\mathfrak{n}$ and its chiral partner $\mathfrak{n}^*$.
The $\sigma$ field modifies the masses $M_\pm$ while the vector fields produce  constant shifts (i.e.\ independent of the momentum) of the single particle energies (opposite for particles and antiparticles). To make things clearer, we set $\varepsilon^{\pm}_\p=(\p^2+M_\pm^2)^{1/2}$. The energies $E^{\tau\pm}_\p$ of the particles and $\bar{E}^{\tau\pm}_\p$ of the antiparticles are then given respectively by 
\begin{IEEEeqnarray}{rCl}\label{eq:spenergy}
E^{\tau\pm}_\p & = & \varepsilon^\pm_\p+g_v\omega+g_w\tau\rho,\\[0.1cm]
\bar E^{\tau\pm}_\p & = & \varepsilon^\pm_\p-g_v\omega-g_w\tau\rho.
\end{IEEEeqnarray}

We readily recover the singlet model by dropping the 
parity-odd fermion $\psi_{b}$ in Eq.~(\ref{eq:actionF}) and setting $m_{0} =0$. We may consequently identify $\psi_{+} \equiv \psi_{a}$. The nucleon mass then reduces to $M_{+} = y_{a}\sigma \equiv y\sigma = M$. 

We now complete the discussion of the Lagrangian of the model, by specifying its 
 mesonic part ${\cal L}_{\rm B}$. Since we treat the meson fields in the classical approximation, only the potential terms of ${\cal L}_{\rm B}$ are relevant. We take the same mass for both vector fields for simplicity, and set
\beq\label{eq:mesonL}
{\cal L}_{\rm B}= -V\!\left(\varphi^{2}\right) +h\left(\sigma - f_{\pi}\right)+ \frac{1}{2} m_v^2 \left( \omega^2+\rho^2 \right),
\eeq
where $\varphi^2\equiv\sigma^2+ \bm{\pi}^2$. 
Following previous works \cite{Floerchinger_2012,brandes_fluctuations_2021, Eser:2023oii}, we express the potential $V(\varphi^{2})$ as a fourth-order polynomial in $\varphi^2-f_{\pi}^{2}$,
\begin{equation}
	V\!\left(\varphi^{2}\right) = \sum_{n\, =\, 1}^{4} 
	\frac{\alpha_{n}}{2^{n} n!} \left(\varphi^{2} - f_{\pi}^{2}\right)^{n},
	\label{eq:potential}
\end{equation}
where $f_{\pi}$ the pion decay constant in vacuum, and $\alpha_{n}$ is referred to as a Taylor coefficient. The term $h\sigma$ accounts for the explicit symmetry breaking in the direction of the $\sigma$ field. It confers the pion a finite mass, corresponding to $h = m_{\pi}^{2} f_{\pi}$ for the physical pion mass $m_{\pi}$. It also prevents chiral symmetry to be strictly restored at high temperature and density. We shall also frequently refer to the ``chiral limit,'' which is obtained by letting $h\to 0$, while keeping all other parameters fixed.

The mean-field approximation that is used in this paper consists in treating the mesonic fields as classical fields, while keeping the fermion fluctuations to order one-loop. These fermion fluctuations are functions of the $\sigma$ field, and contribute therefore to the mesonic effective potential, in particular via logarithmic corrections. We call
\begin{equation}
    U(\sigma,\omega,\rho)=U(\sigma)-\frac{1}{2} m_v^2 (\omega^2+\rho^2)
\end{equation}
the full resulting effective potential in vacuum, noticing that  the classical vector fields are non vanishing only in the presence of matter (see below).  Note that in contrast to previous works (see e.g.\ Ref.~\cite{Zschiesche:2006zj}) we do not include in the meson Lagrangian any self-interactions of the vector fields.

The renormalisation conditions of the fermionic one-loop contribution are chosen such that the first and second derivatives of $U$ with respect to $\sigma^{2}$ vanish in vacuum, i.e.\ at $\sigma^{2} = f_{\pi}^{2}$. A detailed discussion of the renormalisation procedure was given in (I), where the explicit expression of the potential $U$ can be found.\footnote{The only modification compared to (I) is the substitution, in the final expression of the potential $U$, $\frac{1}{2} m_v^2 \omega^2 \mapsto \frac{1}{2}m_v^2 (\omega^2+ \rho^2)$, that takes into account the contribution of the vector-isovector field $\rho$.}

\subsection{Thermodynamics}

In this paper, we explore the properties of equilibrium states as a function of the baryon density and the neutron excess, eventually reaching pure neutron matter. To do so, we introduce a baryon-chemical potential $\mu_{\mathcal B}$ coupled to the baryon density $n_{\mathcal{B}}$:
\beq
 n_{\mathcal{B}}&=& \left\langle\bar{\psi}_{a}\gamma^{0}\psi_{a} 
	+ \bar{\psi}_{b}\gamma^{0}\psi_{b} \right\rangle\nn[0.1cm]
	&=&
	\left\langle\bar{\psi}_{+}\gamma^{0}\psi_{+} 
	+ \bar{\psi}_{-}\gamma^{0}\psi_{-} \right\rangle .
	\eeq
In addition, we control the neutron excess by an isospin-chemical potential $\mu_{\mathcal{I}}$ coupled to the isospin density $n_{\mathcal{I}}$:
	\beq
	n_{\mathcal{I}} & = & \left\langle
	\bar{\psi}_{a}\gamma^{0} \tau_{3} \psi_{a} 
	+ \bar{\psi}_{b}\gamma^{0} \tau_{3} \psi_{b} \right\rangle \nn[0.1cm]
	& = & \left\langle \bar{\psi}_{+}\gamma^{0} \tau_{3} \psi_{+} 
	+ \bar{\psi}_{-}\gamma^{0} \tau_{3} \psi_{-} \right\rangle .
	\eeq
Defining $n_{\mathfrak{p}}$ the density of protons, $n_{\mathfrak{n}}$ that of neutrons, and $n_{\mathfrak{p}^{\ast}}$ as well as $n_{\mathfrak{n}^{\ast}}$ the densities of their respective chiral partners, we can rewrite  $n_{\mathcal{B}}$ and $n_{\mathcal{I}}$ as 
\begin{IEEEeqnarray}{rCl}
	n_{\mathcal{B}}  & = & n_{\mathfrak{p}} + n_{\mathfrak{n}} 
	+ n_{\mathfrak{p}^{\ast}} + n_{\mathfrak{n}^{\ast}}  \equiv n_{\mathsf{P}} + n_{\mathsf{N}}, \\[0.1cm]
	n_{\mathcal{I}} & = & n_{\mathfrak{p}} - n_{\mathfrak{n}} 
	+ n_{\mathfrak{p}^{\ast}} - n_{\mathfrak{n}^{\ast}} \equiv n_{\mathsf{P}} - n_{\mathsf{N}},
\end{IEEEeqnarray}
where
\begin{equation}
    n_{\mathsf{P}} = n_{\mathfrak{p}} + n_{\mathfrak{p}^{\ast}} , \qquad
    n_{\mathsf{N}} = n_{\mathfrak{n}} + n_{\mathfrak{n}^{\ast}} . 
\end{equation}
Thus the numbers $n_{\mathsf{P}} $ and $ n_{\mathsf{N}}$ quantify the total amount of particles with isospin $+ \frac{1}{2}$ and those with isospin $- \frac{1}{2}$, respectively. Note that
\begin{equation}
	\mu_{\mathcal{B}}n_ \mathcal{B} + \mu_{\mathcal{I}} n_\mathcal{I}
	\equiv \mu_{\mathsf{P}} n_{\mathsf{P}} + \mu_{\mathsf{N}} n_{\mathsf{N}},
\end{equation}
where we introduced the chemical potentials
\begin{equation}
    \mu_{\mathsf{P},\mathsf{N}} = \mu_{\mathcal{B}} \pm \mu_{\mathcal{I}},
\end{equation}
which we refer to as the proton and neutron-chemical potentials throughout this work.
	
The grand canonical potential density $\Omega$ contains, in addition to the vacuum contribution $U$, a matter contribution. The latter is the contribution of independent fermion quasiparticles whose energies depend on the mesonic fields. We have
\begin{equation}\label{eq:OmegaUBMat}
	\Omega = U -2T\sum_{i,r,\tau\,=\,\pm 1} \int_{\p} \ln\left[ 1+\rme^{-\beta \left(\varepsilon^i_\p-r\tilde \mu_{\tau}\right)}  \right], \,
\end{equation}
where the index $i$ runs over the two  parity states ($i = \pm 1$), $r$ refers to particles and antiparticles ($r = \pm 1$), and $\tau$ refers to isospin ($\tau = \pm 1$). The overall factor 2 accounts for the sum over spin. The momentum integration is denoted as
\begin{equation}
    \int_{\p} = \int \frac{\rmd^{3} p}{(2\pi)^{3}}.
\end{equation}
The chemical potential $\tilde \mu_{\tau}$ (with $\tau = +1$ meaning proton-like and $\tau = -1$ meaning neutron-like, as already used in the definition of the energy spectrum) represents shifted chemical potentials defined as
\beq
 \tilde {\mu}_{\mathsf{P},\mathsf{N}} = \mu_{\mathsf{P},\mathsf{N}} - g_{v} \omega \mp g_{w} \rho.
 \eeq
 In the equation (\ref{eq:OmegaUBMat}) above, the term proportional to $T$ has a finite limit as $T\to 0$, equal to  $\tilde{{\cal E}}_{\rm qp}-\mu_{\mathcal B} n_{\mathcal B}-\mu_{\mathcal I} n_{\mathcal I}$, where $\tilde{{\cal E}}_{\rm qp}$ denotes the quasiparticle contribution to the energy density, the total energy density being ${\cal E}=U+\tilde{{\cal E}}_{\rm qp}$. The quasiparticle contribution reads
\begin{equation}\label{eq:tildesepsilon}
    \tilde{\mathcal{E}}_{\mathrm{qp}} = 2 \sum_{i,\tau\, =\, \pm 1} \int_{\p} \Theta(\tilde{\mu}_{\tau} - \varepsilon_{\p}^{i}) E_{\p}^{\tau i} ,
\end{equation}
and with $\mathcal{E}_{\mathrm{qp}}$ we shall denote the closely related formula without the vector contribution. That is, $\mathcal{E}_{\mathrm{qp}}$ is obtained by substituting $E_{\p}^{\tau i} \mapsto \varepsilon_{\p}^{i}$ in Eq.~(\ref{eq:tildesepsilon}).
 
The  grand canonical potential density $\Omega$ is a function of the chemical potentials $\mu_{\mathcal{B}}$ and $\mu_{\mathcal{I}}$, and of the temperature $T$. In addition, it depends on the values of the classical fields $\sigma$,  $\omega$, and $\rho$. These constant classical fields are to be considered as internal variables that need to be determined, for given $T$ and $\mu\equiv\{\mu_{\mathcal{B}},\mu_{\mathcal{I}}\}$ by the requirement that $\Omega$ be stationary with respect to their variations. This leads to the equations 
\begin{equation}\label{Omegamin}
	\left. \frac{\partial\Omega}{\partial\omega}\right|_{\mu,T;\, \sigma,\rho}=\left. \frac{\partial\Omega}{\partial\rho}\right|_{\mu,T;\, \sigma,\omega}=0, \quad \left.\frac{\partial\Omega}{\partial\sigma}\right|_{\mu,T;\, \omega,\rho} = 0.
\end{equation}
The first two equations (\ref{Omegamin}) are  essentially the equations of motion for the fields $\omega$ and $\rho$ in the classical approximation where all derivatives of the field vanish:
\begin{IEEEeqnarray}{rCl}
g_v\omega & = & G_v n_{\mathcal B} ,\qquad G_v\equiv \frac{g_v^2}{m_v^2}, \label{eq:Gv} \\[0.1cm]
g_w\rho & = & G_w n_{\mathcal I} ,\qquad G_w\equiv \frac{g_w^2}{m_v^2}. \label{eq:Gw}
\end{IEEEeqnarray}
These equations  relate the  fields $\omega$ and $\rho$ to their sources,  respectively the baryon density $n_{\mathcal{B}}$ and the isospin density $n_{\mathcal{I}}$.

The third equation (\ref{Omegamin}) is akin to a gap equation. It  can be written as 
\beq\label{eq:gap}
	&&\left.\frac{\partial U}{\partial \sigma} \right|_{\omega,\rho}
=- y_{+} \,n_{\rm s}^{+} 
	- y_{-}\,n_{\rm s}^{-},\label{eq:MF_eq_sigma}
	\eeq
where 	the scalar densities are given by 
\begin{multline}
	n_{\rm s}^{\pm} = \left.\frac{\partial\Omega}{\partial M_{\pm}} \right|_{\mu,T;\, \omega,\rho} \\[0.2cm]
	= 2 \sum_{r,\tau\, =\, \pm 1} \int_{\p}  \frac{M_{\pm}}{\varepsilon^{\pm}_\p}
	n_{\rm F}(\varepsilon^{\pm}_\p - r\tilde{\mu}_{\tau}) ,  \quad
\end{multline}
where  $n_{\rm F}$ is  the Fermi-Dirac distribution
\beq
	n_{\rm F}( \varepsilon_\p ) = \left(\rme^{\beta  \varepsilon_\p} + 1 \right)^{-1}.
	\eeq
In writing Eq.~(\ref{eq:MF_eq_sigma})  we have set 
\begin{equation}\label{eq:derivsgimaterm}
	\frac{\mathrm{d}M_{\pm}(\sigma)}{\mathrm{d}\sigma} \equiv y_{\pm}(\sigma),\end{equation}
with
\begin{equation}
	y_{\pm} = \frac{1}{2} \left[\pm \left(y_{a} - y_{b}\right)
	+ \frac{\sigma (y_{a} + y_{b})^{2}}
	{\rule[0ex]{0ex}{2ex}\sqrt{\rule[0ex]{0ex}{2ex}\sigma^{2} (y_{a} + y_{b})^{2} 
	+ 4 m_{0}^{2}}}\right].
	\label{eq:yukawa_phys}
\end{equation}
Let us now finally recall that the value of $\Omega$ calculated with the fields $\omega$, $\rho$, and $\sigma$ that solve equations (\ref{Omegamin}) is equal to $-P(\mu_{\mathcal{B}}, \mu_{\mathcal{I}},T)$, where $P$ is the thermodynamic equilibrium pressure.

As we have seen, the baryon density and the isospin density can be expressed in terms of the densities
\beq
n_{\mathsf{P},\mathsf{N}} = - \frac{\partial\Omega}{\partial\mu_{\mathsf{P},\mathsf{N}}} 
	= \begin{cases} n_{\mathfrak{p}} + n_{\mathfrak{p}^{\ast}} \\[0.1cm]
	n_{\mathfrak{n}} + n_{\mathfrak{n}^{\ast}} \end{cases} \hspace{-0.3cm} .
\eeq
We have
\beq \label{nBpm}
n_{\mathcal B}  
	=2   \sum_{i,r,\tau\, =\, \pm 1} r \int_{\p}  n_{\rm F}( \varepsilon^{i}_\p - r\tilde{\mu}_{\tau}) ,
\eeq
which also splits into the densities of positive-parity baryons and negative-parity baryons, 
\begin{IEEEeqnarray}{rCl}\label{eq:nBpm}
    n_{\mathcal{B}} & = & n_{\mathsf{P}} + n_{\mathsf{N}} = n_{\mathcal{B}}^{+} + n_{\mathcal{B}}^{-}, \\[0.1cm]
    n_{\mathcal{B}}^{\pm} & = & \begin{cases} n_{\mathfrak{p}} + n_{\mathfrak{n}} \\[0.1cm] 
	n_{\mathfrak{p}^{\ast}} + n_{\mathfrak{n}^{\ast}} \end{cases} \hspace{-0.3cm} .
\end{IEEEeqnarray}
Note that only the  baryon density $n_{\mathcal B}$ and the isospin density $n_{\mathcal I}$ are controlled by the chemical potentials $\mu_{\mathcal{B}}$ and $\mu_{\mathcal{I}}$. However, in the present approximation, these densities naturally split into the separate contributions coming from each species of baryons, with the relative sizes of each contribution being determined by the different single particle energies and the corresponding chemical potentials.

At $T = 0$, the densities of the different fermion species are given by the following momentum integrals:
\begin{IEEEeqnarray}{rCl}
    n_{\mathfrak{p}, \mathfrak{p}^{\ast}} & = & 2 \int_{\p} \Theta\!\left(\mu_{\mathsf{P}} - E_{\p}^{+\pm}\right), \label{eq:defdensities1} \\[0.2cm]
    n_{\mathfrak{n}, \mathfrak{n}^{\ast}} & = & 2 \int_{\p} \Theta\!\left(\mu_{\mathsf{N}} - E_{\p}^{-\pm}\right). \label{eq:defdensities2}
\end{IEEEeqnarray}
The derivatives of any of these densities with respect to the corresponding chemical potentials play an important role in the forthcoming discussions. As an example, consider $\partial n_{\mathfrak{p}}/\partial \mu_{\mathsf{P}}$:
\begin{IEEEeqnarray}{rCl}\label{eq:delndelmu}
    \frac{\partial n_{\mathfrak{p}}}{\partial \mu_{\mathsf{P}}} & = & 2 \int_{\p} \delta\!\left(\mu_{\mathsf{P}} - E_{\p}^{++}\right) \nonumber\\[0.1cm]
    & & \qquad \times \left(1 - \frac{\partial E_{\p}^{++}}{\partial n_{\mathcal{B}}} \frac{\partial n_{\mathcal{B}}}{\partial \mu_{\mathsf{P}}} - \frac{\partial E_{\p}^{++}}{\partial n_{\mathcal{I}}} \frac{\partial n_{\mathcal{I}}}{\partial \mu_{\mathsf{P}}}\right) \nonumber\\[0.2cm]
    & \equiv & N_{0}^{\mathfrak{p}} \left(1 - f_{0}^{\mathfrak{p}} \frac{\partial n_{\mathcal{B}}}{\partial \mu_{\mathsf{P}}} - f_{0}^{\prime\mathfrak{p}} \frac{\partial n_{\mathcal{I}}}{\partial \mu_{\mathsf{P}}}\right),
\end{IEEEeqnarray}
where $N_{0}^{\mathfrak{p}}$ the density of proton single particle states at the corresponding Fermi surface,
\begin{equation}
    N_{0}^{\mathfrak{p}} = 2 \int_{\p} \delta\!\left(\mu_{\mathsf{P}} - E_{\p}^{++}\right) = \frac{p_{\mathfrak{p}} M_{\ast}^{\mathfrak{p}}}{\pi^{2}} , \label{eq:denstate}
\end{equation}
with $p_{\mathfrak{p}} = (3\pi^{2} n_{\mathfrak{p}})^{1/3}$ the proton Fermi momentum. In Eq.~(\ref{eq:delndelmu}), $f_{0}^{\mathfrak{p}}$ and $f_{0}^{\prime\mathfrak{p}}$ are Fermi-liquid parameters defined as follows (see e.g.\ Ref.~\cite{Friman:2019ncm}):
\begin{IEEEeqnarray}{rCl}
    f_{0}^{\mathfrak{p}} & = & \left.\frac{\partial E_{\p}^{++}}{\partial n_{\mathcal{B}}}\right|_{|\p|\, =\, p_{\mathfrak{p}}} 
    = G_{v} + y_{+} \frac{M_{+}}{M_{\ast}^{\mathfrak{p}}} \frac{\partial\sigma}{\partial n_{\mathcal{B}}}, \label{eq:f0} \\[0.2cm]
    f_{0}^{\prime\mathfrak{p}} & = & \left.\frac{\partial E_{\p}^{++}}{\partial n_{\mathcal{I}}}\right|_{|\p|\, =\, p_{\mathfrak{p}}}
    = G_{w} + y_{+} \frac{M_{+}}{M_{\ast}^{\mathfrak{p}}} \frac{\partial\sigma}{\partial n_{\mathcal{I}}} . \label{eq:f0prime}
\end{IEEEeqnarray}
The Landau effective mass $M_{\ast}^{\mathfrak{p}}$ in Eq.~(\ref{eq:denstate}) is defined by the relation
\begin{equation}
    \left.\frac{\partial E_{\p}^{++}}{\partial p}\right|_{p_{\mathfrak{p}}} = \frac{p_{\mathfrak{p}}}{M_{\ast}^{\mathfrak{p}}}.
\end{equation}
It allows us to express the chemical potentials at zero temperature in the following way:
\begin{IEEEeqnarray}{rCl}
    \mu_{\mathcal{B}} & = & \frac{1}{2} \left(M_{\ast}^{\mathsf{P}} + M_{\ast}^{\mathsf{N}}\right) + G_{v} n_{\mathcal{B}}, \label{eq:muB} \\[0.2cm]
    \mu_{\mathcal{I}} & = & \frac{1}{2} \left(M_{\ast}^{\mathsf{P}} - M_{\ast}^{\mathsf{N}}\right) + G_{w} n_{\mathcal{I}}, \label{eq:muI}
\end{IEEEeqnarray}
where
\begin{IEEEeqnarray}{rCl}
    M_{\ast}^{\mathsf{P}} & \equiv & M_{\ast}^{\mathfrak{p}} = \sqrt{p_{\mathfrak{p}}^{2} + M_{+}^{2}}
    = M_{\ast}^{\mathfrak{p}^{\ast}} = \sqrt{p_{\mathfrak{p}^{\ast}}^{2} + M_{-}^{2}} , \qquad \label{eq:constr1} \\[0.2cm]
    M_{\ast}^{\mathsf{N}} & \equiv & M_{\ast}^{\mathfrak{n}} = \sqrt{p_{\mathfrak{n}}^{2} + M_{+}^{2}}
    = M_{\ast}^{\mathfrak{n}^{\ast}} = \sqrt{p_{\mathfrak{n}^{\ast}}^{2} + M_{-}^{2}} . \label{eq:constr2}
\end{IEEEeqnarray}
The equalities above come from the fact that both the proton and neutron densities and respectively the densities of the chiral partners are controlled by a single chemical potential, either $\mu_{\mathsf{P}}$ or $\mu_{\mathsf{N}}$, e.g.\
\begin{equation}
    \frac{\partial\mathcal{E}}{\partial n_{\mathfrak{p}}} \equiv \frac{\partial\mathcal{E}}{\partial n_{\mathfrak{p}^{\ast}}} \equiv \frac{\partial\mathcal{E}}{\partial n_{\mathsf{P}}} = \mu_{\mathsf{P}}.
\end{equation}
Note that the equalities above only hold in the presence of the chiral partners, i.e.\ $p_{\mathfrak{p}^{\ast}} > 0$ or $p_{\mathfrak{n}^{\ast}} > 0$, respectively. In turn, for $p_{\mathfrak{p}^{\ast}} = 0$ or $p_{\mathfrak{n}^{\ast}} = 0$, these equalities define the thresholds
\begin{equation}
    M_{\ast}^{\mathfrak{p}, \mathfrak{n}} = M_{-}, \label{eq:thresholds}
\end{equation}
that need to be fulfilled for the onset of populating the respective chiral partners of protons and neutrons.

For the cross derivative $\partial n_{\mathfrak{p}}/\partial \mu_{\mathsf{N}}$ we analogously get
\begin{equation}
    \frac{\partial n_{\mathfrak{p}}}{\partial \mu_{\mathsf{N}}} = - N_{0}^{\mathfrak{p}} \left(f_{0}^{\mathfrak{p}} \frac{\partial n_{\mathcal{B}}}{\partial \mu_{\mathsf{N}}} + f_{0}^{\prime\mathfrak{p}} \frac{\partial n_{\mathcal{I}}}{\partial \mu_{\mathsf{N}}}\right) .
\end{equation}
Using these derivatives the corresponding derivative of the total baryon density $n_{\mathcal{B}}$ with respect to the baryon-chemical potential $\mu_{\mathcal{B}}$ is calculated as
\begin{IEEEeqnarray}{rCl}
    \frac{\partial n_{\mathcal{B}}}{\partial \mu_{\mathcal{B}}} & = & \sum_{i\, \in\, \lbrace \mathfrak{p}, \mathfrak{n}, \mathfrak{p}^{\ast}, \mathfrak{n}^{\ast} \rbrace} \left(\frac{\partial}{\partial \mu_{\mathsf{P}}} + \frac{\partial}{\partial \mu_{\mathsf{N}}} \right) n_{i} \nonumber\\[0.1cm]
    & = & \sum_{i} N_{0}^{i} \left(1 - f_{0}^{i} \frac{\partial n_{\mathcal{B}}}{\partial \mu_{\mathcal{B}}} - f_{0}^{\prime i} \frac{\partial n_{\mathcal{I}}}{\partial \mu_{\mathcal{B}}}\right) , \label{eq:dnBdnmu}
\end{IEEEeqnarray}
with the corresponding densities of states (using Eqs.~(\ref{eq:constr1}) and (\ref{eq:constr2}))
\begin{IEEEeqnarray}{rCl}
    & & N_{0}^{\mathfrak{p}} = \frac{p_{\mathfrak{p}} M_{\ast}^{\mathsf{P}}}{\pi^{2}}, \qquad
    N_{0}^{\mathfrak{p}^{\ast}} = \frac{p_{\mathfrak{p}^{\ast}} M_{\ast}^{\mathsf{P}}}{\pi^{2}} , \\[0.2cm]
    & & N_{0}^{\mathfrak{n}} = \frac{p_{\mathfrak{n}} M_{\ast}^{\mathsf{N}}}{\pi^{2}}, \qquad
    N_{0}^{\mathfrak{n}^{\ast}} = \frac{p_{\mathfrak{n}^{\ast}} M_{\ast}^{\mathsf{N}}}{\pi^{2}} ,
\end{IEEEeqnarray}
as well as definitions of $f_{0}^{i}$ and $f_{0}^{\prime i}$ analogous to Eqs.~(\ref{eq:f0}) and (\ref{eq:f0prime}). We may now similarly express the derivatives of the baryon and isospin-chemical potentials (\ref{eq:muB}) and (\ref{eq:muI}) with respect to $n_{\mathcal{B}}$ and $n_{\mathcal{I}}$ as follows:
\begin{IEEEeqnarray}{rCl}
    \frac{\partial \mu_{\mathcal{B}}}{\partial n_{\mathcal{B}}} & = & \frac{1}{2} \left(\frac{1}{N_{0}^{\mathfrak{p}}} \frac{\partial n_{\mathfrak{p}}}{\partial n_{\mathcal{B}}} + \frac{1}{N_{0}^{\mathfrak{n}}} \frac{\partial n_{\mathfrak{n}}}{\partial n_{\mathcal{B}}} \right) + \frac{1}{2} \left( f_{0}^{\mathfrak{p}} + f_{0}^{\mathfrak{n}} \right) , \label{eq:dmuBdnB} \\[0.2cm]
    \frac{\partial\mu_{\mathcal{I}}}{\partial n_{\mathcal{I}}} & = & \frac{1}{2} \left(\frac{1}{N_{0}^{\mathfrak{p}}} \frac{\partial n_{\mathfrak{p}}}{\partial n_{\mathcal{I}}} - \frac{1}{N_{0}^{\mathfrak{n}}} \frac{\partial n_{\mathfrak{n}}}{\partial n_{\mathcal{I}}} \right) + \frac{1}{2} \left( f_{0}^{\prime \mathfrak{p}} - f_{0}^{\prime \mathfrak{n}} \right) . \label{eq:dmuIdnI} \qquad
\end{IEEEeqnarray}
In the absence of the chiral partners, $n_{\mathfrak{p}^{\ast}} = n_{\mathfrak{n}^{\ast}} = 0$, the derivatives above simplify as
\begin{IEEEeqnarray}{rCl}
    \frac{\partial \mu_{\mathcal{B}}}{\partial n_{\mathcal{B}}} & = & \frac{1}{4} \left(\frac{1}{N_{0}^{\mathfrak{p}}} + \frac{1}{N_{0}^{\mathfrak{n}}} \right) + \frac{1}{2} \left( f_{0}^{\mathfrak{p}} + f_{0}^{\mathfrak{n}} \right) , \label{eq:dmuBdnBsimp} \\[0.2cm]
    \frac{\partial\mu_{\mathcal{I}}}{\partial n_{\mathcal{I}}} & = & \frac{1}{4} \left(\frac{1}{N_{0}^{\mathfrak{p}}} + \frac{1}{N_{0}^{\mathfrak{n}}} \right) + \frac{1}{2} \left( f_{0}^{\prime \mathfrak{p}} - f_{0}^{\prime \mathfrak{n}} \right) , \label{eq:dmuIdnIsimp}
\end{IEEEeqnarray}
where we used that $n_{\mathfrak{p},\mathfrak{n}} = (n_{\mathcal{B}} \pm n_{\mathcal{I}})/2$ in this particular case. These equations are also valid in the corresponding singlet model.

In the case of symmetric matter ($n_{\mathcal{I}} \equiv 0$) the derivative (\ref{eq:dnBdnmu}) can be written as
\begin{equation}
    \frac{\rmd n_{\mathcal{B}}}{\rmd \mu_{\mathcal{B}}} = \frac{N_{0}}{1 + F_{0}} ,
\end{equation}
where $F_{0} = N_{0}^{+} f_{0}^{+} + N_{0}^{-} f_{0}^{-}$ a Landau Fermi-liquid parameter, and
\begin{IEEEeqnarray}{rCl}
    N_{0} & = & N_{0}^{+} + N_{0}^{-}, \qquad N_{0}^{\pm} = \frac{2 p_{\pm} M_{\ast}}{\pi^{2}}, \\[0.2cm]
    f_{0}^{\pm} & = & G_{v} + y_{\pm} \frac{M_{\pm}}{M_{\ast}} \frac{\rmd\sigma}{\rmd n_{\mathcal{B}}},
\end{IEEEeqnarray}
with $\pm$ referring to the totals of positive and negative-parity baryons (compare again the detailed discussion of symmetric matter presented in (I)). Note that the Landau effective mass $M_{\ast}$ is now the same for all particle species, if the respective densities are nonzero. Evaluating the derivatives (\ref{eq:dmuBdnBsimp}) and (\ref{eq:dmuIdnIsimp}) at the symmetric point we consistently get
\begin{IEEEeqnarray}{rCl}
    \left.\frac{\partial \mu_{\mathcal{B}}}{\partial n_{\mathcal{B}}}\right|_{n_{\mathcal{I}}\, =\, 0} & = & \frac{1}{N_{0}^{+}} + \frac{1}{2} \left( f_{0}^{\mathfrak{p}} + f_{0}^{\mathfrak{n}} \right) 
    \equiv \frac{1 + F_{0}}{N_{0}} , \\[0.2cm]
    \left.\frac{\partial \mu_{\mathcal{I}}}{\partial n_{\mathcal{I}}}\right|_{n_{\mathcal{I}}\, =\, 0} & = & \frac{1}{N_{0}^{+}} + \frac{1}{2} \left( f_{0}^{\prime \mathfrak{p}} - f_{0}^{\prime \mathfrak{n}} \right) 
    \equiv \frac{1}{N_{0}} + G_{w}, \qquad \label{eq:dmuIdnIsimpsym}
\end{IEEEeqnarray}
where the last equation will enter the upcoming discussion of the symmetry energy, while the former essentially determines the compression modulus \cite{Eser:2023oii}.

Another important quantity is the ratio
\begin{equation}\label{eq:protonfrac}
    x = \frac{n_{\mathsf{P}}}{n_{\mathcal{B}}} = \frac{n_{\mathfrak{p}} + n_{\mathfrak{p}^{\ast}}}{n_{\mathcal{B}}}
    = 1 - \frac{n_{\mathfrak{n}} + n_{\mathfrak{n}^{\ast}}}{n_{\mathcal{B}}}
    = 1 - \frac{n_{\mathsf{N}}}{n_{\mathcal{B}}},
\end{equation}
which describes the fraction of the proton-like population. We shall simply call $x$ the ``proton fraction,'' keeping in mind that it may also involve the chiral partner of the proton. Neutron excess in the system is then realized for $x < \frac{1}{2}$.

To compute the thermodynamics of the parity-doublet model, and of the corresponding singlet model, in the special case of $T = 0$,  we solve a system of coupled differential equations based on derivatives with respect to the baryon density $n_{\mathcal{B}}$, for fixed proton fraction $x$. These differential equations are derived from (thus equivalent to) the gap equation (\ref{eq:gap}) and the thresholds (\ref{eq:thresholds}), see Appendix~\ref{sec:diffeqs} for a detailed discussion. From the integrated solutions we finally ``cut out'' unphysical parts, yielding then some of the physical results presented here in the main text.

\section{Determination of parameters}
\label{sec:paramters}

The parameters of the model are determined as in (I). We assume that the densities of the chiral partners $\mathfrak{p}^{\ast}$ and $\mathfrak{n}^{\ast}$ vanish at small to moderate values of the total baryon density $n_{\mathcal{B}}$, that is, we set $n_{\mathcal{B}}^-=0$ (see Eq.~(\ref{eq:nBpm})). We use the physical input parameters listed in Table \ref{tab:parameters}. The other model parameters are then derived from these values. Only the isoscalar mass $m_{\sigma}$, the Landau effective mass $M_{\ast}$ (for symmetric matter), and the chiral-invariant mass $m_{0}$ are initially undetermined by these input parameters. The values of $m_{\sigma}$, $m_{0}$, and $M_{\ast}$ (which is related to the vector coupling $G_{v}$) are eventually adjusted so as to reproduce (in particular) the critical endpoint of the liquid-gas transition, as outlined in (I) in great detail. Moreover, the choice of parameters adopted in (I) yields reasonable values for the compression modulus and the nuclear surface tension. 
\begin{table}[ht]
	\caption{\label{tab:parameters}Input parameters for the initialization of
	the model \cite{ParticleDataGroup:2022pth, Eser:2023oii}.}
	\begin{ruledtabular}
		\begin{tabular}{lr}
		Parameter & Numerical value\\
		\colrule\\[-0.25cm]
        Pion decay constant $f_{\pi}\ [\mathrm{MeV}]$ & $93$ \\[0.05cm]
        Pion mass $m_{\pi}\ [\mathrm{MeV}]$ & $138$ \\[0.05cm]
        Nucleon mass in vacuum $M_{N}\ [\mathrm{MeV}]$ & $939$ \\[0.05cm]
        Mass of the chiral partner $M_{N^{\ast}}\ [\mathrm{MeV}]$ & $1510$ \\[0.05cm]
		Nuclear saturation density $n_{0}\ [\mathrm{fm}^{-3}]$ & $0.16$ \\[0.05cm]
		Binding energy $E_{\mathrm{bind}}\ [\mathrm{MeV}]$ & $-16$ \\[0.05cm]
        Symmetry energy $E_{\mathrm{sym}}\ [\mathrm{MeV}]$ & $32$
		\end{tabular}
	\end{ruledtabular}
\end{table}

In the case of asymmetric matter, we have to determine an additional parameter, the vector-isovector coupling $G_{w}$. This is directly related to the symmetry energy, as we discuss in the following. For the sake of completeness, all parameters of the doublet and singlet models are summarized in Tables~\ref{tab:parameter_choice_PD} and \ref{tab:parameter_choice_WT}, respectively.
\begin{table}[ht]
	\caption{\label{tab:parameter_choice_PD}Set of chosen parameter values 
	(if not stated otherwise) in the parity-doublet model. The Taylor coefficients $\alpha_{1}$ and $\alpha_{2}$ are fixed by the bosonic masses $m_{\pi}$ and $m_{\sigma}$, cf.\ also Ref.~\cite{Eser:2023oii}.}
	\begin{ruledtabular}
		\begin{tabular}{lr}
		Parameter & Numerical value\\
		\colrule\\[-0.25cm]
        Chiral-invariant mass $m_{0}\ [\mathrm{MeV}]$ & $800$ \\[0.05cm]
        Isoscalar mass $m_{\sigma}\ [\mathrm{MeV}]$ & $340$ \\[0.05cm]
        Landau effective mass $M_{\ast}$ & $0.93 \times M_{N}$ \\[0.05cm]
		Yukawa coupling $y_{a}$ & $6.9$\\[0.05cm]
		Yukawa coupling $y_{b}$ & $13.0$\\[0.05cm]
		Taylor coefficient $\alpha_{3}\ [\mathrm{MeV}^{-2}]$ & $4.4 \times 10^{-1}$\\[0.05cm]
		Taylor coefficient $\alpha_{4}\ [\mathrm{MeV}^{-4}]$ & $-7.8 \times 10^{-5}$\\[0.05cm]
		Vector-isoscalar coupling $G_{v}\ [\mathrm{fm}^{2}]$ & $1.58$ \\[0.05cm]
        Vector-isovector coupling $G_{w}\ [\mathrm{fm}^{2}]$ & $1.19$ \\[0.05cm]
        Compression modulus $K\ [\mathrm{MeV}]$ & 242.8
		\end{tabular}
	\end{ruledtabular}
\end{table}
\begin{table}[ht]
	\caption{\label{tab:parameter_choice_WT}Set of chosen parameter values 
    in the singlet model (if not stated otherwise). See also Ref.~\cite{Eser:2023oii}.}
	\begin{ruledtabular}
		\begin{tabular}{lr}
		Parameter & Numerical value\\
		\colrule\\[-0.25cm]
        Isoscalar mass $m_{\sigma}\ [\mathrm{MeV}]$ & $640$ \\[0.05cm]
        Landau effective mass $M_{\ast}$ & $0.8 \times M_{N}$ \\[0.05cm]
		Yukawa coupling $y_{a} \equiv y$ & $10.1$\\[0.05cm]
		Taylor coefficient $\alpha_{3}\ [\mathrm{MeV}^{-2}]$ & $2.2 \times 10^{-1}$\\[0.05cm]
		Taylor coefficient $\alpha_{4}\ [\mathrm{MeV}^{-4}]$ & $-4.3 \times 10^{-5}$\\[0.05cm]
		Vector-isoscalar coupling $G_{v}\ [\mathrm{fm}^{2}]$ & $5.44$\\[0.05cm]
        Vector-isovector coupling $G_{w}\ [\mathrm{fm}^{2}]$ & $1.06$\\[0.05cm]
        Compression modulus $K\ [\mathrm{MeV}]$ & 299.2
		\end{tabular}
	\end{ruledtabular}
\end{table}

\subsection{Symmetry energy}
\label{sec:symmetry_energy}

The expansion of the energy per particle $\mathcal{E}/n_{\mathcal{B}}$ around symmetric nuclear matter defines the density-dependent symmetry energy $S(n_{\mathcal{B}})$,
\begin{equation}\label{eq:definitionSnB}
    \frac{\mathcal{E}}{n_{\mathcal{B}}}(n_{\mathcal{B}}, x) \simeq \frac{\mathcal{E}}{n_{\mathcal{B}}}\left(n_{\mathcal{B}}, \frac{1}{2}\right) + S(n_{\mathcal{B}}) (2x - 1)^{2},
\end{equation}
where  $x$ is the proton fraction (\ref{eq:protonfrac}) and $S(n_{0}) \equiv E_{\mathrm{sym}}$, with $n_0$ the saturation density of (symmetric) nuclear matter (cf.\ again Table~\ref{tab:parameters}). A simple calculation yields
\begin{equation}\label{eq:SnB}
   S(n_{\mathcal{B}})= \frac{1}{8} \left.\frac{\partial^{2} \mathcal{E}/n_{\mathcal{B}}}{\partial x^{2}}\right|_{x\, =\, \frac{1}{2}}
   = \frac{1}{2}  n_{\mathcal{B}} \left.\frac{\partial\mu_{\mathcal{I}}}{\partial n_{\mathcal{I}}}\right|_{n_{\mathcal{I}}\, =\, 0} ,
\end{equation}
where $n_{\mathcal{I}} = (2x - 1) n_{\mathcal{B}}$, and for densities of the order of the saturation density, we have $n_{\mathcal{B}}^{-} = 0$, as mentioned earlier.

Setting now $n_{\mathcal{B}}=n_0$, and using Eq.~(\ref{eq:dmuIdnIsimpsym}), one obtains the symmetry energy in the form 
\begin{equation}\label{eq:EsymF0}
    E_{\mathrm{sym}} = \frac{p_{F}^{2}}{6 M_{\ast}} \left(1 + F_{0}'\right), 
\end{equation}
where $F_0'$ is the Landau Fermi-liquid parameter related to the vector-isovector coupling $G_w$ by 
\begin{equation}
    F_0'= N_0 G_{w} ,\qquad N_0 = \frac{2p_F M_*}{\pi^2},
\end{equation}
with $N_0$ the density of states of the nucleons, $N_{0} \equiv N_{0}^{+}$, and $p_F$ the respective Fermi momentum. The relation (\ref{eq:EsymF0}) provides then a simple linear relation between $E_{\rm sym}$ and $G_w$, that can be used to fix the value of $G_w$. Adopting a commonly accepted value $E_{\rm sym} =32$ MeV, one gets $G_w \approx 1.19\ \mathrm{fm}^{2}$ in the doublet model and $G_w \approx 1.06\ \mathrm{fm}^{2}$ in the singlet model\footnote{It is instructive to compare with the results of Ref.~\cite{Alford_2022}, which uses a model similar to our singlet model. Translating the vector couplings into dimensionless couplings (see Eqs.~(\ref{eq:Gv}) and (\ref{eq:Gw})), one finds $g_{v} = 9.25$ and $g_{w} = 3.97$. While the value of $g_v$  matches that of Ref.~\cite{Alford_2022}, this is not the case for $g_w$, which is about four times smaller than that obtained in Ref.~\cite{Alford_2022}. We note however that in Ref.~\cite{Alford_2022} an additional contribution to the symmetry energy comes from a $\rho$-$\omega$ coupling term,  which we ignore in our singlet model. This is presumably the reason why, in spite of such different values of $g_w$, one gets the same value of the symmetry energy in both models. Note finally that the ratio of vector couplings in Ref.~\cite{Alford_2022} is not too far from that expected in the constituent quark model, namely $g_v=3g_w$, see e.g.\ Ref.~\cite{chamseddine_relativistic_2023}, while the values used in the present work are comparable to those determined in Ref.~\cite{brandes_fluctuations_2021}.} (for recent reviews on the symmetry-energy determination see e.g.\ Refs.~\cite{Baldo:2016jhp,sun2023compiled,Zhang_2023}; see also Refs.~\cite{danielewicz_symmetry_2009,danielewicz_symmetry_2014} for a broad discussion concerning the calculation of the symmetry energy).

\begin{figure}[ht]
	\centering
	\includegraphics[scale=1.0]{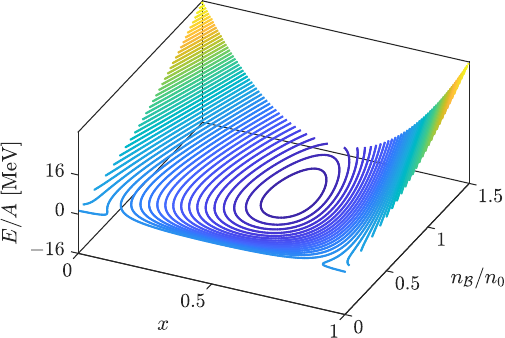}
	\caption{Three-dimensional contour plot of the energy per particle $E/A = \mathcal{E}/n_{\mathcal{B}} - M_{N}$ in the doublet model, as a function of the total baryon density $n_{\mathcal{B}}$ and the proton fraction $x = n_{\mathsf{P}}/n_{\mathcal{B}}$.}
	\label{fig:EA_x}
\end{figure}

\begin{figure}[ht]
	\centering
	\includegraphics[scale=1.0]{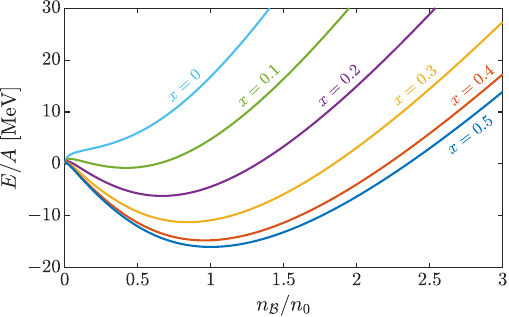}
	\caption{Energy per particle $E/A = \mathcal{E}/n_{\mathcal{B}} - M_{N}$ (at $T = 0$) for different proton fractions $x = n_{\mathsf{P}}/n_{\mathcal{B}}$.}
	\label{fig:energy_per_particle}
\end{figure}

Figure~\ref{fig:EA_x} shows contours of constant energy per particle in the parity-doublet model, as a function of the baryon density $n_{\mathcal{B}}$ and the proton fraction $x$. The plot exhibits a valley around $n_{\mathcal{B}} = n_{0}$ and $x = \frac{1}{2}$, which corresponds to the energy simultaneously being minimum at the saturation point and for symmetric matter. The minimum corresponds to the binding energy $E_{\mathrm{bind}} = -16\ \mathrm{MeV}$ of self-bound nuclear matter.
The parabolic shape with respect to $x$ (e.g.\ clearly visible at $n_{\mathcal{B}} = 1.5 \, n_{0}$) justifies the expansion (\ref{eq:definitionSnB}). Moreover, at $x = 0$, the energy per particle steadily increases with increasing density. This behavior is also illustrated in Fig.~\ref{fig:energy_per_particle}, which contains cuts of the three-dimensional picture of Fig.~\ref{fig:EA_x}, for constant values of $x$ and up to baryon densities of $3\, n_{0}$. Note that the quadratic $x$ dependence of the energy per particle in Eq.~(\ref{eq:definitionSnB}) holds over a wide range of $x$ values. For instance, in pure neutron matter ($x=0$) at $n_{\mathcal{B}} = n_{0}$, it predicts a value of $16\ \mathrm{MeV}$ ($=E_{\mathrm{bind}} + E_{\mathrm{sym}}$), while the exact result is $16.8\ \mathrm{MeV}$ in the doublet model and $16.9\ \mathrm{MeV}$ in the singlet model. This quadratic behavior was observed long ago in many-body calculations of nuclear matter, see e.g.\ Ref.~\cite{bombaci1991asymmetric},  and also Ref.~\cite{Gandolfi:2015jma} and references therein.

\subsection{Slope parameter of the symmetry energy}

The symmetry energy $S(n_{\mathcal{B}}) $ defined in Eq.~(\ref{eq:definitionSnB}) can be expanded around $n_{\mathcal{B}}=n_0$,
\begin{IEEEeqnarray}{rCl}
    S(n_{\mathcal{B}}) & \simeq & E_{\mathrm{sym}} + \frac{L}{3} \frac{n_{\mathcal{B}} - n_{0}}{n_{0}} 
    + \frac{K_{\mathrm{sym}}}{18} \left(\frac{n_{\mathcal{B}} - n_{0}}{n_{0}}\right)^{2} \nonumber\\[0.2cm]
    & & +\ \frac{Q_{\mathrm{sym}}}{162} \left(\frac{n_{\mathcal{B}} - n_{0}}{n_{0}}\right)^{3}, \label{eq:S_expansion}
\end{IEEEeqnarray}
where the parameter $L$ is commonly referred to as the ``slope parameter,'' and $K_{\mathrm{sym}}$ the symmetry incompressibility, as well as $Q_{\mathrm{sym}}$ the symmetry skewness \cite{Baldo:2016jhp, Lattimer:2023rpe}.


While the value of $E_{\rm sym}$ is rather well determined (with an uncertainty of about 2 MeV), this is not the case for the value of $L$, and even more so for the values of $K_{\mathrm{sym}}$ or $Q_{\mathrm{sym}}$. All these quantities enter the determination of the speed of sound in neutron-star matter, as discussed for instance in Ref.~\cite{Zhang_2023}. For completeness, we shall therefore present here the estimates obtained in the doublet and singlet models. The slope parameter $L$ is computed from Eqs.~(\ref{eq:SnB}) and (\ref{eq:dmuIdnIsimpsym}) as
\begin{IEEEeqnarray}{rCl}
    L & = & 3 n_{0} \left.\frac{\rmd S(n_{\mathcal{B}})}{\rmd n_{\mathcal{B}}}\right|_{n_{0}} = \frac{3 n_{0}}{2} 
    \bigg\lbrace \frac{1 + F_{0}'}{N_{0}} \nonumber \\[0.2cm]
    & - & \frac{1}{3} \left[ \frac{p_{F}^{2}}{M_{\ast}^{2}} 
    \left(\frac{1}{N_{0}} + y_{+} \frac{M_{+}}{M_{\ast}} \frac{\rmd \sigma}{\rmd n_{\mathcal{B}}} \right) + \frac{1}{N_{0}} \right] \bigg\rbrace \bigg|_{n_{0},\, \sigma\, =\, \sigma_{0}} \nonumber\\[0.2cm]
    & = & \frac{3 n_{0}}{2} \bigg( \frac{1 + F_{0}'}{N_{0}} - \frac{1}{3} \bigg\lbrace \frac{p_{F}^{2}}{M_{\ast}^{2}} 
    \bigg[\frac{1}{N_{0}} \nonumber\\[0.2cm]
    & & \qquad\qquad - \left. \frac{(y_{+}M_{+})^{2}}{m_{\sigma}^{2} M_{\ast}^{2}} \bigg] + \frac{1}{N_{0}} \bigg\rbrace \bigg)
    \right|_{n_{0},\, \sigma\, =\, \sigma_{0}} ,
\end{IEEEeqnarray}
where $\sigma_{0}$ the isoscalar condensate in the medium ($n_{\mathcal{B}} = n_{0}$). In the equation above we made use of the differential equation for $\sigma$ (in symmetric matter, and in absence of the chiral partners),
\begin{equation}
    \frac{\rmd\sigma}{\rmd n_{\mathcal{B}}} = - \frac{y_{+}}{m_{\sigma}^{2}} \frac{M_{+}}{M_{\ast}} ,
\end{equation}
which is obtained by differentiating the gap equation (\ref{eq:MF_eq_sigma}). See also Appendix~\ref{sec:diffeqs} for a detailed derivation. The sigma mass in this expression is defined as
\begin{equation}
    m_{\sigma}^{2} = \frac{\rmd^{2} U}{\rmd \sigma^{2}} + n_{\mathrm{s}}^{+} \frac{\rmd^{2} M_{+}}{\rmd\sigma^{2}} + y_{+} \left.\frac{\partial n_{\mathrm{s}}^{+}}{\partial\sigma} \right|_{n_{\mathcal{B}}} \label{eq:msigmasym}
\end{equation}
in the parity-doublet model, whereas in the singlet model the second term in Eq.~(\ref{eq:msigmasym}) vanishes. The higher coefficients $K_{\mathrm{sym}}$ and $Q_{\mathrm{sym}}$ are determined as
\begin{IEEEeqnarray}{rCl}
    K_{\mathrm{sym}} & = & 9 n_{0}^2 \left.\frac{\rmd^{2} S(n_{\mathcal{B}})}{\rmd n_{\mathcal{B}}^{2}}\right|_{n_{0}}, \\[0.2cm]
    Q_{\mathrm{sym}} & = & 27 n_{0}^{3} \left.\frac{\rmd^{3} S(n_{\mathcal{B}})}{\rmd n_{\mathcal{B}}^{3}}\right|_{n_{0}} ,
\end{IEEEeqnarray}
for which we do not derive analytic formulae here, and restrict ourselves to the numerical computation.

The numerical values of the parameters of the symmetry energy are collected in Table~\ref{tab:Sparameters}, for both the doublet and singlet models. The value of $L$ in the doublet model is at the upper edge of the range of values determined by taking into account constraints coming from neutron stars and recent nuclear physics experiments on the neutron skin, see Ref.~\cite{Zhang_2023}. It is close to those obtained in Ref.~\cite{Motohiro:2015taa} in a parity-doublet model comparable to the present one, while Ref.~\cite{Minamikawa:2023eky}, in a similar doublet model, uses $L = 57.7\ \mathrm{MeV}$ as input parameter. The values of $K_{\mathrm{sym}}$ and $Q_{\mathrm{sym}}$ are, to within the large uncertainties, in the range of values considered in the analysis of Ref.~\cite{Zhang_2023} and in the more recent one of Ref.~\cite{lopes2024decoding}. Further discussion of the slope parameter is given in Appendix~\ref{sec:slope}.

Correlations between the parameters that characterize the density dependence of the symmetry energy have been much studied (see e.g.\ Ref.~\cite{li_curvature-slope_2020}). This is in particular the case for the correlation between $L$ and the  symmetry energy $E_{\mathrm{sym}}$  (see e.g.\ Refs.~\cite{danielewicz_symmetry_2014, carlson_low-energy_2023,boukari_constraining_2024}). In the models that are discussed here it follows from the one-to-one correspondence (\ref{eq:EsymF0}) between the vector-isovector coupling $G_{w}$ and $E_{\mathrm{sym}}$ that the following simple relation holds:
\begin{equation}\label{eq:relationLEsym}
    \frac{\rmd L}{\rmd E_{\mathrm{sym}}} = 3,
\end{equation}
see also Appendix~\ref{sec:slope} for more details.  This relation is in qualitative agreement with most analyses, although the factor 3 appears to underestimate the actual correlation observed in most calculations. Note however that the relation (\ref{eq:relationLEsym}) relies on the one-to-one connection between the symmetry energy and the vector coupling $G_w$. As pointed out earlier there could be additional contributions to the symmetry energy which, without affecting the properties of symmetric matter, could modify the relation (\ref{eq:relationLEsym}). An example was provided earlier (after Eq.~(\ref{eq:EsymF0})), namely the $\rho$-$\omega$ coupling which is included in the model of Ref.~\cite{Alford_2022}.

We return in Sec.~\ref{sec:neutron_matter} to this discussion of the symmetry energy, in the more specific context of neutron matter and neutron-star physics. 


\begin{table}
	\caption{\label{tab:Sparameters}Expansion parameters of the symmetry energy.}
	\begin{ruledtabular}
		\begin{tabular}{lcc}
		  Parameter & Parity-doublet model & Singlet model\\
		\colrule\\[-0.25cm]
        $L\ [\mathrm{MeV}]$ & $\phantom{-}83.8$ & $89.1$ \\[0.05cm]
        $K_{\mathrm{sym}}\ [\mathrm{MeV}]$ & $-29.2$ & $-2.3\phantom{1}$ \\[0.05cm]
        $Q_{\mathrm{sym}}\ [\mathrm{MeV}]$ & $\phantom{-}94.0$ & $47.1$ 
		\end{tabular}
	\end{ruledtabular}
\end{table}


\section{Asymmetric matter}
\label{sec:asym_matter}

Asymmetric nuclear matter can undergo a liquid-gas transition at low baryon densities, and a chiral transition at large densities, depending on the temperature $T$ and the neutron excess (measured in terms of $x$ or $\mu_{\mathcal{I}}$) being present in the system. If the transitions are first order, there is phase coexistence. Regarding the liquid-gas transition, we have a gaseous phase of low baryon density coexisting with a liquid phase of larger density. At the chiral transition the two coexisting phases distinguish themselves also by the fact that chiral symmetry is broken in the low-density phase and restored in the high-density phase. This restoration of the chiral symmetry is signaled by the vanishing of the scalar field, $\sigma \rightarrow 0$, which plays the role of an order parameter.

We review in this section the variation of the liquid-gas transition as one moves away from symmetric matter, e.g.\ by increasing the isospin-chemical potential, and then turn to the chiral transition in the parity-doublet and singlet models, eventually stressing similarities and differences between the two phenomena, and between the two models. Finally, we work out the general structure of the solutions of the gap equation, and discuss the parity symmetry energy that plays an important role in the equilibration of parity partners across the chiral transition of the parity-doublet model. 

\subsection{Physical solution of the gap equation in asymmetric matter}

\begin{figure}[ht]
    \centering
	\includegraphics[scale=1.0]{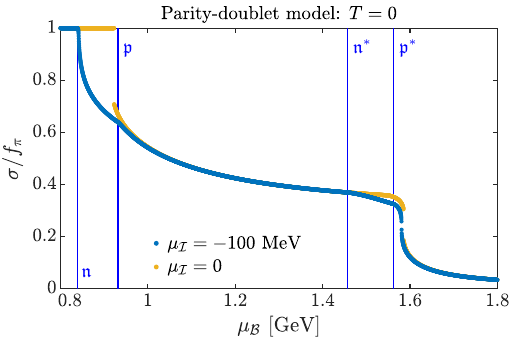}
	\caption{Condensate $\sigma$ as a function of $\mu_{\mathcal{B}}$ for two values of $\mu_{\mathcal{I}}$ in the parity-doublet model. The onsets of the different fermion species in the case of $\mu_{\mathcal{I}} = -100\ \mathrm{MeV}$ are marked as blue vertical lines. Although the curves look continuous, they are obtained from individual calculations for values of $\mu_{\mathcal{B}}$ incremented in steps of $0.25\ \mathrm{MeV}$ in the range $0.8\ \mathrm{GeV} \le \mu_{\mathcal{B}} \le 1.8\ \mathrm{GeV}$.}
	\label{fig:phase_diagram_chempot_PD}
\end{figure}

We start the discussion by considering the $\sigma$ field as a function of the baryon-chemical potential $\mu_{\mathcal{B}}$ for a fixed isospin-chemical potential $\mu_{\mathcal{I}}$, both in the doublet and singlet models: Figure~\ref{fig:phase_diagram_chempot_PD} illustrates the corresponding $\sigma$ variation in the doublet model at zero temperature. The $\sigma$ field that solves the gap equation (\ref{eq:gap}) obviously features two phase transitions, the liquid-gas transition at $\mu_{\mathcal{B}} \sim 0.9\ \mathrm{GeV}$ and the chiral transition at $\mu_{\mathcal{B}} \sim 1.6\ \mathrm{GeV}$. In the case of symmetric matter ($\mu_{\mathcal{I}}=0$), the two vertical drops of $\sigma$ around the aforementioned chemical potentials indicate that the two transitions are first order, as it has already been verified in (I). Concerning the case of neutron-rich matter, $\mu_{\mathcal{I}} = -100\ \mathrm{MeV}$, both transitions become weaker, in the sense that the vertical drops of $\sigma$ become either narrower (chiral transition) or even disappear (liquid-gas transition), while the true transition orders are still to be determined in the upcoming subsections.

Having a closer look at the behavior of $\sigma$ in the case of neutron-rich matter ($\mu_{\mathcal{I}} = -100\ \mathrm{MeV}$), we recognize four kinks in Fig.~\ref{fig:phase_diagram_chempot_PD} that significantly change the $\sigma$ evolution. Every kink can be identified as the onset of the population of a given fermion species, which is indicated as the blue vertical lines. The first occurs once we reach the baryon-chemical potential $\mu_{\mathcal{B}} = 839\ \mathrm{MeV} \equiv M_{N} + \mu_{\mathcal{I}}$, and then start populating the vacuum with neutrons. The second kink is associated with the onset of the proton population, where the onset threshold is determined by the equation
\begin{equation}
    M_{\ast}^{\mathfrak{n}} = M_{+} - 2 \left(\mu_{\mathcal{I}} + G_{w} n_{\mathfrak{n}}\right) , 
    \label{eq:onset_protons}
\end{equation}
which follows from Eqs.~(\ref{eq:defdensities1}) and (\ref{eq:defdensities2}), and yields $\mu_{\mathcal{B}} \approx 931.6\ \mathrm{MeV}$. Eventually, the onsets of the populations of the $\mathfrak{n}^{\ast}$ and $\mathfrak{p}^{\ast}$ (in this order) at even larger chemical potentials prior to the chiral transition are determined by the respective thresholds (\ref{eq:thresholds}), $\mu_{\mathcal{B}} \approx 1456\ \mathrm{MeV}$ ($\mathfrak{n}^{\ast}$) and $\mu_{\mathcal{B}} \approx 1562\ \mathrm{MeV}$ ($\mathfrak{p}^{\ast}$). Let us remark that this connection between the kinks in the solution and the threshold conditions holds because the  solution is continuous when $\mu_{\mathcal{I}} = -100\ \mathrm{MeV}$. When $\mu_{\mathcal{I}} = 0$, one encounters the discontinuous first-order liquid-gas transition (see again Fig.~\ref{fig:phase_diagram_chempot_PD}), at $\mu_{\mathcal{B}} = \mu_{0}=M_N+E_{\rm bind}$, i.e.\ before  $\mu_{\mathcal{B}} = M_{N}$, with $ M_N$ the vacuum nucleon mass. At the transition the baryon density jumps from zero to $n_{0}$ (by model construction), and the $\sigma$ field drops from $f_{\pi}$ to $\sigma_{0} \approx 65.9\ \mathrm{MeV}$.
Interestingly, the $\sigma$ field in the intermediate regime between the two transitions is almost independent of the isospin-chemical potential, with the two sets of data points roughly overlapping between the thresholds of the $\mathfrak{p}$ and the $\mathfrak{n}^{\ast}$. As we shall see explicitly later, the $\sigma$ field (as a function of the baryon density $n_{\mathcal{B}}$) is also nearly independent of the proton fraction $x$ in this intermediate plateau-like regime, and its value does not change too much.

\begin{figure}[ht]
    \centering
	\includegraphics[scale=1.0]{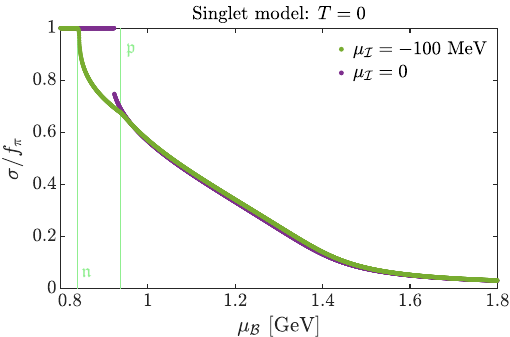}
	\caption{Condensate $\sigma$ as a function of $\mu_{\mathcal{B}}$ for two values of $\mu_{\mathcal{I}}$ in the singlet model. The onsets of the different fermion species in the case of $\mu_{\mathcal{I}} = -100\ \mathrm{MeV}$ are marked as light green vertical lines. Data points are given in steps of $0.25\ \mathrm{MeV}$ in the range $0.8\ \mathrm{GeV} \le \mu_{\mathcal{B}} \le 1.8\ \mathrm{GeV}$.}
	\label{fig:phase_diagram_chempot_WT}
\end{figure}

Figure~\ref{fig:phase_diagram_chempot_WT} shows the corresponding $\sigma$ variation in the singlet model. We consider again the two cases $\mu_{\mathcal{I}} = 0$ (symmetric matter) and $\mu_{\mathcal{I}} = -100\ \mathrm{MeV}$ (as an example of neutron-rich matter), in analogy to Fig.~\ref{fig:phase_diagram_chempot_PD}. The two plots exhibit strong similarities concerning the liquid-gas transition at $\mu_{\mathcal{B}} \sim 0.9\ \mathrm{GeV}$, but substantial differences concerning the chiral transition. The former again disappears (and becomes smeared out as compared to $\mu_{\mathcal{I}} = 0$) in the neutron-rich case, while the chiral transition is a mere crossover for both values of $\mu_{\mathcal{I}}$, which is in contrast to Fig.~\ref{fig:phase_diagram_chempot_PD}. In addition, the decrease of $\sigma$ towards zero is rather linear in the intermediate regime of $1\ \mathrm{GeV} \lesssim \mu_{\mathcal{B}} \lesssim 1.4\ \mathrm{GeV}$, and there is no such plateau behavior as in the doublet model. These findings do not come as a surprise, as this difference was already reported in (I), and stays valid in the asymmetric case.

The onsets of the proton and neutron populations in the singlet model follow the same rules as in the doublet model, with the $\mathfrak{n}$-onset  taking place at $\mu_{\mathcal{B}} = M_{N} + \mu_{\mathcal{I}}$. The $\mathfrak{p}$-onset is again determined by Eq.~(\ref{eq:onset_protons}) with the replacement $M_{+} \mapsto M$, leading to $\mu_{\mathcal{B}} \approx 937.2\ \mathrm{MeV}$. In the case of $\mu_{\mathcal{I}} = -100\ \mathrm{MeV}$, these separate onsets translate into two (small) kinks in the evolution of the $\sigma$ field (see the light green vertical lines in Fig.~\ref{fig:phase_diagram_chempot_WT}), whereas in symmetric matter the simultaneous onset of the $\mathfrak{p}$ and $\mathfrak{n}$ is overlaid by the first-order liquid-gas transition, where the isoscalar condensate drops to $\sigma_{0} \approx 69.7\ \mathrm{MeV}$.

\begin{figure}[ht]
    \centering
	\includegraphics[scale=1.0]{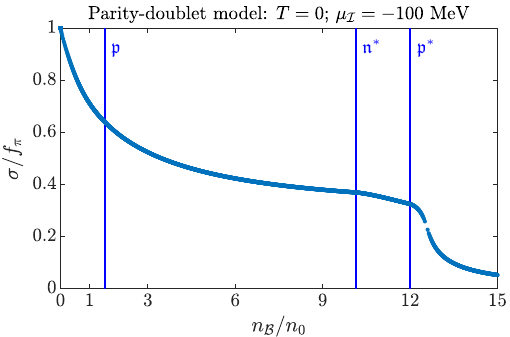}
	\caption{Condensate $\sigma$ as a function of the baryon density $n_{\mathcal{B}}$ for $\mu_{\mathcal{I}} = -100\ \mathrm{MeV}$ in the parity-doublet model, with the data points taken from Fig.~\ref{fig:phase_diagram_chempot_PD}.}
	\label{fig:phase_diagram_density_PD}
\end{figure}

\begin{figure}[ht]
    \centering
	\includegraphics[scale=1.0]{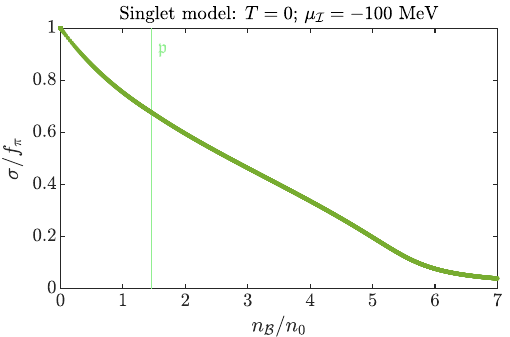}
	\caption{Condensate $\sigma$ as a function of the baryon density $n_{\mathcal{B}}$ for $\mu_{\mathcal{I}} = -100\ \mathrm{MeV}$ in the singlet model, with the data points taken from Fig.~\ref{fig:phase_diagram_chempot_WT}.}
	\label{fig:phase_diagram_density_WT}
\end{figure}

Complementary to these figures, the $\sigma$ field as a function of the baryon density $n_{\mathcal{B}}$ for $\mu_{\mathcal{I}} = -100\ \mathrm{MeV}$ is given in Figs.~\ref{fig:phase_diagram_density_PD} and \ref{fig:phase_diagram_density_WT}, respectively for the doublet and singlet models. Figure~\ref{fig:phase_diagram_density_PD} shows a continuous and smooth curve for small baryon densities of the order of the saturation density $n_{0}$ and below, revealing that the liquid-gas transition is indeed smeared out to a mere crossover. The onset of the protons, which in this picture is located at $n_{\mathcal{B}} \approx 1.53 \, n_{0}$, does not lead to a visible kink in the solution of the gap equation, as it was observed in Fig.~\ref{fig:phase_diagram_chempot_PD}. As we shall explain later, the absence of the $\mathfrak{p}$-kink is a consequence of $\sigma$ being nearly independent of $x$ in the low-density regime. However at large densities, in contrast to the onset of the protons, those of the chiral partners are still clearly visible as kinks, with the onset of the $\mathfrak{n}^{\ast}$ at $n_{\mathcal{B}} \approx 10.1 \, n_{0}$, as well as the $\mathfrak{p}^{\ast}$ at $n_{\mathcal{B}} \approx 12.0 \, n_{0}$.

In the language of the differential equations (see Appendix \ref{sec:diffeqs}) that we can alternatively use to determine the solutions of the gap equation, we are able to show that the overall evolution of the $\sigma$ field in Fig.~\ref{fig:phase_diagram_density_PD} may be split up into different stages. These stages are described by different sets of differential equations, and the two thresholds of the chiral partners mark the two densities at which we have to change from one set to the other. From this viewpoint it will immediately become clear why the solution for $\sigma$ exhibits such kinks at the thresholds, and how the overall solution is shaped for any value of $x$.

The onsets of the population of the chiral partners in the parity-doublet model are precursors to the chiral transition, where the isoscalar condensate $\sigma$ eventually tends to zero, and the chiral partners become degenerate in mass. The transition is visible as a comparably steep decrease in the evolution of $\sigma$ that sets in after the system passed the $\mathfrak{n}^{\ast}$ and $\mathfrak{p}^{\ast}$-thresholds. Moreover we observe, again in Fig.~\ref{fig:phase_diagram_density_PD}, a rather small gap, that is, a positive jump in $n_{\mathcal{B}}$ and a small drop in $\sigma$, which could be attributed, given the finite resolution of data points, to a weak first-order transition. We shall eventually  determine the order of the chiral transition for any proton fraction (and isospin-chemical potential) by computing the convex envelope of the corresponding free energy density in Sec.~\ref{sec:chiral}. Technical details of this computation are summarized in Appendix~\ref{sec:phasecoex}.

In the singlet model, at $\mu_{\mathcal{I}} = -100\ \mathrm{MeV}$, the liquid-gas transition is also a smooth crossover, see Fig.~\ref{fig:phase_diagram_density_WT}, and the onset of the protons expresses itself again as a tiny kink at $n_{\mathcal{B}} \approx 1.46 \, n_{0}$ (hardly visible on the plot). The chiral transition in the singlet model is also smooth, and the $\sigma$ condensate approaches zero asymptotically. This happens much earlier than in the doublet model, where the plateau behavior, which defers the chiral transition to densities typically beyond $10 \, n_{0}$, was identified as an effect  of the chiral-invariant mass $m_{0}$ \cite{Eser:2023oii}.

In the case of a first-order phase transition there are two distinct solutions of the gap equation that exhibit the same thermodynamic pressure $P$ and the same chemical potentials $\lbrace \mu_{\mathcal{B}}, \mu_{\mathcal{I}} \rbrace$ (or equivalently $\lbrace \mu_{\mathsf{P}}, \mu_{\mathsf{N}} \rbrace$), but different $n_{\mathcal{B}}$ and $x$ (or $n_{\mathcal{I}}$). These two solutions represent the two coexisting phases that we precisely determine in the next subsections, also at nonzero temperatures.

\subsection{The liquid-gas transition}

\begin{figure}[ht]
    \centering
	\includegraphics[scale=1.0]{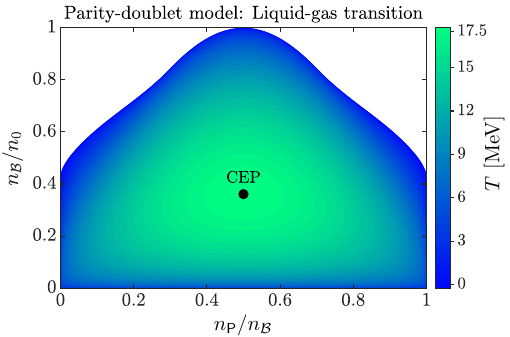}
	\caption{Temperature-dependent coexistence region corresponding to the liquid-gas transition, as a function of the proton fraction $x = n_{\mathsf{P}}/n_{\mathcal{B}}$ and the baryon density $n_{\mathcal{B}}$. The black dot shows the critical endpoint of the liquid-gas transition for symmetric matter ($x = \frac{1}{2}$) at $(n_{c}, T_{c}) \approx (0.06\ \mathrm{fm}^{-3}, 18\ \mathrm{MeV})$ (see also the quoted values in Ref.~\cite{Eser:2023oii}).}
	\label{fig:liquid_gas_T_den}
\end{figure}

At low baryon densities, we have seen that the parity-doublet and singlet models undergo a first-order liquid-gas transition, if the neutron excess is not too large. Figure~\ref{fig:liquid_gas_T_den} shows the region of phase coexistence in the doublet model that is associated with the liquid-gas transition, as a function of the proton fraction $x$ and the baryon density $n_{\mathcal{B}}$, and for various temperatures up to the critical endpoint (``CEP''), with $T_{c} \approx 18\ \mathrm{MeV}$. By construction the coexistence region spans the range of $0 \le n_{\mathcal{B}} \le n_{0}$ at $x = \frac{1}{2}$ and $T = 0$, reflecting that symmetric matter with a density  $n_{0}$ exists as a self-bound system of vanishing pressure. Keeping $x$ fixed, and increasing the temperature, one continues to observe two coexisting phases of nonzero pressure. As one departs from the symmetric case, e.g. as $x$ decreases from $x= \frac{1}{2}$, the density contrast and the maximum temperature of the coexistence region decreases.
Note that the two points obtained by cutting the coexistence volume by the two planes of constant  $x$ and constant $T$ do not necessarily correspond to two coexisting phases, since in general the corresponding proton fractions  differ. The only exception holds for $x = \frac{1}{2}$. Note also that  the coexistence volume of Fig.~\ref{fig:liquid_gas_T_den} is symmetric with respect to the vertical plane at $x = \frac{1}{2}$, as the system is symmetric under the exchange of protons and neutrons.

\begin{figure}[ht]
	\centering
	\includegraphics[scale=1.0]{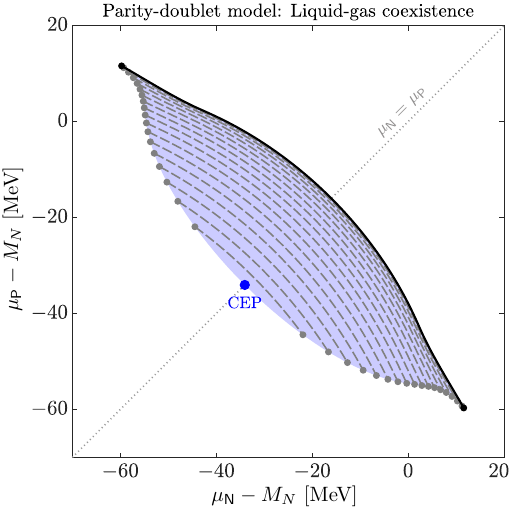}
	\caption{Coexistence lines of the liquid-gas transition. The lines correspond to constant temperatures, with the black solid line corresponding to $T = 0$ and the gray dashed lines to nonzero temperatures, $1\ \mathrm{MeV} \le T \le 17\ \mathrm{MeV}$, in steps of $1\ \mathrm{MeV}$ (towards the lower left corner). The blue shaded area represents the general region of coexistence. The black and gray dots show the respective endpoints of each line, converging to the critical endpoint (blue dot) for symmetric matter at $T \approx 18\ \mathrm{MeV}$ and $\mu_{\mathsf{P}} = \mu_{\mathsf{N}} \approx 905\ \mathrm{MeV}$ \cite{Eser:2023oii}.}
	\label{fig:liquid_gas_coex}
\end{figure}

If we map the coexistence volume of Fig.~\ref{fig:liquid_gas_T_den} onto the plane of the proton and neutron-chemical potentials, we find the light blue zone displayed in Fig.~\ref{fig:liquid_gas_coex}, which comprises the data for all temperatures up to the critical temperature $T_{c}$. On top of the light blue zone we furthermore plotted different lines corresponding to different constant temperatures. The critical endpoint is shown as the blue dot denoted again ``CEP.'' The length of the lines shrinks with increasing temperature, and eventually dwindles away (to the single blue point) at the critical temperature. In this picture, the coexistence region roughly spans the range of $ -60\ \mathrm{MeV} \lesssim \mu_{\mathsf{N}} - M_{N} \lesssim 12\ \mathrm{MeV}$, and the same holds for the $\mu_{\mathsf{P}}$-axis, since the system is symmetric with respect to the diagonal dotted line ($\mu_{\mathsf{P}} = \mu_{\mathsf{N}}$).

With the help of Fig.~\ref{fig:liquid_gas_coex} we can eventually confirm that the liquid-gas transition that was shown in Figs.~\ref{fig:phase_diagram_chempot_PD} and \ref{fig:phase_diagram_density_PD} is of crossover type in the asymmetric case, since the maximum absolute distance between proton and neutron-chemical potentials, which follows from the areal extent of the light blue coexistence region, is far less than $200\ \mathrm{MeV}$ (corresponding to $\mu_{\mathcal{I}} = -100\ \mathrm{MeV}$). The same is true regarding the singlet model, where the numerical results are very similar to those of Figs.~\ref{fig:liquid_gas_T_den} and \ref{fig:liquid_gas_coex} (thus omitted here).

\begin{figure}[ht]
	\centering
	\includegraphics[scale=1.0]{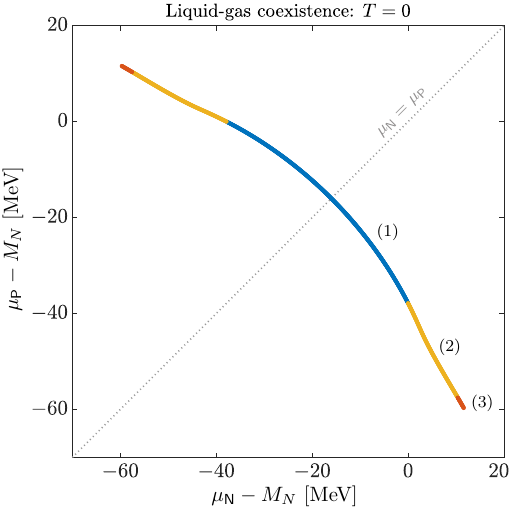}
	\caption{Line of phase coexistence corresponding to the liquid-gas transition at zero temperature. The color code indicates different types of coexistence, also labeled with (1), (2), and (3); an explanation is given in the main text.}
	\label{fig:liquid_gas_coex_Tzero}
\end{figure}

Consistent with the analysis of Ref.~\cite{Ducoin:2005aa}, which was carried out in a similar ``singlet'' model, we find three different types of phase coexistence at zero temperature: (1) self-bound matter in coexistence with vacuum, (2) two coexisting phases of nonzero densities, but a vanishing proton or neutron density in the gaseous phase, and (3) nonzero proton and neutron densities in both coexisting phases. Figure \ref{fig:liquid_gas_coex_Tzero} shows the different cases at zero temperature in terms of colors: (1) in blue, (2) in orange, and (3) in red. In the part of the figure below the diagonal we find neutron-rich matter. The blue part refers to self-bound asymmetric matter with $n_{\mathfrak{n}} > n_{\mathfrak{p}} > 0$, according to the description above. In the orange part we then have $n_{\mathfrak{n}} > 0$ and $n_{\mathfrak{p}} = 0$ in the gaseous phase, while both proton and neutron densities are non vanishing in the low-density gas of the red region.

\subsection{The chiral transition}
\label{sec:chiral}

\subsubsection{The region of phase coexistence}

\begin{figure}[ht]
    \centering
	\includegraphics[scale=1.0]{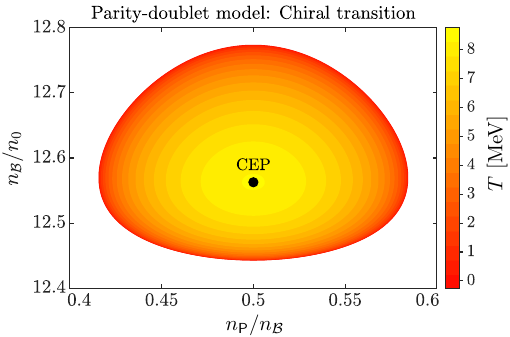}
	\caption{Temperature-dependent coexistence region corresponding to the chiral transition (for physical pion mass), as a function of the proton fraction $x = n_{\mathsf{P}}/n_{\mathcal{B}}$ and the baryon density $n_{\mathcal{B}}$. The black dot shows the critical endpoint (``CEP'') of the chiral transition for symmetric matter ($x = \frac{1}{2}$) at $(n_{c}, T_{c}) \approx (2.01\ \mathrm{fm}^{-3}, 8.5\ \mathrm{MeV})$ (see also the coexistence region in Ref.~\cite{Eser:2023oii}).}
	\label{fig:chiral_transition_T_den}
\end{figure}

Similar to the liquid-gas transition, the chiral transition in the parity-doublet model is of first order, if both the neutron excess and the temperature are not too large. In analogy to Fig.~\ref{fig:liquid_gas_T_den} and the liquid-gas transition, Fig.~\ref{fig:chiral_transition_T_den} shows the region of phase coexistence associated with the first-order chiral phase transition. The overall shape of this coexistence region differs from the one in Fig.~\ref{fig:liquid_gas_T_den}: it appears to be more rounded, and consists again of all sets of points corresponding to coexisting phases, as already described above. Its width with respect to the proton fraction $x$ is much smaller as compared to the liquid-gas transition, where we even found that e.g.\ a dilute neutron gas ($x = 0$) may coexist with a mixed phase ($x > 0$) at $T = 0$. Across the chiral transition, in contrast, the values of $x$  in both coexisting phases differ by at most 30\%.

\begin{figure}[ht]
	\centering
	\includegraphics[scale=1.0]{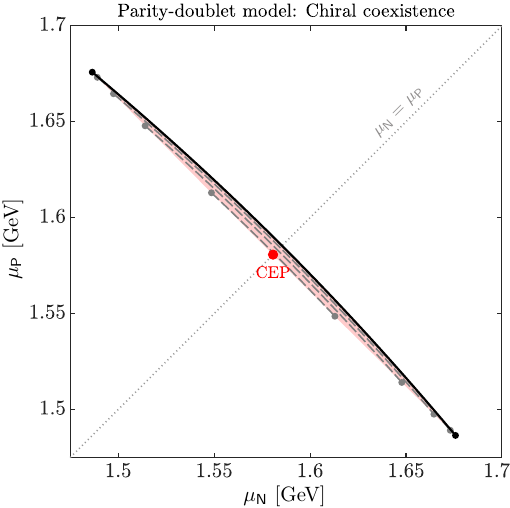}
    \caption{Coexistence lines of the chiral transition. The lines correspond to constant temperatures, with the black solid line corresponding to $T = 0$ and the gray dashed lines to nonzero temperatures, $2\ \mathrm{MeV} \le T \le 8\ \mathrm{MeV}$, in steps of $2\ \mathrm{MeV}$ (towards the lower left corner). The red shaded area represents the general region of coexistence. The black and gray dots show the respective endpoints of each line, converging to the critical endpoint (red dot) for symmetric matter at $T \approx 8.5\ \mathrm{MeV}$ and $\mu_{\mathsf{P}} = \mu_{\mathsf{N}} \approx 1580\ \mathrm{MeV}$ \cite{Eser:2023oii}.}
    \label{fig:chiral_coex}
\end{figure}

Mapping again the coexistence region given in Fig.~\ref{fig:chiral_transition_T_den} onto the plane of $\mu_{\mathsf{N}}$ and $\mu_{\mathsf{P}}$ we find the light red zone displayed in Fig.~\ref{fig:chiral_coex}. The coexistence points of the first-order chiral phase transition for constant temperatures are illustrated again as lines on top of this light red zone (at zero temperature and higher temperatures up to the critical temperature around $8.5\ \mathrm{MeV}$). The region of chiral coexistence appears to be rather shallow in comparison with the liquid-gas transition (cf.\ again the light blue zone in Fig.~\ref{fig:liquid_gas_coex}). The chiral transition roughly occurs at chemical potentials of the order of $1.5\ \mathrm{GeV}$ to $1.7\ \mathrm{GeV}$, corresponding to baryon densities typically above $12 \, n_{0}$.

\begin{figure}[ht]
	\centering
    \includegraphics[scale=1.0]{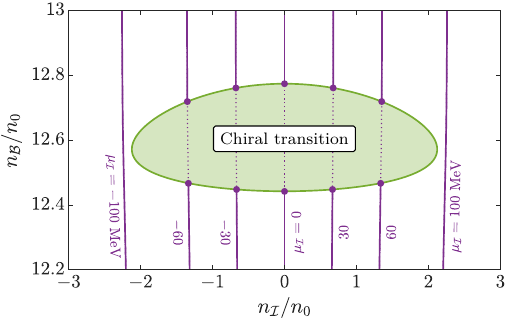}
	\caption{The chiral transition ``gap'' at $T = 0$ (for physical pion mass) in the plane of the isospin density $n_{\mathcal{I}}$ and the baryon density $n_{\mathcal{B}}$. The green area depicts the positive jump in baryon density that the system experiences for $|\mu_{\mathcal{I}}| \lesssim 100\ \mathrm{MeV}$ (magenta dotted lines).}
	\label{fig:chiral_transition_nI_nB}
\end{figure}

From Fig.~\ref{fig:chiral_transition_nI_nB} it clearly follows that the chiral transition is accompanied by a positive jump in the total baryon density $n_{\mathcal{B}}$, as it was already anticipated in the context of Figs.~\ref{fig:phase_diagram_chempot_PD} and \ref{fig:phase_diagram_density_PD} (and observed in (I)). In order to verify the transition order in the respective case of neutron-rich matter ($\mu_{\mathcal{I}} = -100\ \mathrm{MeV}$) we investigate the chiral transition ``gap'' at $T = 0$ more closely. This allows us to draw Fig.~\ref{fig:chiral_transition_nI_nB}, now as a function of the isospin density $n_{\mathcal{I}}$, and featuring lines of constant isospin-chemical potential $\mu_{\mathcal{I}}$. For a rather small absolute value of $\mu_{\mathcal{I}}$ the system undergoes a first-order chiral transition when it hits the green area, e.g.\ following the line of constant $\mu_{\mathcal{I}} = -30\ \mathrm{MeV}$ in the direction of increasing $n_{\mathcal{B}}$. As the system reaches the green area, the physical solution jumps to the point where it leaves again the green area, according to the respective dotted lines.

When increasing or decreasing the isospin density $n_{\mathcal{I}}$ the green area in Fig.~\ref{fig:chiral_transition_nI_nB} shrinks, so that the first-order gap becomes smaller and smaller and, at the densities where it eventually vanishes, the chiral transition becomes second order. In particular, for $\mu_{\mathcal{I}} = -100\ \mathrm{MeV}$, the transition is then finally a mere crossover, which allows us to conclude that the tiny gap in the decrease of $\sigma$ that we observed in Fig.~\ref{fig:phase_diagram_density_PD} is only an effect of a limited numerical resolution.

Regarding the singlet model there is no such analysis needed, since the chiral transition (for physical pion mass) is of the crossover type, both in symmetric and asymmetric matter, see Figs.~\ref{fig:phase_diagram_chempot_WT} and \ref{fig:phase_diagram_density_WT} (and consider once again (I)). Another interesting study (although perhaps somewhat academic)  is that of the chiral limit, since then the chiral crossover becomes a true phase transition. The corresponding analysis is deferred to Appendix~\ref{sec:chiral_limit}. In the following we now provide a complete overview of the composition of dense asymmetric matter as it is predicted by the parity-doublet model.

\subsubsection{The composition  of dense matter}

\begin{figure}[ht]
	\centering
    \includegraphics[scale=1.0]{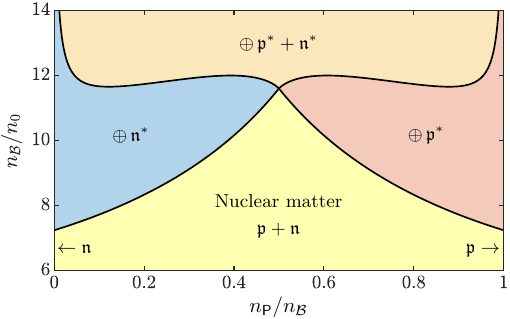}
	\caption{The composition of matter (for physical pion mass and $T = 0$) as a function of the proton fraction $x=n_{\mathsf{P}}/n_{\mathcal{B}}$ and the total baryon density $n_{\mathcal{B}}$, according to the thresholds (\ref{eq:thresholds}). The onset of the $\mathfrak{n}^{\ast}$ is depicted by the black line that starts in the lower left part and ends in the upper right part; the respectively mirrored black line depicts the onset of the $\mathfrak{p}^{\ast}$.}
	\label{fig:thresholds_diagram}
\end{figure}

The composition of dense matter within the parity-doublet model is determined by the two threshold functions (\ref{eq:thresholds}). To map out these thresholds in the plane of the proton fraction $x=n_{\mathsf{P}}/n_{\mathcal{B}}$ and the baryon density $n_{\mathcal{B}}$ we solve Eq.~(\ref{eq:muI}) together with Eq.~(\ref{eq:thresholds}), for $x$ and $n_{\mathcal{B}}$, at a given isospin-chemical potential $\mu_{\mathcal{I}}$. So for a given $\mu_{\mathcal{I}}$ we obtain two pairs $(x, n_{\mathcal{B}})$ corresponding respectively to the $\mathfrak{p}^{\ast}$ and the $\mathfrak{n}^{\ast}$-onsets. By varying the chemical potential $\mu_{\mathcal{I}}$ we then determine the onset lines shown in Fig.~\ref{fig:thresholds_diagram}. These lines delineate various regions of the phase diagram as a function of the baryon density and the neutron excess, which illustrates the associated changes in the composition of matter.

The lines corresponding to the onsets of the population of neutron and proton  chiral partners intersect at $x = \frac{1}{2}$ (corresponding to $\mu_{\mathcal{I}} = 0$), and  split when $n_{\mathsf{P}} \neq n_{\mathsf{N}}$, as we have seen earlier (see e.g.\ Fig.~\ref{fig:phase_diagram_density_PD}). Below $n_{\mathcal{B}} \approx 7 \, n_{0}$, nuclear matter consists exclusively of protons and neutrons (indicated as ``$\mathfrak{p} + \mathfrak{n}$'' in Fig.~\ref{fig:thresholds_diagram}), as the density is not high enough to populate any chiral partners. As we vary continuously the value of the proton fraction $x$, we change the ratio between protons and neutrons, and at $x = 0$ or $x = 1$ we get pure neutron or pure proton matter, respectively. 

If we now consider neutron-rich matter ($x < \frac{1}{2}$) and increase the total baryon density while keeping $x$ fixed, we start to populate the chiral partner $\mathfrak{n}^{\ast}$ once the threshold line $M_{\ast}^{\mathfrak{n}} = M_{-}$ is crossed. This is denoted in the figure as ``$\oplus\, \mathfrak{n}^{\ast}$,'' which means that matter then also consists of $\mathfrak{n}^{\ast}$-states, in addition to protons and neutrons, but the $\mathfrak{p}^{\ast}$-states still remain unpopulated. Even further increasing $n_{\mathcal{B}}$ eventually allows us to also populate the $\mathfrak{p}^{\ast}$, if the value of $x$ is not too small, then entering the zone ``$\oplus\, \mathfrak{p}^{\ast} + \mathfrak{n}^{\ast}$.'' This happens at the corresponding threshold $M_{\ast}^{\mathfrak{p}} = M_{-}$. However, if the fraction $x$ gets too small, the onset of the $\mathfrak{p}^{\ast}$ is pushed to larger and larger baryon densities, and at $x = 0$ it is never reached (neutron-like matter is devoid of $\mathfrak{p}$ and $\mathfrak{p}^{\ast}$ for all densities).

By considering the onset condition for the protons in the form of $(p_{\mathfrak{p}}^{2} + M_{+}^{2})^{1/2} = M_{-}$, it is straightforward to explain the divergent behavior of the $\mathfrak{p}^{\ast}$-threshold at $x = 0$: the Fermi momentum of the protons vanishes in pure neutron-like matter,
\begin{equation}
    p_{\mathfrak{p}} = \left. \left(3 \pi^{2} x n_{\mathcal{B}}\right)^{\frac{1}{3}} \right|_{x\, =\, 0} = 0 ,
\end{equation}
so that the $\mathfrak{p}^{\ast}$-threshold reduces to $M_{+} = M_{-}$. This is only fulfilled for $\sigma = 0$, cf.\ Eq.~(\ref{eq:massphys}), which is only reached when $n_{\mathcal{B}} \rightarrow \infty$, see Fig.~\ref{fig:phase_diagram_density_PD}.

As the system is symmetric under exchange of protons and neutrons, the diagram shown in Fig.~\ref{fig:thresholds_diagram} exhibits the same symmetry. This means that we can repeat our arguments above for proton-rich matter, simply by exchanging protons and neutrons. Moreover, as mentioned, the two onset lines for $\mathfrak{p}^{\ast}$ and $\mathfrak{n}^{\ast}$ intersect at $x = \frac{1}{2}$, i.e.\ at the point of symmetric nuclear matter with $n_{\mathsf{P}} = n_{\mathsf{N}}$. In the case of $x = \frac{1}{2}$, the system directly enters the region populated with both  $\mathfrak{p}^{\ast}$ and  $\mathfrak{n}^{\ast}$, with $n_{\mathfrak{p}^{\ast}} = n_{\mathfrak{n}^{\ast}}$ (see again the detailed discussion of symmetric matter in (I)).

\subsubsection{The parity symmetry energy}
\label{sec:more_insight}

Crossing both threshold lines of the chiral partners is a precursor to the chiral transition, which we confirm by comparing the respective density ranges in Figs.~\ref{fig:chiral_transition_nI_nB} and \ref{fig:thresholds_diagram}. From the physics point of view it is reasonable that the phase space allows for a nonzero population of the chiral partners before the chiral transition occurs, where the masses of the doublet, and consequently their densities,  become (approximately) degenerate. However, one may recognize in the equilibration of the populations of the positive and negative-parity baryons the role of a kind of symmetry energy among the chiral partners \cite{Eser:2023oii}, which we refer to as a ``parity symmetry energy'' in order to distinguish it from the more familiar isospin symmetry energy. 

To identify this mechanism we evaluate the energy density $\mathcal{E} \equiv \mathcal{E}\left(n_{\mathsf{P}}, n_{\mathsf{P}}', n_{\mathsf{N}}, n_{\mathsf{N}}'\right)$ for  densities which differ slightly from their equilibrium values,  denoted respectively by $\bar{n}_{\mathsf{P}}$, $\bar{n}_{\mathsf{P}}'$, $\bar{n}_{\mathsf{N}}$, and $\bar{n}_{\mathsf{N}}'$. We set
\begin{equation}
    \delta\mathcal{E} \equiv \mathcal{E}\left(n_{\mathsf{P}}, n_{\mathsf{P}}', n_{\mathsf{N}}, n_{\mathsf{N}}'\right) 
    - \mathcal{E}\left(\bar{n}_{\mathsf{P}}, \bar{n}_{\mathsf{P}}', \bar{n}_{\mathsf{N}}, \bar{n}_{\mathsf{N}}'\right) , 
    \label{eq:defvariation}
\end{equation}
where the ``parity densities'' $n_{\mathsf{P},\mathsf{N}}'$ are given by
\begin{equation}
    n_{\mathsf{P}}' = n_{\mathfrak{p}} - n_{\mathfrak{p}^{\ast}} , \qquad
    n_{\mathsf{N}}' = n_{\mathfrak{n}} - n_{\mathfrak{n}^{\ast}} .
\end{equation}
We are interested in variations $\delta\mathcal{E}$ such that the total baryon density remains unchanged, $\delta n_{\mathcal{B}} = 0$, and that the proton fraction is also constant, $\delta x = 0$. It follows that
\begin{IEEEeqnarray}{rCl}
    \delta n_{\mathsf{P}} & = & \delta \left(x n_{\mathcal{B}}\right) = 0, \\[0.1cm]
    \delta n_{\mathsf{N}} & = & \delta \left[\left(1 - x\right) n_{\mathcal{B}}\right] = 0,
\end{IEEEeqnarray}
and, additionally,
\begin{IEEEeqnarray}{rCl}
    \delta n_{\mathsf{P}}' & = & \delta n_{\mathfrak{p}} - \delta n_{\mathfrak{p}^{\ast}} \equiv \delta (n_{\mathsf{P}} - n_{\mathfrak{p}^{\ast}}) - \delta n_{\mathfrak{p}^{\ast}} = - 2 \delta n_{\mathfrak{p}^{\ast}}, \qquad\ \\[0.1cm]
    \delta n_{\mathsf{N}}' & = & \delta n_{\mathfrak{n}} - \delta n_{\mathfrak{n}^{\ast}} \equiv \delta (n_{\mathsf{N}} - n_{\mathfrak{n}^{\ast}}) - \delta n_{\mathfrak{n}^{\ast}} = - 2 \delta n_{\mathfrak{n}^{\ast}}. 
\end{IEEEeqnarray}

Expanding Eq.~(\ref{eq:defvariation}) to linear order in $\delta n_{\mathsf{P}}' = n_{\mathsf{P}}' - \bar{n}_{\mathsf{P}}'$ and $\delta n_{\mathsf{N}}' = n_{\mathsf{N}}' - \bar{n}_{\mathsf{N}}'$ , taking into account that $\delta n_{\mathsf{P}}=\delta n_{\mathsf{N}}=0, $ one gets
\beq\label{eq:linercalE}
\delta \mathcal{E}\simeq\mu_{\mathsf{P}}' \delta n_{\mathsf{P}}'+\mu_{\mathsf{N}}' \delta n_{\mathsf{N}}',
\eeq
with the ``parity-chemical potentials'' $\mu_{\mathsf{P},\mathsf{N}}'$ given by
\begin{equation}
    \mu_{\mathsf{P}}' = \frac{1}{2} \left(\mu_{\mathfrak{p}} - \mu_{\mathfrak{p}^{\ast}}\right), \qquad
    \mu_{\mathsf{N}}' = \frac{1}{2} \left(\mu_{\mathfrak{n}} - \mu_{\mathfrak{n}^{\ast}}\right),
\end{equation}
where
\begin{equation}
    \mu_{\mathfrak{p},\mathfrak{p}^{\ast}} = \frac{\partial \mathcal{E}}{\partial n_{\mathfrak{p},\mathfrak{p}^{\ast}}} , \qquad
    \mu_{\mathfrak{n},\mathfrak{n}^{\ast}} = \frac{\partial \mathcal{E}}{\partial n_{\mathfrak{n},\mathfrak{n}^{\ast}}} .
\end{equation}
For the system in  equilibrium at a given baryon density $n_{\mathcal{B}}$ and a given proton fraction $x$, we have $\mu_{\mathfrak{p}} = \mu_{\mathfrak{p}^{\ast}} \equiv \mu_{\mathsf{P}}$ and $\mu_{\mathfrak{n}} = \mu_{\mathfrak{n}^{\ast}} \equiv \mu_{\mathsf{N}}$, hence $\mu_{\mathsf{P}}' = \mu_{\mathsf{N}}' = 0$. Therefore the linear variation (\ref{eq:linercalE}) vanishes.

The expansion of Eq.~(\ref{eq:defvariation}) up to second order in $\delta n_{\mathsf{P}}' $ and $\delta n_{\mathsf{N}}'$ gives access to the parity symmetry energy density. We obtain
\begin{IEEEeqnarray}{rCl}
    \delta^{2} \mathcal{E} & \simeq & \frac{1}{2} \Bigg[ \left.\frac{\partial \mu_{\mathsf{P}}'}{\partial n_{\mathsf{P}}'}\right|_{\bar{n}_{\mathsf{P}}',\bar{n}_{\mathsf{N}}'} \left(\delta n_{\mathsf{P}}'\right)^{2} + \left.\frac{\partial \mu_{\mathsf{N}}'}{\partial n_{\mathsf{N}}'}\right|_{\bar{n}_{\mathsf{P}}',\bar{n}_{\mathsf{N}}'} \left(\delta n_{\mathsf{N}}'\right)^{2} 
    \nonumber\\[0.2cm]
    & & \quad  + \left(\left.\frac{\partial \mu_{\mathsf{P}}'}{\partial n_{\mathsf{N}}'}\right|_{\bar{n}_{\mathsf{P}}',\bar{n}_{\mathsf{N}}'} + \left.\frac{\partial \mu_{\mathsf{N}}'}{\partial n_{\mathsf{P}}'}\right|_{\bar{n}_{\mathsf{P}}',\bar{n}_{\mathsf{N}}'}\right) \delta n_{\mathsf{P}}' \delta n_{\mathsf{N}}' \Bigg] . \qquad
\end{IEEEeqnarray}
 In analogy to Eqs.~(\ref{eq:dmuIdnIsimp}), (\ref{eq:definitionSnB}), and (\ref{eq:SnB}) we rewrite this as follows (leaving out the arguments $\bar{n}_{\mathsf{P}}$ and $\bar{n}_{\mathsf{N}}$):
\begin{IEEEeqnarray}{rCl}
    \delta^{2} \mathcal{E} & \simeq & \frac{1}{n_{\mathcal{B}}} 
    \Big[ \tilde{S}_{\mathsf{P}}\left(\bar{n}_{\mathsf{P}}',\bar{n}_{\mathsf{N}}'\right) \left(\delta n_{\mathsf{P}}'\right)^{2}  
    + \tilde{S}_{\mathsf{N}}\left(\bar{n}_{\mathsf{P}}',\bar{n}_{\mathsf{N}}'\right) \left(\delta n_{\mathsf{N}}'\right)^{2} \nonumber\\[0.1cm]
    & & \qquad  +\, 2 \tilde{S}_{\mathsf{PN}}\left(\bar{n}_{\mathsf{P}}',\bar{n}_{\mathsf{N}}'\right) \delta n_{\mathsf{P}}' \delta n_{\mathsf{N}}' \Big] ,
\end{IEEEeqnarray}
with the (parity) symmetry energies $\tilde{S}_{\mathsf{P}}$, $\tilde{S}_{\mathsf{N}}$, and $\tilde{S}_{\mathsf{PN}}$ defined as functions of the equilibrium parity densities  $\bar{n}_{\mathsf{P}}'$ and $\bar{n}_{\mathsf{N}}'$ by
\begin{IEEEeqnarray}{rCl}
    \tilde{S}_{\mathsf{P}}\left(\bar{n}_{\mathsf{P}}',\bar{n}_{\mathsf{N}}'\right) & = & \frac{1}{2} n_{\mathcal{B}} \left.\frac{\partial \mu_{\mathsf{P}}'}{\partial n_{\mathsf{P}}'}\right|_{\bar{n}_{\mathsf{P}}',\bar{n}_{\mathsf{N}}'}, \\[0.2cm]
    \tilde{S}_{\mathsf{N}}\left(\bar{n}_{\mathsf{P}}',\bar{n}_{\mathsf{N}}'\right) & = & \frac{1}{2} n_{\mathcal{B}} \left.\frac{\partial \mu_{\mathsf{N}}'}{\partial n_{\mathsf{N}}'}\right|_{\bar{n}_{\mathsf{P}}',\bar{n}_{\mathsf{N}}'}, \\[0.2cm]
    \tilde{S}_{\mathsf{PN}}\left(\bar{n}_{\mathsf{P}}',\bar{n}_{\mathsf{N}}'\right) & = & \frac{1}{2} n_{\mathcal{B}} \left.\frac{\partial \mu_{\mathsf{P}}'}{\partial n_{\mathsf{N}}'}\right|_{\bar{n}_{\mathsf{P}}',\bar{n}_{\mathsf{N}}'}
    \equiv \frac{1}{2} n_{\mathcal{B}} \left.\frac{\partial \mu_{\mathsf{N}}'}{\partial n_{\mathsf{P}}'}\right|_{\bar{n}_{\mathsf{P}}',\bar{n}_{\mathsf{N}}'} ,  \qquad
\end{IEEEeqnarray}
and where the last identity must hold due to the intrinsic $\mathsf{P} \leftrightarrow \mathsf{N}$ symmetry of the model (at equilibrium). These expressions further evaluate as 
\begin{IEEEeqnarray}{rCl}
    \tilde{S}_{\mathsf{P}} & = & \frac{n_{\mathcal{B}}}{4}
    \left[\frac{1}{2} \left(\frac{1}{N_{0}^{\mathfrak{p}}} + \frac{1}{N_{0}^{\mathfrak{p}^{\ast}}}\right)
    + \tilde{f}_{\mathsf{P}}^{\prime\mathfrak{p}} - \tilde{f}_{\mathsf{P}}^{\prime\mathfrak{p}^{\ast}} \right] , \label{eq:SP1} \\[0.2cm]
    \tilde{S}_{\mathsf{N}} & = & \frac{n_{\mathcal{B}}}{4}
    \left[\frac{1}{2} \left(\frac{1}{N_{0}^{\mathfrak{n}}} + \frac{1}{N_{0}^{\mathfrak{n}^{\ast}}}\right)
    + \tilde{f}_{\mathsf{N}}^{\prime\mathfrak{n}} - \tilde{f}_{\mathsf{N}}^{\prime\mathfrak{n}^{\ast}} \right] , \label{eq:SN1} \\[0.2cm]
    \tilde{S}_{\mathsf{PN}} & = & \frac{n_{\mathcal{B}}}{4}
    \left(\tilde{f}_{\mathsf{N}}^{\prime\mathfrak{p}} - \tilde{f}_{\mathsf{N}}^{\prime\mathfrak{p}^{\ast}} \right) 
    \equiv \frac{n_{\mathcal{B}}}{4}
    \left(\tilde{f}_{\mathsf{P}}^{\prime\mathfrak{n}} - \tilde{f}_{\mathsf{P}}^{\prime\mathfrak{n}^{\ast}} \right) , \qquad \label{eq:SPN1}
\end{IEEEeqnarray}
with $n_{\mathsf{P}}' = \bar{n}_{\mathsf{P}}'$ and $n_{\mathsf{N}}' = \bar{n}_{\mathsf{N}}'$. Here we introduced the Landau parameters 
\begin{IEEEeqnarray}{rCl}
    \tilde{f}_{\mathsf{P}}^{\prime \mathfrak{p},\mathfrak{p}^{\ast}} & = & \left.\frac{\partial E_{\p}^{+\pm}}{\partial n_{\mathsf{P}}'}\right|_{|\p|\, =\, p_{\mathfrak{p},\mathfrak{p}^{\ast}}} = y_{\pm} \frac{M_{\pm}}{M_{\ast}^{\mathfrak{p},\mathfrak{p}^{\ast}}} \frac{\partial\sigma}{\partial n_{\mathsf{P}}'} , \label{eq:ftildeprime1} \\[0.2cm]
    \tilde{f}_{\mathsf{N}}^{\prime \mathfrak{p},\mathfrak{p}^{\ast}} & = & \left.\frac{\partial E_{\p}^{+\pm}}{\partial n_{\mathsf{N}}'}\right|_{|\p|\, =\, p_{\mathfrak{p},\mathfrak{p}^{\ast}}} = y_{\pm} \frac{M_{\pm}}{M_{\ast}^{\mathfrak{p},\mathfrak{p}^{\ast}}} \frac{\partial\sigma}{\partial n_{\mathsf{N}}'} , \label{eq:ftildeprime2} \\[0.2cm]
    \tilde{f}_{\mathsf{P}}^{\prime \mathfrak{n},\mathfrak{n}^{\ast}} & = & \left.\frac{\partial E_{\p}^{-\pm}}{\partial n_{\mathsf{P}}'}\right|_{|\p|\, =\, p_{\mathfrak{n},\mathfrak{n}^{\ast}}} = y_{\pm} \frac{M_{\pm}}{M_{\ast}^{\mathfrak{n},\mathfrak{n}^{\ast}}} \frac{\partial\sigma}{\partial n_{\mathsf{P}}'} , \label{eq:ftildeprime3} \\[0.2cm]
    \tilde{f}_{\mathsf{N}}^{\prime \mathfrak{n},\mathfrak{n}^{\ast}} & = & \left.\frac{\partial E_{\p}^{-\pm}}{\partial n_{\mathsf{N}}'}\right|_{|\p|\, =\, p_{\mathfrak{n},\mathfrak{n}^{\ast}}} = y_{\pm} \frac{M_{\pm}}{M_{\ast}^{\mathfrak{n},\mathfrak{n}^{\ast}}} \frac{\partial\sigma}{\partial n_{\mathsf{N}}'} . \quad \label{eq:ftildeprime4}
\end{IEEEeqnarray}
By using the derivative
\begin{IEEEeqnarray}{rCl}
    \frac{\partial \sigma}{\partial n_{\mathsf{P}}'} & = & \frac{1}{2} \left(\frac{\partial \sigma}{\partial n_{\mathfrak{p}}} - \frac{\partial \sigma}{\partial n_{\mathfrak{p}^{\ast}}}\right) \nonumber\\[0.2cm]
    & = & - \frac{1}{2 m_{\sigma}^{2}} \left( \frac{y_{+} M_{+}}{M_{\ast}^{\mathfrak{p}}} - \frac{y_{-} M_{-}}{M_{\ast}^{\mathfrak{p}^{\ast}}} \right) ,
\end{IEEEeqnarray}
and the analogous derivative $\partial\sigma /\partial n_{\mathsf{N}}'$, we eventually get
\begin{IEEEeqnarray}{rCl}
    \tilde{S}_{\mathsf{P}} & = & \frac{n_{\mathcal{B}}}{8}
    \left[\frac{1}{N_{0}^{\mathfrak{p}}} + \frac{1}{N_{0}^{\mathfrak{p}^{\ast}}}
    - \frac{(y_{+} M_{+} - y_{-} M_{-})^{2}}{m_{\sigma}^{2} (M_{\ast}^{\mathsf{P}})^{2}} \right] , \\[0.2cm]
    \tilde{S}_{\mathsf{N}} & = & \frac{n_{\mathcal{B}}}{8} 
    \left[\frac{1}{N_{0}^{\mathfrak{n}}} + \frac{1}{N_{0}^{\mathfrak{n}^{\ast}}}
    - \frac{(y_{+} M_{+} - y_{-} M_{-})^{2}}{m_{\sigma}^{2} (M_{\ast}^{\mathsf{N}})^{2}} \right] , \label{eq:SN2} \qquad\quad  \\[0.2cm]
    \tilde{S}_{\mathsf{PN}} & = & - \frac{n_{\mathcal{B}}}{8}
    \frac{(y_{+} M_{+} - y_{-} M_{-})^{2}}{m_{\sigma}^{2} M_{\ast}^{\mathsf{P}} M_{\ast}^{\mathsf{N}}} .
\end{IEEEeqnarray}
These parity symmetry energies enter the differential equations that we use to compute the thermodynamics of the parity-doublet model, as  discussed in detail in Appendix~\ref{sec:diffeqs}.

We may now rewrite the energy per particle in terms of the parity symmetry energies
\begin{IEEEeqnarray}{rCl}
    \frac{\delta^{2}\mathcal{E}}{n_{\mathcal{B}}} \simeq 4 & \Big[ & x^{2} \tilde{S}_{\mathsf{P}} \left(\delta x_{\mathsf{P}}'\right)^{2} + (1 - x)^{2} \tilde{S}_{\mathsf{N}} \left(\delta x_{\mathsf{N}}'\right)^{2} \nonumber\\
    & & +\, 2x(1 - x) \tilde{S}_{\mathsf{PN}} \delta x_{\mathsf{P}}' \delta x_{\mathsf{N}}' \Big] ,
    \label{eq:EAparity}
\end{IEEEeqnarray}
where we introduced the dimensionless parity ratios
\begin{equation}
    x_{\mathsf{P}}' = \frac{n_{\mathfrak{p}^{\ast}}}{n_{\mathsf{P}}}, \qquad
    x_{\mathsf{N}}' = \frac{n_{\mathfrak{n}^{\ast}}}{n_{\mathsf{N}}}; \qquad
    x_{\mathsf{P}}', x_{\mathsf{N}}' \le \frac{1}{2}.
\end{equation}
This general formula allows us to assess the change in the energy per particle when moving away from the respective equilibrium point (along a path of fixed proton fraction $x$, and fixed baryon density $n_{\mathcal{B}}$). For instance, if the chiral partners $\mathfrak{p}^{\ast}$ and $\mathfrak{n}^{\ast}$ are populated at the same rate, i.e., $\delta x_{\mathsf{P}}' = \delta x_{\mathsf{N}}' \equiv \delta x'$, then Eq.~(\ref{eq:EAparity}) becomes
\begin{IEEEeqnarray}{rCl}
    \frac{\delta^{2}\mathcal{E}}{n_{\mathcal{B}}} \simeq 4 \tilde{S}_{x} \left(\delta x'\right)^{2} ,
\end{IEEEeqnarray}
with the $x$-dependent parity symmetry energy
\begin{equation}\label{eq:paritysymenergy}
    \tilde{S}_{x} = x^{2} \tilde{S}_{\mathsf{P}} + (1 - x)^{2} \tilde{S}_{\mathsf{N}} + 2x(1 - x) \tilde{S}_{\mathsf{PN}} .
\end{equation}
A further simplification occurs if the equilibrium point additionally lies at $\bar{x}' = \frac{1}{2}$, which means that the densities of the nucleons and their respective chiral partners are equal, $\bar{n}_{\mathsf{P}}' = \bar{n}_{\mathsf{N}}' = 0$. This equality implies that the $\sigma$ field vanishes, as it is required that $M_{+} = M_{-} = m_{0}$ in order to achieve the pair-wise equilibration of the densities of the parity doublet. Then we have 
\begin{equation}
    \frac{\delta^{2}\mathcal{E}}{n_{\mathcal{B}}} \simeq \tilde{E}_{\mathrm{sym}}^{x} \left(2 x' - 1\right)^{2} ,
    \label{eq:parabola_parity}
\end{equation}
with $x' \le \frac{1}{2}$, and
\begin{equation}\label{eq:EsymFx}
    \tilde{E}_{\mathrm{sym}}^{x} = \frac{1}{6} \left[x \frac{p_{\mathsf{P}}^{2}}{M_{\ast}^{\mathsf{P}}} + (1 - x) \frac{p_{\mathsf{N}}^{2}}{M_{\ast}^{\mathsf{N}}}\right] \left(1 + \tilde{F}_{x}'\right) , 
\end{equation}
where the Fermi momenta $p_{\mathsf{P}}$ and $p_{\mathsf{N}}$ are given by
\begin{equation}
    p_{\mathsf{P}} = \left(3\pi^{2} x \frac{n_{\mathcal{B}}}{2}\right)^{\frac{1}{3}}, \quad 
    p_{\mathsf{N}} = \left[3\pi^{2} \left(1 - x\right) \frac{n_{\mathcal{B}}}{2}\right]^{\frac{1}{3}} . \label{eq:Fermi_momenta_x}
\end{equation}
The parity symmetry energy $\tilde{E}_{\mathrm{sym}}^{x}$ bears a strong similarity with the isospin symmetry energy of Eq.~(\ref{eq:EsymF0}), the first contribution in Eq.~(\ref{eq:EsymFx}) being the respective kinetic energies of proton-like and neutron-like quasiparticles, weighted by the proton fraction $x$. The second contribution corresponds to the effect of the interactions, expressed in terms of (properly normalized) Fermi-liquid parameters,
\begin{IEEEeqnarray}{rCl}
    \tilde{F}_{x}' & = & \frac{1}{2} \bigg[ x^{2} \left(\tilde{f}_{\mathsf{P}}^{\prime \mathfrak{p}} - \tilde{f}_{\mathsf{P}}^{\prime \mathfrak{p}^{\ast}}\right) + (1 - x)^{2} \left(\tilde{f}_{\mathsf{N}}^{\prime \mathfrak{n}} - \tilde{f}_{\mathsf{N}}^{\prime \mathfrak{n}^{\ast}}\right) \nonumber\\[0.2cm]
    & & \quad\ +\, 2 x (1 - x) \left(\tilde{f}_{\mathsf{P}}^{\prime \mathfrak{n}} - \tilde{f}_{\mathsf{P}}^{\prime \mathfrak{n}^{\ast}} \right) \bigg] \left\slash \left[\frac{x^{2}}{N_{0}^{\mathsf{P}}} + \frac{(1 - x)^{2}}{N_{0}^{\mathsf{N}}}\right] \right. \nonumber\\[0.2cm]
    & \equiv & - \left. \frac{(y_{a} - y_{b})^{2} m_{0}^{2}}{4 m_{\sigma}^{2} (M_{\ast}^{x})^{2}}
    \right\slash \left[\frac{x^{2}}{N_{0}^{\mathsf{P}}} + \frac{(1 - x)^{2}}{N_{0}^{\mathsf{N}}}\right]. 
\end{IEEEeqnarray}
In this equation $N_{0}^{\mathsf{P}} = N_{0}^{\mathfrak{p}} + N_{0}^{\mathfrak{p}^{\ast}}$ and $N_{0}^{\mathsf{N}} = N_{0}^{\mathfrak{n}} + N_{0}^{\mathfrak{n}^{\ast}}$ are densities of states, and 
\begin{equation}
    M_{\ast}^{x} = \left(\frac{x}{M_{\ast}^{\mathsf{P}}} + \frac{1 - x}{M_{\ast}^{\mathsf{N}}}\right)^{-1} 
\end{equation}
is the Landau effective mass. 
Moreover, we  used that
\begin{equation}
    y_{+}(\sigma = 0) = \frac{y_{a} - y_{b}}{2} \equiv - y_{-}(0) .
\end{equation}
Finally, an even simpler expression is found in symmetric matter ($x = \frac{1}{2}$),
\begin{equation}
    \tilde{E}_{\mathrm{sym}}^{\frac{1}{2}} = \frac{p_{F}^{2}}{6 M_{\ast}} \left(1 + \tilde{F}_{\frac{1}{2}}'\right) ,
\end{equation}
with
\begin{equation}
    \tilde{F}_{\frac{1}{2}}' = - N_{0} \frac{(y_{a} - y_{b})^{2} m_{0}^{2}}{4 m_{\sigma}^{2} M_{\ast}^{2}} 
    = - \frac{p_{F}}{\pi^{2}} \frac{(y_{a} - y_{b})^{2} m_{0}^{2}}{m_{\sigma}^{2} M_{\ast}} ,
\end{equation}
and $N_{0} = N_{0}^{+} + N_{0}^{-} \equiv 4 p_{F} M_{\ast}/\pi^{2}$,  $p_{F} = (3\pi^{2} n_{\mathcal{B}}/4)^{1/3}$.

\begin{figure}[ht]
	\centering
	\includegraphics[scale=1.0]{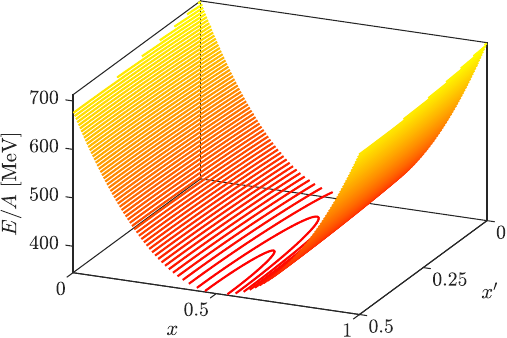}
	\caption{Three-dimensional contour plot of the energy per particle in the chiral limit, $E/A = \mathcal{E}/n_{\mathcal{B}} - M_{N}$, at the baryon density $n_{\mathcal{B}} = 15 \, n_{0}$, and as a function of the proton fraction $x$ and the parity fraction $x' = n_{\mathfrak{p}^{\ast}}/n_{\mathsf{P}} = n_{\mathfrak{n}^{\ast}}/n_{\mathsf{N}}$.}
	\label{fig:EAasym}
\end{figure}

In Fig.~\ref{fig:EAasym} we plot the energy per particle (in the chiral limit) in the chiral-restored phase at $n_{\mathcal{B}} = 15 \, n_{0}$. The contours show again the parabolic shape of the energy per particle with respect to the proton fraction $x$, with the minimum lying at the point of isospin-symmetric matter, $x = \frac{1}{2}$. Likewise, the contours further reveal a (half-)parabolic shape in the direction of the parity ratio $x' = x_{\mathsf{P}}' = x_{\mathsf{N}}'$, which is much more shallow. Thus the overall minimum of the energy per particle is located at $(x', x) = (\frac{1}{2}, \frac{1}{2})$ (where $\sigma = 0$), and amounts to $E/A \approx 345\ \mathrm{MeV}$. At this equilibrium point the masses of all fermions are degenerate at $m_{0}$, and all fermion densities are equal.

\begin{figure}[ht]
	\centering
	\includegraphics[scale=1.0]{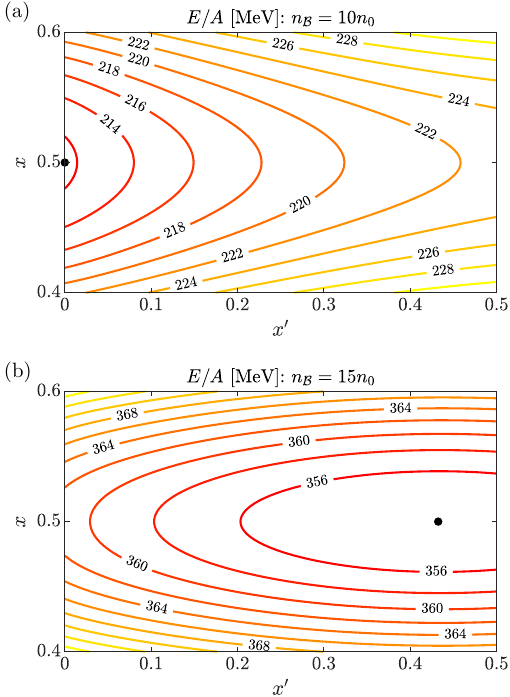}
	\caption{Contour plot of the energy per particle (physical pion mass) as a function of the proton fraction $x$ and the parity fraction $x'$: (a) $n_{\mathcal{B}} = 10 \, n_{0}$; (b) $n_{\mathcal{B}} = 15 \, n_{0}$. The black dots indicate the respective points of minimum energy, i.e.\ the equilibrium points.}
	\label{fig:EAasym_xprime}
\end{figure}

Further insight is obtained by following the location of the equilibrium point as the total baryon density $n_{\mathcal{B}}$ increases. Thus we show in  Fig.~\ref{fig:EAasym_xprime} the energy per particle for two different baryon densities, $n_{\mathcal{B}} = 10\, n_{0}$ and  $n_{\mathcal{B}} = 15\, n_{0}$ (now for the case of physical pion mass). The minimum at $n_{\mathcal{B}} = 10\, n_{0}$ obviously lies at $(x', x) = (0, \frac{1}{2})$, corresponding to $n_{\mathcal{B}}^{-} = 0$. At larger density, $n_{\mathcal{B}} = 15\, n_{0}$, the minimum has moved to the right, close to the parity-symmetric point at $x' = \frac{1}{2}$, as $n_{\mathcal{B}}^{-} > 0$. These findings match the composition diagram in Fig.~\ref{fig:thresholds_diagram}, where in symmetric matter the chiral partners remain unpopulated at $n_{\mathcal{B}} = 10\, n_{0}$ below the thresholds, while at $n_{\mathcal{B}} = 15\, n_{0}$ (in the region ``$\oplus \mathfrak{p}^{\ast} + \mathfrak{n}^{\ast}$,'' although beyond the scale of the plot) the densities of the chiral partners are nonzero, and actually already close to the densities of the nucleons. 

\begin{figure}[ht]
	\centering
	\includegraphics[scale=1.0]{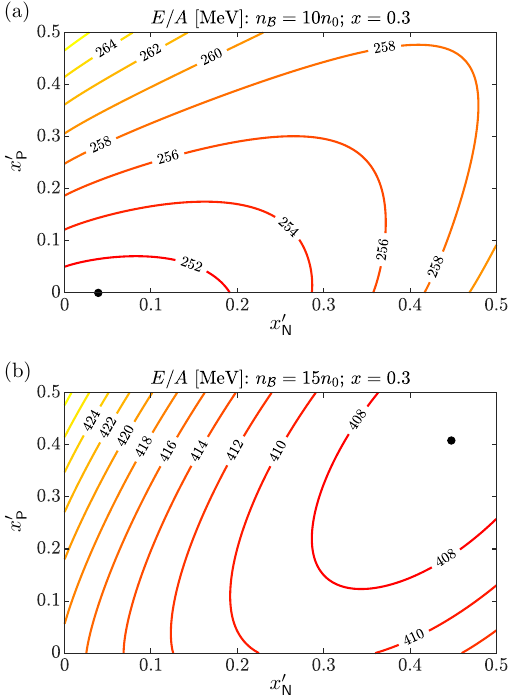}
	\caption{Contour plot of the energy per particle (physical pion mass) for fixed proton fraction $x = 0.3$, as a function of the neutron parity ratio $x_{\mathsf{N}}'$ and the proton parity ratio $x_{\mathsf{P}}'$: (a) $n_{\mathcal{B}} = 10 \, n_{0}$; (b) $n_{\mathcal{B}} = 15 \, n_{0}$. The black dots indicate the equilibrium points.}
	\label{fig:EAasym_x}
\end{figure}

We consider now a fixed value of the proton fraction, e.g.\ $x = 0.3$, and plot in Fig.~\ref{fig:EAasym_x} the energy per particle as a function of the neutron and proton parity fractions,  respectively $x_{\mathsf{N}}'$ and $x_{\mathsf{P}}'$. The two panels represent again the two baryon densities $n_{\mathcal{B}} = 10\, n_{0}$ and $n_{\mathcal{B}} = 15\, n_{0}$. In the former case the equilibrium point lies at $(x_{\mathsf{N}}',x_{\mathsf{P}}') \approx (0.04, 0)$, meaning that we already passed the threshold of the $\mathfrak{n}^{\ast}$, but not yet the one of the $\mathfrak{p}^{\ast}$, which is again consistent with Fig.~\ref{fig:thresholds_diagram}. When increasing the total baryon density to $n_{\mathcal{B}} = 15 \, n_{0}$, the equilibrium point moves to the upper right corner, as we then substantially increased the populations of both the $\mathfrak{n}^{\ast}$ and the $\mathfrak{p}^{\ast}$ beyond the chiral transition, coming closer to the parity-symmetric point $(x_{\mathsf{N}}',x_{\mathsf{P}}') = (\frac{1}{2}, \frac{1}{2})$. 
These figures illustrate how the surface of the energy per particle $E/A$ arranges for a unique minimum that corresponds to the physical solution. When $n_{\mathcal{B}}$ increases, the surface changes its shape and ``guides'' the system towards a minimum that corresponds to a chirally symmetric state ($x_{\mathsf{N}}' \rightarrow \frac{1}{2}$ and $x_{\mathsf{P}}' \rightarrow \frac{1}{2}$).

\subsubsection{Lower bound for the chiral transition density}

Another important characteristic of the parity-doublet model is that the threshold functions for the onsets of the populations of the parity partners do not depend on any bosonic parameters. The bosonic potential as well as the pion mass (controlled by the parameter $h$) only decide on where the threshold lines are crossed. To be more specific, the two threshold conditions $M_{\ast}^{\mathfrak{p},\mathfrak{n}} = M_{-}$ determine two functions $\sigma_{\mathrm{thresh}}^{x}(n_{\mathcal{B}})$ that do not depend on the bosonic parameters of the model, since neither the Landau masses nor the fermion masses $M_{\pm}$ depend on any other model parameters than the fermionic parameters $m_{0}$, $y_{a}$, and $y_{b}$ (once $x$ is fixed). The only place where the bosonic parameters come into play is the actual solution of the gap equation, $\sigma (n_{\mathcal{B}})$, so that at threshold we have $\sigma(n_{\mathrm{thresh}}) = \sigma_{\mathrm{thresh}}^{x}(n_{\mathrm{thresh}})$.

\begin{figure}[ht]
	\centering
    \includegraphics[scale=1.0]{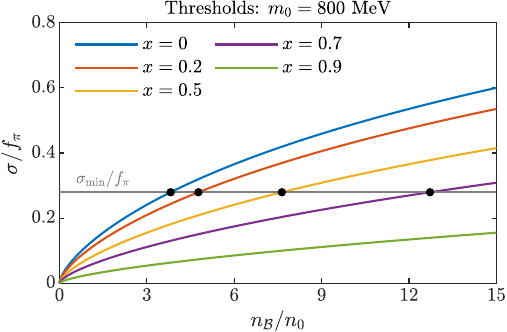}
	\caption{Threshold lines according to the $\mathfrak{n}^{\ast}$-onset condition $M_{\ast}^{\mathfrak{n}} = M_{-}$, for different proton fractions $x$, and for the chosen mass parameter $m_{0} = 800\ \mathrm{MeV}$. The black dots show intersections of the threshold lines with $\sigma = \sigma_{\mathrm{min}}$ (gray horizontal line), the corresponding baryon densities being determined by Eq.~(\ref{eq:onset_min}).}
	\label{fig:thresholds_sigma_min}
\end{figure}

In Fig.~\ref{fig:thresholds_sigma_min} we have drawn the threshold lines corresponding to the $\mathfrak{n}^{\ast}$-onset condition, for different values of $x$ (the lines would be identical for the $\mathfrak{p}^{\ast}$-onset, with the replacement $x \mapsto 1 - x$). For a given $x$, the solution of the gap equation (e.g.\ the one shown in Fig.~\ref{fig:phase_diagram_density_PD})  hits the threshold line at a specific baryon density $n_{\mathrm{thresh}}$, in the direction of increasing $n_{\mathcal{B}}$, thereby opening up  the phase space that allows for the  population of the $\mathfrak{n}^{\ast}$. These crossing points ($n_{\mathrm{thresh}}$ for different $x$) between the solution of the gap equation and the threshold functions are exactly the onset lines shown in Fig.~\ref{fig:thresholds_diagram}.

A special crossing point with the threshold function occurs, if we consider the specific value $\sigma = \sigma_{\mathrm{min}}$, at which the nucleon mass becomes minimum, i.e.
\begin{equation}\label{eq:Yukawazero}
    y_{+}(\sigma_{\mathrm{min}}) = \left. \frac{\rmd M_{+}}{\rmd \sigma} \right|_{\sigma_{\mathrm{min}}} = 0.
\end{equation}
The crossing points of the threshold lines with the value of $\sigma_{\mathrm{min}}$ are shown in Fig.~\ref{fig:thresholds_sigma_min}. As described above, the corresponding density $n_{\mathrm{thresh}}$ increases with increasing $x$, and since the threshold lines tend to zero identically as $x \rightarrow 1$, the crossing point is eventually pushed to infinity. With the help of Eq.~(\ref{eq:sigmamin}) we can compute the threshold density as follows: The  $\mathfrak{n}^{\ast}$-threshold line, $M_{\ast}^{\mathfrak{n}} = M_{-}$, when evaluated at $\sigma = \sigma_{\mathrm{min}}$ yields
\begin{equation}\label{eq:thresh_sigmamin}
    p_{\mathfrak{n}}  = \sigma_{\mathrm{min}} (y_{a} + y_{b}) = \frac{y_{b} - y_{a}}{\sqrt{y_{a} y_{b}}} m_{0}.
\end{equation}
The relation between $x$ and $n_{\mathcal{B}}$ that fulfills the threshold condition (\ref{eq:thresh_sigmamin}) manifestly  depends only on the fermionic parameters $m_{0}$, $y_{a}$, and $y_{b}$, and not on the bosonic potential, and in particular not on the pion mass. If we additionally eliminate $y_{a}$ and $y_{b}$ in favor of the physical masses $M_{N}$ and $M_{N^{\ast}}$, we get  
\begin{equation}
    p_{\mathrm{min}}(m_{0}) = \frac{y_{b} - y_{a}}{\sqrt{y_{a} y_{b}}} m_{0}
    \equiv \frac{(M_{N^{\ast}} - M_{N}) m_{0}}{\sqrt{M_{N^{\ast}} M_{N} - m_{0}^{2}}} , 
    \label{eq:p_min}
\end{equation}
such that
\begin{equation}
    n_{\mathcal{B}} = \frac{p_{\mathrm{min}}(m_{0})^{3}}{3 \pi^{2} (1 - x)} .
    \label{eq:onset_min}
\end{equation}
In that case, the thresholds, which constitute crucial ingredients in the solution of the gap equations, depend solely  on $m_{0}$. Note that both $\sigma_{\rm min}(m_0)$ and $p_{\mathrm{min}}(m_{0})$ are increasing functions of $m_0$ so that both quantities can be lowered by decreasing $m_0$ \cite{Zschiesche:2006zj,gao2024exploringfirstorderphasetransition}.

This simple formula determines the (minimum) onset baryon density of the $\mathfrak{n}^{\ast}$ for the given value of $m_{0}$, and for any value of $x$, if the $\sigma$ value at threshold equals $\sigma_{\mathrm{min}}$. Equation~(\ref{eq:onset_min}) also dictates the minimum onset baryon density for the $\mathfrak{p}^{\ast}$, by replacing $x \mapsto 1 - x$. The formula yields $n_{\mathcal{B}} \approx 3.82 \, n_{0}$ for neutron matter and $n_{\mathcal{B}} \approx 7.63 \, n_{0}$ for symmetric matter,  significantly lower than the values of the chiral transition quoted in the previous sections. We recall however that these estimates correspond to lower bounds for the chiral transition density \cite{Eser:2023oii}.

\subsection{Summary on asymmetric matter}

We now summarize the results of this section about asymmetric matter. In both the parity-doublet model and its corresponding singlet model, we observe a first-order liquid-gas transition, which becomes a mere crossover in the case of pure neutron matter. Because the parity partners play no role at the densities where the transition occurs,  both models yield nearly identical results. In fact, by parameter adjustments, the critical endpoint of the liquid-gas transition is located in both models at $(n_{c}, T_{c}) \approx (0.06\ \mathrm{fm}^{-3}, 18\ \mathrm{MeV})$ (for symmetric matter), thereby also matching experimental data on various critical observables \cite{Elliott:2013pna, Eser:2023oii}.

For larger baryon densities the behaviors of the doublet and singlet models are entirely different. In the parity-doublet model the chiral transition occurs at densities beyond $10\, n_{0}$, the transition being first order for a proton fraction roughly in the range of $0.42 \lesssim x \lesssim 0.58$, and temperatures below $T_{c} \approx 8.5\ \mathrm{MeV}$. In the singlet model, the chiral transition  is a mere crossover, with the isoscalar condensate $\sigma$ smoothly approaching zero at large baryon density. The crossover in the singlet model takes place at about half the critical density of the chiral transition in the doublet model.
In the parity-doublet model,  the parity symmetry energy triggers the equilibration of the densities of the chiral partners. This equilibration sets in once the corresponding particle thresholds are passed. These thresholds are determined exclusively  by the physical vacuum masses of the nucleon and the $N^{\ast}(1535)$, as well as the mass parameter $m_{0} = 800\ \mathrm{MeV}$. 

Finally, we note that:
\begin{enumerate}[label=(\roman*)]
    \item The onset density (\ref{eq:onset_min}) is the minimum density at which the phase space opens for the population of the chiral partners. This density derives from the fact that $\sigma_{\mathrm{min}}$ is the lowest value at which the solution of the gap equation, $\sigma(n_{\mathcal{B}})$, may cross the corresponding threshold lines (see Fig.~\ref{fig:thresholds_sigma_min}).
    \item We associate with this minimum density the effective Fermi momentum $p_{\mathrm{min}}$, which is a pure function of the mass parameter $m_{0}$, if we take the physical masses $M_{N}$ and $M_{N^{\ast}}$ as fixed. The momentum $p_{\mathrm{min}}$ is independent of the parameters of the bosonic potential, hence of the value of the pion mass.
    \item The minimum density may be taken as a strict lower bound for the chiral transition density, since the opening of the phase space for the chiral partners is a (necessary) precursor of the chiral transition.
    \item The change in the lower bound due to a change in the chiral-invariant mass $m_{0}$ can be determined directly without the need to follow the entire solution of the gap equation from the vacuum to the thresholds.
\end{enumerate}
The statement (i) follows from the fact that the Yukawa coupling $y_{+}$ vanishes at the point $\sigma = \sigma_{\mathrm{min}}$, cf.\ Eq.~(\ref{eq:Yukawazero}), and we consequently get
\begin{equation}
    \left.\frac{\rmd \sigma}{\rmd n_{\mathcal{B}}}\right|_{x;\,\sigma\, = \, \sigma_{\mathrm{min}}} 
    = - \left.\frac{y_{+}}{m_{\sigma}^{2}} \frac{M_{+}}{M_{\ast}^{x}} \right|_{\sigma\, = \, \sigma_{\mathrm{min}}} 
    = 0, \label{eq:dsigmadn_initial}
\end{equation}
where the differential equation for $\sigma$ is obtained by differentiating the gap equation (\ref{eq:gap}), at a constant proton fraction $x$. Assuming that the solution of this equation is smooth up to the first crossing point with the threshold functions, one sees that the initial decrease of $\sigma$ ($\rmd \sigma /\rmd n_{\mathcal{B}} < 0$) eventually dies out when  $\sigma$ approaches $\sigma_{\mathrm{min}}$ from above.

The other statements (ii) to (iv) then follow directly from the first statement, and from exploiting the relations (\ref{eq:p_min}) and (\ref{eq:onset_min}). Furthermore, the fact that the restoration of chiral symmetry comes along with $\sigma(n_{\mathcal{B}}) \rightarrow 0$, as well as $M_{\pm}\rightarrow m_{0} $, makes  necessary the opening of the phase space for chiral partners prior to the chiral transition, as it is mentioned in statement (iii).

In Appendix~\ref{sec:chiral_limit} we further explore the chiral transition, in particular in the chiral limit where smooth crossover transitions  turn into real phase transitions. The chiral limit also illustrates there the application of the differential equations that we advertised at several places in this work (the related technical aspects are also found in Appendix~\ref{sec:diffeqs}). The subsequent section is dedicated to pure neutron matter.

\section{Neutron matter}
\label{sec:neutron_matter}

 In the context of the parity-doublet model, ``pure neutron matter'' consists of neutrons ($\mathfrak{n}$) at low baryon densities, and their chiral partners ($\mathfrak{n}^{\ast}$) beyond the corresponding threshold at $n_{\mathcal{B}} \approx 7.24 \, n_{0}$ (see Fig.~\ref{fig:thresholds_diagram}), whereas the densities of the protons and their chiral partners $\mathfrak{p}^{\ast}$ vanish at all baryon densities, $n_{\mathsf{P}} = 0$. It follows that $n_{\mathcal{B}} = n_{\mathsf{N}}$, as well as $n_{\mathcal{I}} = - n_{\mathcal{B}}$. The energy density (at $T = 0$) reduces to
\begin{equation}
    \mathcal{E}(n_{\mathcal{B}}, x = 0) = - P + \mu_{\mathsf{N}} n_{\mathsf{N}} .
\end{equation}
In this section we investigate  the properties of neutron matter, considered as the limiting case  of asymmetric matter  when $x \rightarrow 0$. We furthermore alleviate here the notation and denote the baryon density simply by $n$ instead of $n_{\mathcal{B}}$, and similarly for the neutron-chemical potential, $\mu_{\mathsf{N}} \mapsto \mu$, and the Landau effective mass, $M_{\ast}^{\mathsf{N}} \mapsto M_{\ast}$. In the singlet model, we have simply $n = n_{\mathfrak{n}}$.

\subsection{Equation of state}

In the parity-doublet model the zero-temperature pressure $P$ is given by
\begin{IEEEeqnarray}{rCl}
    P & = & - \mathcal{E}_{\mathrm{qp}}^{\mathfrak{n}}(n; \sigma) - \mathcal{E}_{\mathrm{qp}}^{\mathfrak{n}^{\ast}}(n; \sigma) - U(\sigma) + M_{\ast} n \nonumber\\[0.1cm]
    & & +\, \frac{1}{2} \left( G_{v} + G_{w} \right) n^{2} , \label{eq:Pneutron}
\end{IEEEeqnarray}
with the quasiparticle energy contribution $\mathcal{E}_{\mathrm{qp}}^{\mathfrak{n},\mathfrak{n}^{\ast}}$ as defined in Eq.~(\ref{eq:tildesepsilon}),
\begin{equation}
    \mathcal{E}_{\mathrm{qp}}^{\mathfrak{n},\mathfrak{n}^{\ast}}(n; \sigma) = 2 \int_{|p|\, \le\, p_{\mathfrak{n},\mathfrak{n}^{\ast}}} \frac{\rmd^{3}p}{(2\pi)^{3}} \sqrt{p^{2} + M_{\pm}(\sigma)^{2}} . 
\end{equation}
The Fermi momenta of the neutrons and their chiral partners read
\begin{equation}
    p_{\mathfrak{n}} = \left[3 \pi^{2} \left(1 - x'\right) n\right]^{\frac{1}{3}}, \qquad
    p_{\mathfrak{n}^{\ast}} = \left(3\pi^{2} x' n\right)^{\frac{1}{3}} ,
\end{equation}
where we used again the parity fraction $x' = n_{\mathfrak{n}^{\ast}}/n$. 
The neutron-chemical potential $\mu$ takes the form
\begin{equation}\label{eq:muGvGwn}
    \mu = M_{\ast} + \left(G_{v} + G_{w}\right) n . 
\end{equation}
Finally, the $\sigma$ condensate in the expressions above is the corresponding solution of the gap equation (\ref{eq:gap}), at the given value of the baryon density $n$.

\begin{figure}[ht]
	\centering
    \includegraphics[scale=1.0]{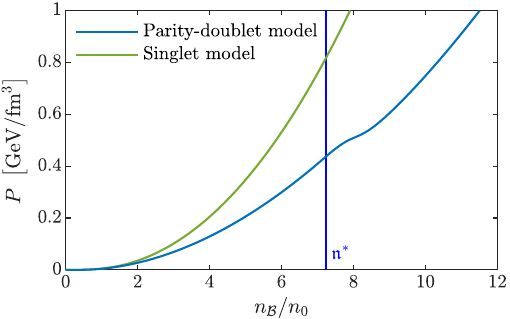}
	\caption{Equation of state (zero-temperature pressure $P$ as a function of the baryon density $n_{\mathcal{B}}$); the blue vertical line represents the onset density of the $\mathfrak{n}^{\ast}$ in the doublet model.}
	\label{fig:EOS_neutron_matter}
\end{figure}

Figure~\ref{fig:EOS_neutron_matter} displays the pressure $P(n)$ over a large range of baryon densities. In both models, $P(n)$ increases steadily, while being systematically larger in the singlet model. Thus, for instance, $P(n)$ reaches a value of $1\ \mathrm{GeV}/\mathrm{fm}^{3}$ at about $8 \, n_{0}$ in the singlet model, while in the doublet model  the same value is reached only above $11 \, n_{0}$.  To some extent, the observed difference between the pressures of the two models reflects the difference in the corresponding  values of the compression modulus in each model (see Tables~\ref{tab:parameter_choice_PD} and \ref{tab:parameter_choice_WT}).  
A special feature of the doublet model is the visible inflection of the pressure growth beyond the onset of the presence of the chiral partners of the neutrons. We shall see later that this small inflection in the pressure yields a significant drop in the speed of sound in the range of density where it occurs (see Fig.~\ref{fig:speed_of_sound} below). 

\begin{figure}[ht]
	\centering
    \includegraphics[scale=1.0]{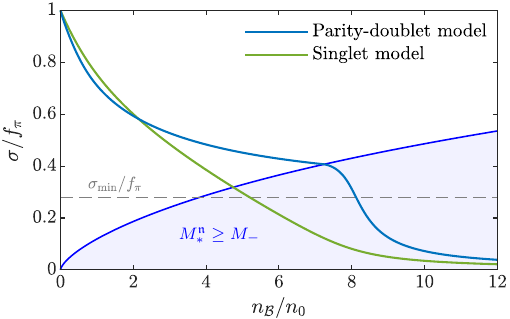}
	\caption{Solution of the gap equation for $\sigma$ in neutron matter, in both the doublet and singlet models. The blue line and the light blue zone indicate the threshold and the density regime where $M_{\ast}^{\mathfrak{n}} \ge M_{-}$.}
	\label{fig:sigma_neutron_matter}
\end{figure}

In Fig.~\ref{fig:sigma_neutron_matter} we present the solutions of the gap equations for $\sigma(n)$. At small density, the sigma field drops more rapidly in the doublet model than in the singlet model, reflecting the different values of the respective pion-nucleon sigma terms (see Eq.~(\ref{eq:sigma_term2}) below). The drop of $\sigma(n)$ in the doublet model slows down as $\sigma(n)$ approaches the value $\sigma_{\mathrm{min}}$ from above, as expected. In fact,  $\sigma(n)$ would eventually converge to $\sigma_{\mathrm{min}}$ at large density, if  $\sigma(n)$ did not hit  the $\mathfrak{n}^{\ast}$-threshold line, where new degrees of freedom appear and their equilibration sets in. This equilibration of densities of the neutrons and their chiral partners accompanies the chiral crossover, with $\sigma(n)$ further decreasing rapidly towards (almost) zero. In the singlet model, in contrast, the $\sigma$ condensate smoothly approaches zero, without any particular feature in its density dependence.

In neutron matter the lower bound of the chiral transition is predicted by Eq.~(\ref{eq:onset_min}) to be $n = 3.82 \,n_{0}$, as mentioned earlier. This lower bound is  explicitly visible in Fig.~\ref{fig:sigma_neutron_matter}, at the point where the horizontal dashed line corresponding to $\sigma_{\mathrm{min}}$ crosses the blue line and enters the light blue area (where $M_{\ast}^{\mathfrak{n}} \ge M_{-}$). The actual solution of the gap equation, $\sigma(n)$, however enters the blue area at a density well above this lower bound, namely at the $\mathfrak{n}^{\ast}$-onset indicated in Fig.~\ref{fig:EOS_neutron_matter}. This discrepancy (by almost a factor $2$) is typically due to the ``slow'' (logarithmic) decrease of $\sigma(n)$, as it was explained in (I) (see also Appendix~\ref{sec:diffeqs}). Let us recall that this slow decrease of $\sigma(n)$ is a consequence of the presence of the chiral-invariant mass $m_{0}$ in the doublet model.

\begin{figure}[ht]
	\centering
    \includegraphics[scale=1.0]{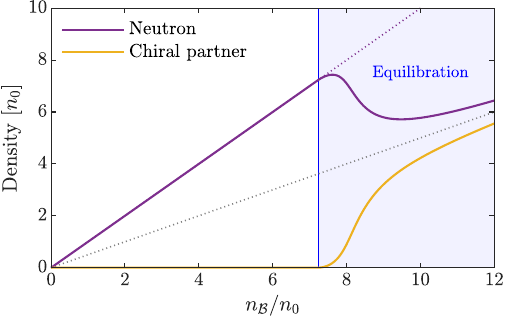}
	\caption{Equilibration of the densities of the neutrons and their chiral partners beyond the $\mathfrak{n}^{\ast}$-threshold (indicated by the vertical line), as it is induced by the parity symmetry energy.}
	\label{fig:densities_neutron_matter}
\end{figure}

The parity symmetry energy leads to the equilibration of the $\mathfrak{n}$ and $\mathfrak{n}^{\ast}$ densities, as illustrated in Fig.~\ref{fig:densities_neutron_matter}. These densities approach each other once the threshold of the chiral partner is crossed, with the density of the neutrons initially dropping before following the increase of the $\mathfrak{n}^{\ast}$.  The system releases energy through the initial reshuffling of neutron states into $\mathfrak{n}^{\ast}$-states, according to the parity symmetry energy $\tilde{S}_{\mathsf{N}}$ given in Eqs.~(\ref{eq:SN1}) and (\ref{eq:SN2}), as well as Eq.~(\ref{eq:EAparity}), 
\begin{IEEEeqnarray}{rCl}
    \frac{\delta^{2} \mathcal{E}}{n} & \simeq & 4 \tilde{S}_{\mathsf{N}} \left(\delta x'\right)^{2} 
    = \frac{p_{\mathsf{N}}^{2}}{6 M_{\ast}} \Bigg\lbrace \left[\left(\frac{4}{x'}\right)^{\frac{1}{3}} + \left(\frac{4}{1 - x'}\right)^{\frac{1}{3}}\right] \nonumber\\[0.2cm]
    & & \qquad -\, \frac{2p_{\mathsf{N}}}{\pi^{2}} \frac{(y_{+} M_{+} - y_{-} M_{-})^{2}}{m_{\sigma}^{2} M_{\ast}} \Bigg\rbrace \left(\delta x'\right)^{2} ,
\end{IEEEeqnarray}
where we recognize the kinetic contribution in the first line and, in the second line, the interaction contribution which depends on the strength of the $\sigma$ field. The Fermi momentum $p_{\mathsf{N}}$ was defined in Eq.~(\ref{eq:Fermi_momenta_x}). If the equilibrium point is attained at $x' = \frac{1}{2}$ (which is the case in the restored phase in the chiral limit), we consistently get
\begin{equation}
    \tilde{E}_{\mathrm{sym}}^{0} = \frac{p_{\mathsf{N}}^{2}}{6 M_{\ast}} \left(1 + \tilde{F}_{0}'\right), \label{eq:Esym_neutron_matter}
\end{equation}
with
\begin{equation}
    \tilde{F}_{0}' = - \frac{p_{\mathsf{N}}}{\pi^{2}} \frac{(y_{a} - y_{b})^{2} m_{0}^{2}}{2 m_{\sigma}^{2} M_{\ast}} .
\end{equation}
The equilibration mechanism that is shown in Fig.~\ref{fig:densities_neutron_matter} typically involves a negative contribution to the compressibility of the system, in the region  where the neutron density shrinks in favor of the density of the $\mathfrak{n}^{\ast}$ \cite{Marczenko_2023, koch2023fluctuations, Eser:2023oii}, so that in this region $\rmd n_{\mathfrak{n}}/\rmd \mu < 0$. This negative contribution is however overcompensated by the positive contribution from the chiral partner, so that the total compressibility always stays positive, as it should. We shall verify this statement shortly, since the derivative $\rmd n/\rmd \mu$ enters the computation of the speed of sound in neutron matter, which we consider next.

\subsection{Speed of sound and compressibility}

\begin{figure}[ht]
	\centering
    \includegraphics[scale=1.0]{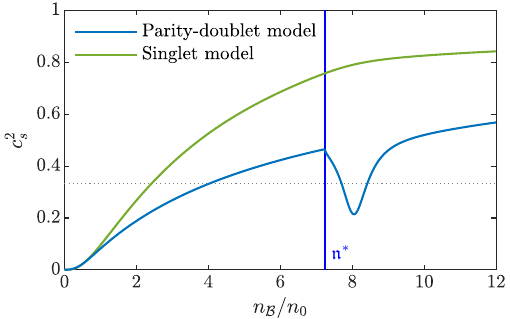}
	\caption{Squared speed of sound $c_{s}^{2}$ in neutron matter as a function of the baryon density $n_{\mathcal{B}}$. The blue vertical line indicates the onset of the chiral partner of the neutron in the doublet model. The dotted gray horizontal line indicates the conformal limit for $n_{\mathcal{B}} \rightarrow \infty$, $c_{s}^{2} = \frac{1}{3}$.}
	\label{fig:speed_of_sound}
\end{figure}

Based on the equation of state of the parity-doublet model, we can compute the speed of sound $c_{s}$ in neutron matter as
\begin{equation}\label{eq:cssqareddef}
    c_{s}^{2} = \frac{\rmd P}{\rmd \mathcal{E}} = \frac{n}{\mu} \frac{\rmd \mu}{\rmd n} ,
\end{equation}
which we can further evaluate by using the relation
\begin{IEEEeqnarray}{rCl}
    \frac{\rmd n}{\rmd \mu} & = & 2 \sum_{i\, =\, \pm 1} \int_{\p} \delta(\mu - E_{\p}^{-i}) \left( 1 - \frac{\rmd E_{\p}^{-i}}{\rmd n} \frac{\rmd n}{\rmd \mu} \right) \nonumber\\[0.2cm]
    & \equiv & \sum_{i\, \in\, \lbrace \mathfrak{n},\mathfrak{n}^{\ast} \rbrace} N_{0}^{i} \left[1 - \left(f_{0}^{i} - f_{0}^{\prime i}\right) \frac{\rmd n}{\rmd \mu} \right] ,
\end{IEEEeqnarray}
with
\begin{equation}
    \frac{\rmd E_{\p}^{-\pm}}{\rmd n} = y_{\pm} \frac{M_{\pm}}{M_{\ast}} \frac{\rmd \sigma}{\rmd n} + G_{v} + G_{w} 
    \equiv \begin{cases}
        f_{0}^{\mathfrak{n}} - f_{0}^{\prime \mathfrak{n}} \\[0.1cm]
        f_{0}^{\mathfrak{n}^{\ast}} - f_{0}^{\prime \mathfrak{n}^{\ast}} 
    \end{cases} \hspace{-0.3cm} . \ 
\end{equation}
We then have
\begin{IEEEeqnarray}{rCl}
    \frac{\rmd n}{\rmd \mu} & = & \frac{N_{0}^{\mathfrak{n}} + N_{0}^{\mathfrak{n}^{\ast}}}{1 + N_{0}^{\mathfrak{n}} (f_{0}^{\mathfrak{n}} - f_{0}^{\prime \mathfrak{n}}) + N_{0}^{\mathfrak{n}^{\ast}} (f_{0}^{\mathfrak{n}^{\ast}} - f_{0}^{\prime \mathfrak{n}^{\ast}})} \nonumber\\[0.2cm]
    & \equiv & \frac{N_{0}}{1 + F_{0} - F_{0}'} ,
\end{IEEEeqnarray}
so that the speed of sound simply becomes
\begin{equation}
    c_{s}^{2} = \frac{n}{\mu} \frac{1 + F_{0} - F_{0}'}{N_{0}} .
\end{equation}
Here we employed the usual notation
\begin{IEEEeqnarray}{rCl}
    N_{0} & = & N_{0}^{\mathfrak{n}} + N_{0}^{\mathfrak{n}^{\ast}} , \\[0.15cm]
    F_{0} & = & N_{0}^{\mathfrak{n}} f_{0}^{\mathfrak{n}} + N_{0}^{\mathfrak{n}^{\ast}} f_{0}^{\mathfrak{n}^{\ast}}, \\[0.15cm]
    F_{0}' & = & N_{0}^{\mathfrak{n}} f_{0}^{\prime \mathfrak{n}} + N_{0}^{\mathfrak{n}^{\ast}} f_{0}^{\prime \mathfrak{n}^{\ast}} .
\end{IEEEeqnarray}
In the singlet model the second terms in these expressions vanish. The numerical estimate of $c_{s}^{2}$ is plotted in Fig.~\ref{fig:speed_of_sound} as a function of the baryon density. In line with the behavior of the pressure displayed in Fig.~\ref{fig:EOS_neutron_matter}, we observe that the value of $c_{s}^{2}$  is smaller in the doublet model than in the singlet model. The curve corresponding to the singlet model is close to those obtained  in Ref.~\cite{Alford_2022} (in particular with the sets RMF3 and RMF4). We also observe a dip after the $\mathfrak{n}^{\ast}$-onset. This dip corresponds to the softening of $P(n)$ beyond  the $\mathfrak{n}^{\ast}$-onset, as  can be  seen in Fig.~\ref{fig:EOS_neutron_matter}. We note however that the density range, where the dip occurs, appears to be much beyond the densities that are relevant for the physics of neutron stars. Finally, we note that at asymptotic values of the baryon density, the speed of sound reaches the value $c_s=1$. This can easily be seen from Eq.~(\ref{eq:cssqareddef}), using the expression (\ref{eq:muGvGwn}) for the chemical potential, which implies that at large $n$,  $n/\mu \rightarrow 1/(G_{v} + G_{w})$, and $\rmd \mu /\rmd n \rightarrow G_{v} + G_{w}$.

Regarding the compressibility
\begin{IEEEeqnarray}{rCl}
    \chi & = & \frac{1}{n} \left( \frac{\rmd P}{\rmd n} \right)^{-1} = \frac{1}{n^{2}} \frac{\rmd n}{\rmd \mu}
    \equiv \frac{1}{\mu n} \frac{1}{c_{s}^{2}} \nonumber\\[0.2cm]
    & = & \frac{1}{n^{2}} \frac{N_{0}}{1 + F_{0} - F_{0}'} ,
\end{IEEEeqnarray}
we find that $\chi > 0$ throughout the entire density range, since $c_{s}^{2} \ge 0$ as well as $\mu > 0$. It is therefore clear that a negative contribution $\rmd n_{\mathfrak{n}}/\rmd \mu$ in the derivative
\begin{equation}
    \frac{\rmd n}{\rmd \mu} = \frac{\rmd n_{\mathfrak{n}}}{\rmd \mu} + \frac{\rmd n_{\mathfrak{n}^{\ast}}}{\rmd \mu} 
\end{equation}
is overcompensated during the equilibration phase of the densities, so that $\rmd n/\rmd\mu > 0$ is always positive.


\subsection{Constraints from neutron stars}

\begin{figure}[ht]
	\centering
    \includegraphics[scale=1.0]{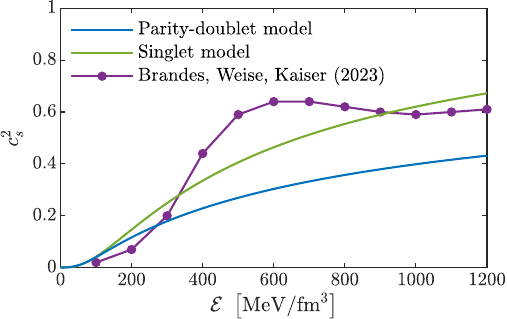}
	\caption{Squared speed of sound $c_{s}^{2}$ in neutron matter as a function of the energy density $\mathcal{E}$; the reference values are taken from Ref.~\cite{Brandes_2023} and are indicated as ``Brandes, Weise, Kaiser (2023).''}
	\label{fig:comparison_Weise_cs}
\end{figure}

\begin{figure}[ht]
	\centering
    \includegraphics[scale=1.0]{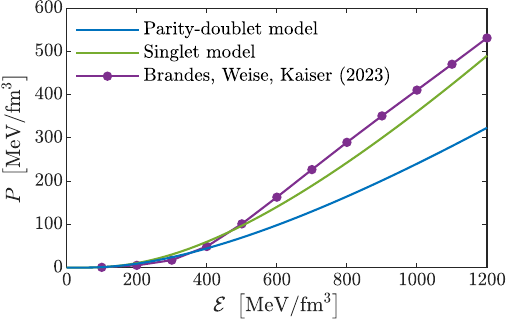}
	\caption{Pressure $P$ in neutron matter as a function of the energy density $\mathcal{E}$; the reference values are taken from Ref.~\cite{Brandes_2023}.}
	\label{fig:comparison_Weise_pressure}
\end{figure}

\begin{figure}[ht]
	\centering
    \includegraphics[scale=1.0]{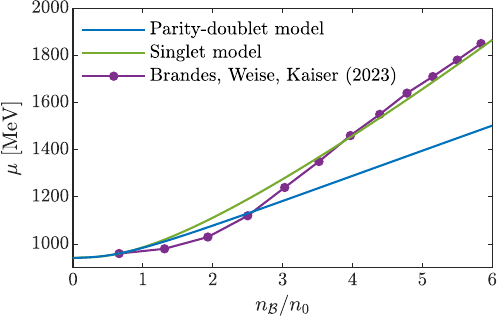}
	\caption{Neutron-chemical potential $\mu$ as a function of the baryon density $n_{\mathcal{B}}$ (in units of $n_{0}$); the reference values are taken from Ref.~\cite{Brandes_2023}.}
	\label{fig:comparison_Weise_mu}
\end{figure}

We finally confront our results on pure neutron matter to constraints obtained from neutron-star observations. Thus, in Figs.~\ref{fig:comparison_Weise_cs}, \ref{fig:comparison_Weise_pressure}, and \ref{fig:comparison_Weise_mu}, we plot our results for the speed of sound and the equation of state of neutron matter against the corresponding (tabulated) median values of the recent Bayesian analysis of Ref.~\cite{Brandes_2023}.  We emphasize that this comparison is only semi-quantitative. On the one hand, pure neutron matter is not quite the same as the matter under beta-equilibrium that is found in neutron stars. We do not expect however that this alters significantly the present comparison because of the compensating mechanism discussed in Refs.~\cite{blaschke2016universal,margueron_equation_2018}: the increase of the pressure due to the electrons balances the decrease of the symmetry energy due to the  proton excess (at fixed total baryon density). On the other hand, we plot only the most probable values for the various physical quantities, while the Bayesian analysis comes with error estimates which can be quite large \cite{Brandes_2023}. This comparison is nevertheless instructive, as we shall see. Note that we consider a range of densities which appears relevant for neutron-star physics, namely neutron densities well below the $\mathfrak{n}^{\ast}$-threshold, so that we can ignore  the contribution of the parity partners. 

Figure~\ref{fig:comparison_Weise_cs} displays  the speed of sound squared as a function of the energy density $\mathcal{E}$. Clearly none of the models can reproduce the qualitative trend inferred from the Bayesian analysis of Ref.~\cite{Brandes_2023}: a rapid increase followed by a plateau at $c_{s}^{2} \approx 0.6$ beyond three to four times the saturation density. Other Bayesian analyses yield similar qualitative behaviors for $c_{s}^{2}$ (see e.g.\ Refs.~\cite{drischler_how_2020,altiparmak_sound_2022}). 
In fact, since $c_{s}^{2}$ crosses the conformal limit $c_{s}^{2} = \frac{1}{3}$ roughly between $2\, n_{0}$ and $4\, n_0$ (see Fig.~\ref{fig:speed_of_sound}), and since $c_{s}^{2}$ must reach this value $\frac{1}{3}$ (from below) at asymptotic values of the density, the existence of a peak is expected. Where this peak occurs remains under debate (for a recent discussion of the constraints provided by perturbative-QCD calculations of the  pressure at high density, see e.g.\ Ref.~\cite{komoltsev2024equationstateneutronstardensities} and references therein). In contrast, we note that the present equation of state (for either model)  behaves as most equations of states based on nucleon  or meson degrees of freedom, meaning that it shows a regular increase with the density, without any particular structure, except for the dip in the doublet model near the $\mathfrak{n}^{\ast}$-onset. In both models, $c_s^2\to 1$ when $n\to\infty$, as already mentioned.   

As argued in Ref.~\cite{Kojo_2022}, the dependence of  $c_{s}^{2}$ on the baryon density could naturally reveal the emergence of new degrees of freedom as the density increases. By building an hybrid equation of state that interpolates between a nuclear equation of state for densities below $2\, n_0$ and a quark equation of state for $n\gtrsim 5\, n_0$, one is indeed able to reproduce the expected structure \cite{Kojo_2022}. Such ``hybrid'' approaches have been developed by many groups, see e.g.\ Refs.~\cite{Tak_tsy_2023,https://doi.org/10.48550/arxiv.2302.00825,Baym:2017whm,gao2024exploringfirstorderphasetransition}. Let us also point out  here another perspective on this issue provided by the relation between the speed of sound and the trace anomaly, discussed in Ref.~\cite{fujimoto_trace_2022} (see also Ref.~\cite{Marczenko_2024}).


The deviations from the expected qualitative trends that we observed for the speed of sound is less visible in the plots of the pressure $P$ and the neutron-chemical potential $\mu$ in Figs.~\ref{fig:comparison_Weise_pressure} and \ref{fig:comparison_Weise_mu}, respectively. Indeed,  $P$ and $\mu$ in the singlet model appear to be broadly consistent with the Bayesian values. In fact, the curve of $\mu$ for the singlet model almost overlaps with the median values of Ref.~\cite{Brandes_2023}, in particular for $n > 3 \, n_{0}$, and there are only significant deviations in the range of $n_{0} \lesssim n \lesssim 3 \, n_{0}$. However, the equation of state of the parity-doublet model is too soft, and $P$ tends to substantially undershoot the Bayesian values beyond $\mathcal{E} \approx 500\ \mathrm{MeV}/\mathrm{fm}^{3}$. Similarly, the chemical potential of the doublet model is too low for $n > 3 \, n_{0}$.  

\begin{figure}[ht]
	\centering
    \includegraphics[scale=1.0]{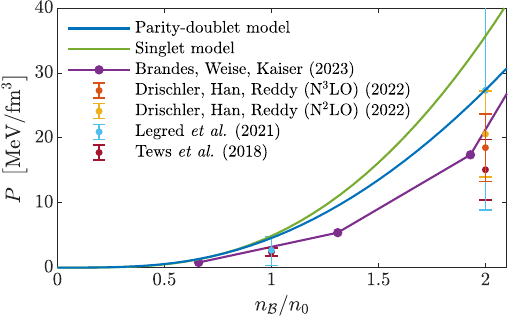}
	\caption{Comparison of the equation of state of the doublet and singlet models to the Bayesian analysis of Ref.~\cite{Brandes_2023}, as well as the microscopic calculations based on chiral effective field theory at next-to-next-to-leading order (``$\mathrm{N}^{2}\mathrm{LO}$'') and next-to-next-to-next-to-leading order (``$\mathrm{N}^{3}\mathrm{LO}$'') \cite{drischler_large_2021}, in combination with constraints from neutron-star observations \cite{Tews:2018kmu, Legred:2021hdx}.}
	\label{fig:comparison_chEFT}
\end{figure}

The equation of state, both in the doublet model and even more so in the singlet model, while being too soft when confronted to the Bayesian analysis, appears to be harder than those obtained from the best available many-body calculations, as illustrated in Fig.~\ref{fig:comparison_chEFT}. Thus for instance, the pressure of neutron matter at twice the saturation density is $P(2n_0)=27.4\ \mathrm{MeV}/\mathrm{fm}^3$ for the doublet model, and  $P(2n_0)=35.6\ \mathrm{MeV}/\mathrm{fm}^3$ for the singlet model. In comparison, values $\lesssim 20\ \mathrm{MeV}/\mathrm{fm}^3$ are quoted in Ref.~\cite{drischler_large_2021} for recent equations of states (see more details and references therein).


This observation is not new and various ways have been explored to produce a stiffening of the equation of state beyond nuclear-matter saturation density. Aside from the hybrid approaches already mentioned, which invoke the use of additional degrees of freedom, in particular quark degrees of freedom (see additionally Refs.~\cite{PhysRevLett.122.122701,Zhao_2020}), there is much flexibility to modify the equation of state while staying with nucleon and meson degrees of freedom. Thus the role of three-body neutron interactions in stiffening the equation of state is reviewed in Ref.~\cite{Gandolfi:2015jma}. Moreover, sophisticated mesonic Lagrangians have been considered, including various (sometimes density-dependent) couplings  between the mesons~\cite{de_tovar_determination_2021,char2023generalised,Dutra_2014,Kumar_2024},  and the  potential effect of the $a_0$ meson (also called $\delta$)  was emphasized in Refs.~\cite{KUBIS1997191, Thakur_2022}. So clearly, the extrapolation from symmetric matter to asymmetric matter, and in particular neutron matter, involves dynamics that is not fully understood.

\section{Conclusions}
\label{sec:summary}

In this paper we have generalised our previous work (I) about isospin-symmetric matter within the parity-doublet model to also account for isospin-asymmetric matter, and in particular pure neutron matter. This generalisation was motivated by the burst of interest in the properties of dense baryonic matter,  triggered in part by recent neutron-star observations.

Within an extended mean-field approximation, we have systematically investigated the phase structure of asymmetric matter, as predicted by the parity-doublet model.  This model predicts two phase transitions: a liquid-gas transition at small density, and a chiral transition at large density. Both transitions are first order for zero or small temperatures, and not too small proton fraction,  ending in a second-order critical point when further increasing $T$. In the corresponding singlet model, which one obtains by neglecting the chiral partner of the nucleon, the chiral transition turns into a mere crossover, occurring at much lower $n_{\mathcal{B}}$ than in the doublet model. The chiral transition in the doublet model also becomes a crossover if the proton fraction $x$ decreases below $x \lesssim 0.4$.

Furthermore, we have underlined the role of the  parity symmetry energy in the equilibration of the populations of the opposite-parity fermions, once the phase space has opened for the population of the chiral partners, an effect which ultimately leads to the restoration of chiral symmetry at large baryon density. We indicated how  the fermionic parameters, in particular the chiral-invariant mass $m_{0}$, exclusively decide on essential model features, such as the composition of matter at large density, or the lower bound of the chiral transition in the direction of increasing $n_{\mathcal{B}}$.

In this work, we have kept the parameters to their values adjusted on symmetric matter, just minimally adding a vector-isovector coupling which is adjusted on the symmetry energy. This results in specific constraints on the density dependence of the symmetry energy,   which one may want to relax in order to better account for  the recent measurements of the neutron skin of nuclei, or some neutron-star observations. Regarding the latter, we have seen that the equations of states, obtained in either the doublet or the singlet models, fail to reproduce the rapid increase of the speed of sound around twice nuclear matter density that seems to be required by the Bayesian analysis of neutron-star observations.  At the same time, for lower baryon densities up to $2 \, n_{0}$, a comparison with the best available many-body calculations reveals that the equations of states of the doublet and singlet models tend to be too stiff, with the doublet model being slightly closer to the many-body calculations. All these observations point to the fact that parametrizing the deviation from symmetric matter by a single parameter, the strength of the vector-isovector coupling, which determines the symmetry energy, is too restrictive. We have pointed out the potential role of other couplings such as the $\rho$-$\omega$ coupling, or that of additional mesons such as the $\delta$ meson. Clearly, more work along these lines would be needed to further asses the ability of the parity-doublet model to provide a reliable equation of state.

Finally, aside from the physics discussion of asymmetric matter properties, which constitutes the main body of the paper, we have reported, in Appendix~\ref{sec:diffeqs}, on a technical development. This concerns the solution of the gap equations using a coupled set of (ordinary) differential equations which describe the flows of the isoscalar condensate and the densities of the various baryons as a function of the total  baryon density (for zero temperature). The virtue of this formulation is that it  allows the flow with the baryon density to be  decomposed into different stages, which can be solved essentially independently from each other, and which can be associated with well identified physical regimes. In this formulation, we have derived important features of the parity-doublet model, such as a lower bound for the chiral restoration density, and the decoupling of the onsets of the chiral partners from the bosonic parameters of the model. Additionally, compared to directly solving the gap equations by root finding, the flow formulation  gives more direct access to the critical behavior at the chiral transition (in the chiral limit), and to the corresponding critical isoscalar mode. Certain aspects of this formalism are model independent, and might thus be exploited in other contexts.

\begin{acknowledgments}
We acknowledge interesting exchanges with J.~Margueron and W.~Weise on several topics discussed in this paper. JE acknowledges funding by the German National Academy of Sciences Leopoldina (through the fellowships LPDS 2020-06 and LPDR 2023-01). JE thanks the IPhT in Saclay and the University of Heidelberg for hospitality. JPB thanks the hospitality of the University of Heidelberg for a stay during which this paper was completed. 
\end{acknowledgments}

\appendix

\section{Estimates of the slope parameter}
\label{sec:slope}

Several methods are typically used to get, from various experimental data, an estimate of the slope parameter $L$ of the symmetry energy. In this appendix we discuss the corresponding determinations of $L$, obtain their numerical values, and compare them to the result quoted in the main text, see Table~\ref{tab:Sparameters}.

If we take the approximation (\ref{eq:definitionSnB}) and solve for $S(n_{\mathcal{B}})$, we get
\begin{equation}
    S(n_{\mathcal{B}}) = \frac{1}{(2x - 1)^{2}} \left[\frac{\mathcal{E}}{n_{\mathcal{B}}}(n_{\mathcal{B}}, x) - \frac{\mathcal{E}}{n_{\mathcal{B}}}\left(n_{\mathcal{B}}, \frac{1}{2}\right)\right] . \label{eq:Salternative}
\end{equation}
Strictly speaking, this approximation is valid in the vicinity of the symmetric point, so that taking the limit $x \rightarrow \frac{1}{2}$ in this Eq.~(\ref{eq:Salternative}) is the natural choice, and precisely leads to the estimate of $L$ that was presented in the main text. However, different choices of $x$ may be used, including  $x = 0$ (neutron matter). By doing so we  obtain different estimates of $L$, which allows us to investigate how strongly $L$ deviates from its original value when moving further and further away from the symmetric point at $x = \frac{1}{2}$.

(1) A first estimate of $L$ is therefore obtained by computing the derivative of Eq.~(\ref{eq:Salternative}), at fixed value of $x$,
\begin{IEEEeqnarray}{rCl}
    L^{(1)} & = & \frac{3 n_{0}}{(2 x - 1)^{2}} \left.\frac{\partial \mathcal{E}/n_{\mathcal{B}}(n_{\mathcal{B}}, x)}{\partial n_{\mathcal{B}}}\right|_{n_{0}} \nonumber\\[0.2cm]
    & = & \left. \frac{1}{(2x - 1)^{2}} \frac{3 P}{n_{0}} \right|_{x,\, n_{\mathcal{B}}\, =\, n_{0}},
\end{IEEEeqnarray}
where we used that the energy per particle of symmetric matter is minimum at $n_{\mathcal{B}} = n_{0}$. The zero-temperature pressure $P$ is given by
\begin{IEEEeqnarray}{rCl}
    P & = & - \mathcal{E}^{\mathfrak{p}}_{\mathrm{qp}}(n_{\mathfrak{p}}; \sigma) - \mathcal{E}^{\mathfrak{n}}_{\mathrm{qp}}(n_{\mathfrak{n}}; \sigma) - U(\sigma) \nonumber\\[0.2cm]
    & & +\, M_{\ast}^{\mathfrak{p}} n_{\mathfrak{p}} + M_{\ast}^{\mathfrak{n}} n_{\mathfrak{n}} 
    + \frac{1}{2} \left( G_{v} n_{\mathcal{B}}^{2} + G_{w} n_{\mathcal{I}}^{2} \right) , \quad
\end{IEEEeqnarray}
being evaluated at $n_{\mathfrak{p}} = x n_{0}$ and $n_{\mathfrak{n}} = (1 - x) n_{0}$. The corresponding quasi-particle energies read
\begin{equation}
    \mathcal{E}_{\mathrm{qp}}^{\mathfrak{p},\mathfrak{n}}(n_{\mathfrak{p},\mathfrak{n}}; \sigma) = 2 \int_{|p|\, \le\, p_{\mathfrak{p},\mathfrak{n}}} \frac{\rmd^{3}p}{(2\pi)^{3}} \sqrt{p^{2} + M_{+}(\sigma)^{2}} .
\end{equation}
The $\sigma$ condensate at $n_{\mathcal{B}} = n_{0}$, and for the chosen value of $x$, is formally given by
\begin{equation}
    \left.\sigma\right|_{x} = \left.\sigma\right|_{\frac{1}{2}}
    + \int^{\frac{1}{2}}_{x} \rmd z \left.\frac{y_{+} M_{+}}{m_{\sigma}^{2}} n_{0} \left(\frac{1}{M_{\ast}^{\mathfrak{p}}} - \frac{1}{M_{\ast}^{\mathfrak{n}}}\right) \right|_{\sigma_{z}} ,
\end{equation}
where $\sigma|_{x\, =\, \frac{1}{2}} \equiv \sigma_{0}$ is the in-medium condensate of symmetric matter.
To obtain this result, we have integrated the differential equation for $\sigma$ at fixed baryon density $n_{\mathcal{B}}$, namely
\begin{equation}
    \left.\frac{\rmd\sigma}{\rmd x}\right|_{n_{\mathcal{B}}} = - \frac{y_{+} M_{+}}{m_{\sigma}^{2}} n_{\mathcal{B}} \left(\frac{1}{M_{\ast}^{\mathfrak{p}}} - \frac{1}{M_{\ast}^{\mathfrak{n}}}\right). \label{eq:dsigmadx}
\end{equation}
Using the relation (\ref{eq:EsymF0}) between $E_{\mathrm{sym}}$ and $G_{w}$ we may write the parameter $L^{(1)}$ as
\begin{IEEEeqnarray}{rCl}
    L^{(1)} & = & 3\, \bigg\lbrace - \frac{\mathcal{E}_{\mathrm{qp}}^{\mathfrak{p}} + \mathcal{E}_{\mathrm{qp}}^{\mathfrak{n}} + U(\sigma)}{(2x - 1)^{2} n_{0}} + \frac{x M_{\ast}^{\mathfrak{p}} + (1 - x) M_{\ast}^{\mathfrak{n}}}{(2x - 1)^{2}} \nonumber\\[0.2cm] 
    & & \quad +\, \frac{1}{2} \left[\frac{G_{v}}{(2x - 1)^{2}} - \frac{1}{N_{0}}\right] n_{0} + E_{\mathrm{sym}} \bigg\rbrace ,
    \label{eq:L1}
\end{IEEEeqnarray}
where $N_{0}$ the density of states at the Fermi surface of symmetric matter. Let us recall that this expression is based on the parabolic approximation (\ref{eq:Salternative}), which is expected to deteriorate slightly as  $x$ moves away from the symmetric point. If we take e.g.\ $x = 0.4$, we get $L^{(1)} = 83.9\ \mathrm{MeV}$ in the doublet model and $L^{(1)} = 89.2\ \mathrm{MeV}$ in the singlet model, values that are still quite close to those displayed in Table~\ref{tab:Sparameters}.

(2) The estimate presented in the main text corresponds to the limit
\begin{equation}
    L^{(2)} \equiv \lim_{x\, \rightarrow\, \frac{1}{2}} L^{(1)} = \frac{3}{8 n_{0}} \left.\frac{\partial^{2} P}{\partial x^{2}}\right|_{x\, =\, \frac{1}{2},\, n_{0}} .
\end{equation}
To prove this statement, we first compute the derivative
\begin{equation}
    \frac{\partial P}{\partial x} = 2 n_{\mathcal{B}} \left(\frac{\partial \mu_{\mathcal{B}}}{\partial n_{\mathcal{I}}} n_{\mathcal{B}} + \frac{\partial \mu_{\mathcal{I}}}{\partial n_{\mathcal{I}}} n_{\mathcal{I}}\right) ,
\end{equation}
where we used $P = - \mathcal{E} + \mu_{\mathcal{B}} n_{\mathcal{B}} + \mu_{\mathcal{I}} n_{\mathcal{I}}$. Note also that $\partial/\partial x = 2 n_{\mathcal{B}} \partial/\partial n_{\mathcal{I}}$, since $n_{\mathcal{I}} = (2x - 1) n_{\mathcal{B}}$. Then
\begin{equation}
    \frac{\partial^{2} P}{\partial x^{2}} = \left(2 n_{\mathcal{B}}\right)^{2} \left(\frac{\partial^{2} \mu_{\mathcal{B}}}{\partial n_{\mathcal{I}}^{2}} n_{\mathcal{B}} + \frac{\partial^{2} \mu_{\mathcal{I}}}{\partial n_{\mathcal{I}}^{2}} n_{\mathcal{I}} + \frac{\partial \mu_{\mathcal{I}}}{\partial n_{\mathcal{I}}}\right) .
\end{equation}
At the symmetric point ($x = \frac{1}{2}$ and $n_{\mathcal{I}} = 0$) we get
\begin{equation}
    \left.\frac{\partial^{2} P}{\partial x^{2}}\right|_{\frac{1}{2}} = 
    \left(2 n_{\mathcal{B}}\right)^{2} \left. \left(\frac{\partial^{2} \mu_{\mathcal{I}}}{\partial n_{\mathcal{B}} \partial n_{\mathcal{I}}} n_{\mathcal{B}} + \frac{\partial \mu_{\mathcal{I}}}{\partial n_{\mathcal{I}}}\right) \right|_{n_{\mathcal{I}}\, =\, 0} ,
\end{equation}
where we used that
\begin{equation}
    \frac{\partial^{2} \mu_{\mathcal{B}}}{\partial n_{\mathcal{I}}^{2}} = \frac{\partial^{3}\mathcal{E}}{\partial n_{\mathcal{I}}^{2} \partial n_{\mathcal{B}}} \equiv \frac{\partial^{3}\mathcal{E}}{\partial n_{\mathcal{B}} \partial n_{\mathcal{I}}^{2}} = \frac{\partial^{2} \mu_{\mathcal{I}}}{\partial n_{\mathcal{B}} \partial n_{\mathcal{I}}} .
\end{equation}
Eventually, with Eq.~(\ref{eq:SnB}), we find
\begin{IEEEeqnarray}{rCl}
    \left.\frac{\partial^{2} P}{\partial x^{2}}\right|_{x\, =\, \frac{1}{2}} & = & \left(2 n_{\mathcal{B}}\right)^{2} \frac{\rmd}{\rmd n_{\mathcal{B}}} \left(\left.\frac{\partial\mu_{\mathcal{I}}}{\partial n_{\mathcal{I}}}\right|_{n_{\mathcal{I}}\, =\, 0} n_{\mathcal{B}}\right) \nonumber\\[0.2cm]
    & \equiv & 8 n_{\mathcal{B}}^{2} \frac{\rmd S(n_{\mathcal{B}})}{\rmd n_{\mathcal{B}}} ,
\end{IEEEeqnarray}
and thus, as stated above,
\begin{equation}
    L^{(2)} = \lim_{x\,\rightarrow\,\frac{1}{2}} L^{(1)} = 3 n_{0} \left.\frac{\rmd S(n_{\mathcal{B}})}{\rmd n_{\mathcal{B}}}\right|_{n_{0}} \equiv L .
\end{equation}

(3) A third estimate is obtained by computing the derivative of the energy per particle of neutron matter at the density $n_{0}$, i.e.
\begin{equation}
    L^{(3)} \equiv \lim_{x\,\rightarrow\, 0} L^{(1)} = 3 n_{0} \left.\frac{\partial\mathcal{E}/n_{\mathcal{B}}(n_\mathcal{B}, 0)}{\partial n_{\mathcal{B}}}\right|_{n_{0}} = \frac{3 P}{n_{0}},
\end{equation}
with $P$ now the pressure of neutron matter, as given in Eq.~(\ref{eq:Pneutron}). Consistently with the first estimate, we finally arrive at
\begin{IEEEeqnarray}{rCl}
    L^{(3)} & = & 3\, \bigg[ - \frac{\mathcal{E}_{\mathrm{qp}}^{\mathfrak{n}} + U(\sigma)}{n_{0}} + M_{\ast}^{\mathfrak{n}} \nonumber\\[0.2cm] 
    & & \quad +\, \frac{1}{2} \left(G_{v} - \frac{1}{N_{0}}\right) n_{0} + E_{\mathrm{sym}} \bigg] .
\end{IEEEeqnarray}
The numerical values for the case of pure neutron matter amount to $L^{(3)} = 85.6\ \mathrm{MeV}$ in the doublet model, and $L^{(3)} = 91.6\ \mathrm{MeV}$ in the singlet model. As expected, these values (corresponding to $x = 0$) differ more strongly from the original estimates than the ones for $x = 0.4$.

(4) A final method that we want to list here consists in computing the symmetry energy at twice the saturation density, $S(2n_{0})$, and then determine the slope parameter as
\begin{equation}
	L^{(4)} = 3 \left[\frac{\mathcal{E}}{n_{\mathcal{B}}}( 2n_{0}, 0) 
    - \frac{\mathcal{E}}{n_{\mathcal{B}}}\left(2 n_{0}, \frac{1}{2}\right) - E_{\mathrm{sym}}\right].
\end{equation}
This formula has been used to extract the value of the slope parameter from data on Au+Au collisions \cite{Russotto:2016ucm, brandes_fluctuations_2021}, with $S(2n_{0}) = (55 \pm 5)\ \mathrm{MeV}$. Regarding the present work, we find about $60\ \mathrm{MeV}$ for $S(2n_{0})$ in the doublet model, and $64\ \mathrm{MeV}$ in the singlet model, with $L^{(4)} = 84.2\ \mathrm{MeV}$ and $L^{(4)} = 94.5\ \mathrm{MeV}$, respectively.

The different estimates of $L$ are summarized in Table~\ref{tab:Lparameter}. The values are closer to each other in the doublet model, as compared to the related singlet model, and the values are smaller. In fact, we observe that the maximum distance between two estimates is only about $1.8\ \mathrm{MeV}$ in the parity-doublet model (difference between the estimates $L^{(2)}$ and $L^{(3)}$), while $5.4\ \mathrm{MeV}$ in the singlet model (here between $L^{(2)}$ and $L^{(4)}$).
\begin{table}[ht]
	\caption{\label{tab:Lparameter}Different estimates of the slope parameter $L$.}
	\begin{ruledtabular}
		\begin{tabular}{lcc}
		Estimate [MeV] & Parity-doublet model & Singlet model\\
		\colrule\\[-0.25cm]
        (1): $x = 0.4$ & $83.9$ & $89.2$ \\[0.05cm]
        (2): $x = 0.5$ & $83.8$ & $89.1$ \\[0.05cm]
		(3): $x = 0$ & $85.6$ & $91.6$ \\[0.05cm]
		(4): $S(2 n_{0})$ & $84.2$ & $94.5$  
		\end{tabular}
	\end{ruledtabular}
\end{table}

In the present doublet and singlet models, it follows from Eq.~(\ref{eq:L1}) that the relation between the slope parameter $L^{(1)}$ (which also holds for $L^{(2)}$ and $L^{(3)}$) and the symmetry energy $E_{\mathrm{sym}}$ is given by the simple derivative
\begin{equation}
    \frac{\rmd L^{(1)}}{\rmd E_{\mathrm{sym}}} = \frac{\rmd L^{(2)}}{\rmd E_{\mathrm{sym}}} = \frac{\rmd L^{(3)}}{\rmd E_{\mathrm{sym}}} = 3 .
\end{equation}
Regarding the estimate $L^{(4)}$ we find as well that 
\begin{equation}
    \frac{\rmd L^{(4)}}{\rmd E_{\mathrm{sym}}} = 3 ,
\end{equation}
since
\begin{equation}
    \frac{\rmd \mathcal{E}/n_{\mathcal{B}}(2 n_{0}, 0)}{\rmd G_{w}} = n_{0}, \qquad
    \frac{\rmd G_{w}}{\rmd E_{\mathrm{sym}}} = \frac{2}{n_{0}} ,
\end{equation}
and the energy per particle in symmetric matter does not depend on $G_{w}$.



\section{Formulation in terms of ordinary differential equations}
\label{sec:diffeqs}

In this appendix we explain the method used in this paper to solve the gap equations, which consists in transforming these gap equations into coupled ordinary differential equations.
These differential equations are derived from two conditions: (i)  the stationarity of the physical pressure  with respect to variation of the internal variable $\sigma$, for any $n_{\mathcal{B}}$,
\begin{equation}
    \left.\frac{\rmd}{\rmd n_{\mathcal{B}}}\right|_{x} \frac{\partial P}{\partial\sigma} = 0, \label{eq:ODEgen1}
\end{equation}
where we keep the proton fraction $x$ fixed, and (ii) the equality of the Landau masses,
\begin{IEEEeqnarray}{rCl}
    \left.\frac{\rmd}{\rmd n_{\mathcal{B}}}\right|_{x} \left(M_{\ast}^{\mathfrak{p}} - M_{\ast}^{\mathfrak{p}^{\ast}}\right) = 0, \label{eq:ODEgen2} \\[0.2cm]
    \left.\frac{\rmd}{\rmd n_{\mathcal{B}}}\right|_{x} \left(M_{\ast}^{\mathfrak{n}} - M_{\ast}^{\mathfrak{n}^{\ast}}\right) = 0. \label{eq:ODEgen3}
\end{IEEEeqnarray}
The resulting differential equations yield continuous solutions for $\sigma$, as well as for $n_{\mathfrak{p}}$ and $n_{\mathfrak{n}}$. Depending on the initial conditions, some of these quantities may  run into divergences, a behavior that may signal the occurrence of a first-order phase transition (if the divergence appears at $\sigma > 0$). Each of the composition equations (\ref{eq:ODEgen2}) and (\ref{eq:ODEgen3}) are only relevant once the baryon density has passed the 
thresholds (\ref{eq:thresholds}) for the population of the chiral partners. For the singlet model, the physical solution is solely determined by Eq.~(\ref{eq:ODEgen1}).

The physical solution of the system, hence the phase structure, exclusively depends on the initialization of the differential equations at $n_{\mathcal{B}} = 0$. For a physical pion mass, the initial value of $\sigma$ is given by the pion decay constant, $\sigma = f_{\pi}$, and for zero pion mass, by its corresponding value in the chiral limit, $\sigma = \sigma_{\chi}$. The densities of the chiral partners are initialized as $n_{\mathfrak{p}^{\ast}} = n_{\mathfrak{n}^{\ast}} = 0$, and  once the respective thresholds are reached they are initialized as a tiny seed number, $n_{\mathfrak{p}^{\ast}}/n_{0} \ll 1$ and $n_{\mathfrak{n}^{\ast}}/n_{0} \ll 1$.

In the next section we derive explicitly the differential equations which are relevant for both asymmetric matter and neutron matter.

\subsection{The system of coupled equations}

\subsubsection{Asymmetric matter}

The differential equation for $\sigma$ that is obtained from Eq.~(\ref{eq:ODEgen1}) reads
\begin{equation}
    \left.\frac{\rmd\sigma}{\rmd n_{\mathcal{B}}}\right|_{x} = - \frac{y_{\pm}}{m_{\sigma}^{2}} \left.\frac{\partial n_{\mathrm{s}}^{\pm}}{\partial n_{\mathcal{B}}}\right|_{x;\, \sigma} , \label{eq:ODE_sigma1}
\end{equation}
where the sigma mass in the doublet model is given by
\begin{equation}
    m_{\sigma}^{2} = \frac{\rmd^{2} U}{\rmd\sigma^{2}} + n_{\mathrm{s}}^{\pm} \frac{\rmd^{2} M_{\pm}}{\rmd\sigma^{2}} + y_{\pm} \left.\frac{\partial n_{\mathrm{s}}^{\pm}}{\partial\sigma}\right|_{n_{\mathfrak{p}}, n_{\mathfrak{n}}, n_{\mathfrak{p}^{\ast}}, n_{\mathfrak{n}^{\ast}}} ,
\end{equation}
with the summation convention $a_{\pm} b_{\pm} = a_{+} b_{+} + a_{-} b_{-}$. For low baryon densities, when $n_{\mathcal{B}}^{-} = 0$, as well as in the case of the singlet model, Eq.~(\ref{eq:ODE_sigma1}) reduces to
\begin{equation}
    \left.\frac{\rmd\sigma}{\rmd n_{\mathcal{B}}}\right|_{x} = - \frac{y_{+}}{m_{\sigma}^{2}} \left.\frac{\partial n_{s}^{+}}{\partial n_{\mathcal{B}}}\right|_{x;\, \sigma} = - \frac{y_{+}}{m_{\sigma}^{2}} \frac{M_{+}}{M_{\ast}^{x}}, \label{eq:ODE_sigma2}
\end{equation}
as already employed in Eq.~(\ref{eq:dsigmadn_initial}). The scalar densities at zero temperature are given by
\begin{equation}
    n_{\mathrm{s}}^{\pm} = 2 M_{\pm} \sum_{\tau \, = \, \pm 1} \int_{|\p| \, \le \, p_{\tau}} \frac{1}{\sqrt{\p^2 + M_{\pm}^{2}}},
\end{equation}
with $p_{\tau}$ the Fermi momentum of proton-like ($\tau = +1$) or neutron-like ($\tau = -1$) states.

Equations~(\ref{eq:ODEgen2}) and (\ref{eq:ODEgen3}) determine the populations of the chiral partners, depending on the total baryon density $n_{\mathcal{B}}$, once the respective thresholds (\ref{eq:thresholds}) are passed. Below the thresholds, the two equations are trivially replaced by the relations
\begin{equation}
    \left.\frac{\rmd n_{\mathfrak{p}}}{\rmd n_{\mathcal{B}}}\right|_{x} = x, \qquad
    \left.\frac{\rmd n_{\mathfrak{n}}}{\rmd n_{\mathcal{B}}}\right|_{x} = 1 - x, \label{eq:ODE_den_simp}
\end{equation}
since by definition $n_{\mathfrak{p}} = xn_{\mathcal{B}}$ and $n_{\mathfrak{n}} = (1 - x) n_{\mathcal{B}}$. Beyond the thresholds, the complete set of the coupled differential equations (\ref{eq:ODEgen1}), (\ref{eq:ODEgen2}), and (\ref{eq:ODEgen3}) explicitly generalises as follows:
\begin{widetext}
\begin{IEEEeqnarray}{rCl}
    \left.\frac{\rmd\sigma}{\rmd n_{\mathcal{B}}}\right|_{x} & = & - \frac{1}{m_{\sigma}^{2}} \left[y_{+} M_{+} \left(\frac{1}{M_{\ast}^{\mathsf{P}}} \left.\frac{\rmd n_{\mathfrak{p}}}{\rmd n_{\mathcal{B}}}\right|_{x} + \frac{1}{M_{\ast}^{\mathsf{N}}} \left.\frac{\rmd n_{\mathfrak{n}}}{\rmd n_{\mathcal{B}}}\right|_{x}\right) + y_{-} M_{-} \left(\frac{1}{M_{\ast}^{\mathsf{P}}} \left.\frac{\rmd n_{\mathfrak{p}^{\ast}}}{\rmd n_{\mathcal{B}}}\right|_{x} + \frac{1}{M_{\ast}^{\mathsf{N}}} \left.\frac{\rmd n_{\mathfrak{n}^{\ast}}}{\rmd n_{\mathcal{B}}}\right|_{x}\right) \right] \nonumber\\[0.2cm]
    & = & - \frac{1}{m_{\sigma}^{2}} \left[ \left(y_{+} M_{+} - y_{-} M_{-}\right) \left(\frac{1}{M_{\ast}^{\mathsf{P}}} \left.\frac{\rmd n_{\mathfrak{p}}}{\rmd n_{\mathcal{B}}}\right|_{x} + \frac{1}{M_{\ast}^{\mathsf{N}}} \left.\frac{\rmd n_{\mathfrak{n}}}{\rmd n_{\mathcal{B}}}\right|_{x}\right) + \frac{y_{-} M_{-}}{M_{\ast}^{x}} \right] , \label{eq:ODE1} \\[0.3cm]
    \left.\frac{\rmd n_{\mathfrak{p}}}{\rmd n_{\mathcal{B}}}\right|_{x} & = & \frac{n_{\mathcal{B}}}{8} \left[ \frac{x}{N_{0}^{\mathfrak{p}^{\ast}}} + x \left(\tilde{f}_{\mathsf{P}}^{\prime\mathfrak{p}} - \tilde{f}_{\mathsf{P}}^{\prime\mathfrak{p}^{\ast}} + \tilde{f}_{\mathsf{P}}^{\mathfrak{p}^{\ast}} - \tilde{f}_{\mathsf{P}}^{\mathfrak{p}}\right) + (1 - x) \left(\tilde{f}_{\mathsf{N}}^{\prime\mathfrak{p}} - \tilde{f}_{\mathsf{N}}^{\prime\mathfrak{p}^{\ast}} + \tilde{f}_{\mathsf{N}}^{\mathfrak{p}^{\ast}} - \tilde{f}_{\mathsf{N}}^{\mathfrak{p}}\right)\right] \bigg\slash \tilde{S}_{\mathsf{P}} - \frac{\tilde{S}_{\mathsf{PN}}}{\tilde{S}_{\mathsf{P}}} \left.\frac{\rmd n_{\mathfrak{n}}}{\rmd n_{\mathcal{B}}}\right|_{x} \nonumber\\[0.2cm]
    & \equiv & \frac{n_{\mathcal{B}}}{8} \left[ \mathfrak{N}_{0}^{\mathsf{P}} + \frac{y_{-} M_{-}}{m_{\sigma}^{2} M_{\ast}^{x}} \frac{y_{+} M_{+} - y_{-} M_{-}}{M_{\ast}^{\mathsf{P}}} + \frac{(y_{+} M_{+} - y_{-} M_{-})^{2}}{m_{\sigma}^{2} M_{\ast}^{\mathsf{P}} M_{\ast}^{\mathsf{N}}} \left.\frac{\rmd n_{\mathfrak{n}}}{\rmd n_{\mathcal{B}}}\right|_{x} \right] \bigg\slash \tilde{S}_{\mathsf{P}} , \label{eq:ODE2} \\[0.3cm]
    \left.\frac{\rmd n_{\mathfrak{n}}}{\rmd n_{\mathcal{B}}}\right|_{x} & = & \frac{n_{\mathcal{B}}}{8} \left[ \frac{1 - x}{N_{0}^{\mathfrak{n}^{\ast}}} + x \left(\tilde{f}_{\mathsf{P}}^{\prime\mathfrak{n}} - \tilde{f}_{\mathsf{P}}^{\prime\mathfrak{n}^{\ast}} + \tilde{f}_{\mathsf{P}}^{\mathfrak{n}^{\ast}} - \tilde{f}_{\mathsf{P}}^{\mathfrak{n}}\right) + (1 - x) \left(\tilde{f}_{\mathsf{N}}^{\prime\mathfrak{n}} - \tilde{f}_{\mathsf{N}}^{\prime\mathfrak{n}^{\ast}} + \tilde{f}_{\mathsf{N}}^{\mathfrak{n}^{\ast}} - \tilde{f}_{\mathsf{N}}^{\mathfrak{n}}\right) \right] \bigg\slash \tilde{S}_{\mathsf{N}} - \frac{\tilde{S}_{\mathsf{PN}}}{\tilde{S}_{\mathsf{N}}} \left.\frac{\rmd n_{\mathfrak{p}}}{\rmd n_{\mathcal{B}}}\right|_{x}  \nonumber\\[0.2cm]
    & \equiv & \frac{n_{\mathcal{B}}}{8} \left[\mathfrak{N}_{0}^{\mathsf{N}} + \frac{y_{-} M_{-}}{m_{\sigma}^{2} M_{\ast}^{x}} \frac{y_{+} M_{+} - y_{-} M_{-}}{M_{\ast}^{\mathsf{N}}} + \frac{(y_{+} M_{+} - y_{-} M_{-})^{2}}{m_{\sigma}^{2} M_{\ast}^{\mathsf{P}} M_{\ast}^{\mathsf{N}}} \left.\frac{\rmd n_{\mathfrak{p}}}{\rmd n_{\mathcal{B}}}\right|_{x}\right] \bigg\slash \tilde{S}_{\mathsf{N}} , \label{eq:ODE3}
\end{IEEEeqnarray}
\end{widetext}
where we used Eq.~(\ref{eq:ODE_sigma1}), the parity symmetry energies (\ref{eq:SP1}), (\ref{eq:SN1}), and (\ref{eq:SPN1}), as well as the relations
\begin{equation}
    \frac{\rmd n_{\mathfrak{p}^{\ast}}}{\rmd n_{\mathcal{B}}} = x - \frac{\rmd n_{\mathfrak{p}}}{\rmd n_{\mathcal{B}}}, \qquad
    \frac{\rmd n_{\mathfrak{n}^{\ast}}}{\rmd n_{\mathcal{B}}} = 1 - x - \frac{\rmd n_{\mathfrak{n}}}{\rmd n_{\mathcal{B}}} .\label{eq:dx_rel}
\end{equation}
Furthermore, analogously to Eqs.~(\ref{eq:ftildeprime1}) to (\ref{eq:ftildeprime4}), we introduced the Landau parameters
\begin{IEEEeqnarray}{rCl}
    \tilde{f}_{\mathsf{P}}^{\mathfrak{p},\mathfrak{p}^{\ast}} & = & \left.\frac{\partial E_{\p}^{+\pm}}{\partial n_{\mathsf{P}}}\right|_{|\p|\, =\, p_{\mathfrak{p},\mathfrak{p}^{\ast}}} \nonumber\\[0.2cm]
    & = & G_{v} + G_{w} + y_{\pm} \frac{M_{\pm}}{M_{\ast}^{\mathfrak{p},\mathfrak{p}^{\ast}}} \frac{\partial\sigma}{\partial n_{\mathsf{P}}} , \\[0.2cm]
    \tilde{f}_{\mathsf{N}}^{\mathfrak{p},\mathfrak{p}^{\ast}} & = & \left.\frac{\partial E_{\p}^{+\pm}}{\partial n_{\mathsf{N}}}\right|_{|\p|\, =\, p_{\mathfrak{p},\mathfrak{p}^{\ast}}} \nonumber\\[0.2cm]
    & = & G_{v} - G_{w} + y_{\pm} \frac{M_{\pm}}{M_{\ast}^{\mathfrak{p},\mathfrak{p}^{\ast}}} \frac{\partial\sigma}{\partial n_{\mathsf{N}}} , \\[0.2cm]
    \tilde{f}_{\mathsf{P}}^{\mathfrak{n},\mathfrak{n}^{\ast}} & = & \left.\frac{\partial E_{\p}^{-\pm}}{\partial n_{\mathsf{P}}}\right|_{|\p|\, =\, p_{\mathfrak{n},\mathfrak{n}^{\ast}}} \nonumber\\[0.2cm] 
    & = & G_{v} - G_{w} + y_{\pm} \frac{M_{\pm}}{M_{\ast}^{\mathfrak{n},\mathfrak{n}^{\ast}}} \frac{\partial\sigma}{\partial n_{\mathsf{P}}} , \\[0.2cm]
    \tilde{f}_{\mathsf{N}}^{\mathfrak{n},\mathfrak{n}^{\ast}} & = & \left.\frac{\partial E_{\p}^{-\pm}}{\partial n_{\mathsf{N}}}\right|_{|\p|\, =\, p_{\mathfrak{n},\mathfrak{n}^{\ast}}} \nonumber\\[0.2cm] 
    & = & G_{v} + G_{w} + y_{\pm} \frac{M_{\pm}}{M_{\ast}^{\mathfrak{n},\mathfrak{n}^{\ast}}} \frac{\partial\sigma}{\partial n_{\mathsf{N}}} , 
\end{IEEEeqnarray}
and the variables
\begin{equation}
    \mathfrak{N}_{0}^{\mathsf{P}} = \frac{x}{N_{0}^{\mathfrak{p}^{\ast}}}, \qquad
    \mathfrak{N}_{0}^{\mathsf{N}} = \frac{1 - x}{N_{0}^{\mathfrak{n}^{\ast}}}.
\end{equation}
The Landau parameters $\tilde{f}_{\mathsf{P},\mathsf{N}}^{i}$, with $i \in \lbrace \mathfrak{p}, \mathfrak{n}, \mathfrak{p}^{\ast}, \mathfrak{n}^{\ast} \rbrace$, are related to the Landau parameters $f_{0}^{i}$ and $f_{0}^{\prime i}$ by
\begin{IEEEeqnarray}{rCl}
    f_{0}^{i} & = & \frac{1}{2} \left(\tilde{f}_{\mathsf{P}}^{i} + \tilde{f}_{\mathsf{N}}^{i}\right) \quad \forall i, \label{eq:PNbasis1} \\[0.2cm]
    f_{0}^{\prime i} & = & \frac{1}{2} \left(\tilde{f}_{\mathsf{P}}^{i} - \tilde{f}_{\mathsf{N}}^{i}\right) \quad \forall i, \label{eq:PNbasis2}
\end{IEEEeqnarray}
corresponding to the basis transformation between baryon and isospin numbers, and proton and neutron numbers, $\lbrace \mathcal{B}, \mathcal{I} \rbrace \leftrightarrow \lbrace \mathsf{P}, \mathsf{N} \rbrace$. Note however that the parameters $\tilde{f}_{\mathsf{P},\mathsf{N}}^{\prime i}$, that account for changes in the parity fractions $x_{\mathsf{P}}'$ and $x_{\mathsf{N}}'$, cannot be expressed in terms of the parameters $f_{0}^{i}$ and $f_{0}^{\prime i}$. We can further disentangle Eqs.~(\ref{eq:ODE2}) and (\ref{eq:ODE3}) by solving for $\rmd n_{\mathfrak{p}}/\rmd n_{\mathcal{B}}$ and $\rmd n_{\mathfrak{n}}/\rmd n_{\mathcal{B}}$, finally yielding
\begin{IEEEeqnarray}{rCl}
    & & \left.\frac{\rmd n_{\mathfrak{p}}}{\rmd n_{\mathcal{B}}}\right|_{x} = \frac{n_{\mathcal{B}}}{8} \frac{\tilde{S}_{\mathsf{N}} \!\left[\mathfrak{N}_{0}^{\mathsf{P}} + \tilde{\mathfrak{f}}_{\mathsf{P}}(x)\right] - \tilde{S}_{\mathsf{PN}} \!\left[\mathfrak{N}_{0}^{\mathsf{N}} + \tilde{\mathfrak{f}}_{\mathsf{N}}(x)\right]}{\tilde{S}_{\mathsf{P}} \tilde{S}_{\mathsf{N}} - \tilde{S}_{\mathsf{PN}}^{2}} , \qquad\quad \label{eq:ODE_np} \\[0.2cm]
    & & \left.\frac{\rmd n_{\mathfrak{n}}}{\rmd n_{\mathcal{B}}}\right|_{x} = \frac{n_{\mathcal{B}}}{8} \frac{\tilde{S}_{\mathsf{P}} \!\left[\mathfrak{N}_{0}^{\mathsf{N}} + \tilde{\mathfrak{f}}_{\mathsf{N}}(x)\right] - \tilde{S}_{\mathsf{PN}} \!\left[\mathfrak{N}_{0}^{\mathsf{P}} + \tilde{\mathfrak{f}}_{\mathsf{P}}(x)\right]}{\tilde{S}_{\mathsf{P}} \tilde{S}_{\mathsf{N}} - \tilde{S}_{\mathsf{PN}}^{2}} , \label{eq:ODE_nn}
\end{IEEEeqnarray}
where
\begin{IEEEeqnarray}{rCl}
    \tilde{\mathfrak{f}}_{\mathsf{P}}(x) & = & x \left(\tilde{f}_{\mathsf{P}}^{\prime\mathfrak{p}} - \tilde{f}_{\mathsf{P}}^{\prime\mathfrak{p}^{\ast}} + \tilde{f}_{\mathsf{P}}^{\mathfrak{p}^{\ast}} - \tilde{f}_{\mathsf{P}}^{\mathfrak{p}}\right) \nonumber\\
    & & +\, (1 - x) \left(\tilde{f}_{\mathsf{N}}^{\prime\mathfrak{p}} - \tilde{f}_{\mathsf{N}}^{\prime\mathfrak{p}^{\ast}} + \tilde{f}_{\mathsf{N}}^{\mathfrak{p}^{\ast}} - \tilde{f}_{\mathsf{N}}^{\mathfrak{p}}\right) , \\
    \tilde{\mathfrak{f}}_{\mathsf{N}}(x) & = & x \left(\tilde{f}_{\mathsf{P}}^{\prime\mathfrak{n}} - \tilde{f}_{\mathsf{P}}^{\prime\mathfrak{n}^{\ast}} + \tilde{f}_{\mathsf{P}}^{\mathfrak{n}^{\ast}} - \tilde{f}_{\mathsf{P}}^{\mathfrak{n}}\right) \nonumber\\
    & & +\, (1 - x) \left(\tilde{f}_{\mathsf{N}}^{\prime\mathfrak{n}} - \tilde{f}_{\mathsf{N}}^{\prime\mathfrak{n}^{\ast}} + \tilde{f}_{\mathsf{N}}^{\mathfrak{n}^{\ast}} - \tilde{f}_{\mathsf{N}}^{\mathfrak{n}}\right) . \qquad
\end{IEEEeqnarray}
The integration of these differential equations gives access to the quantities $\sigma(n_{\mathcal{B}})$, $n_{\mathfrak{p}}(n_{\mathcal{B}})$, and $n_{\mathfrak{n}}(n_{\mathcal{B}})$. For a given $x$, we can then deduce the densities of the chiral partners as $n_{\mathfrak{p}^{\ast}}(n_{\mathcal{B}}) = x n_{\mathcal{B}} - n_{\mathfrak{p}}(n_{\mathcal{B}})$ and $n_{\mathfrak{n}^{\ast}}(n_{\mathcal{B}}) = (1 - x) n_{\mathcal{B}} - n_{\mathfrak{n}}(n_{\mathcal{B}})$, compare again to the relations (\ref{eq:dx_rel}).
The initial conditions for the integration are given by
\begin{equation}
    \sigma(0) = f_{\pi}, \qquad 
    n_{\mathfrak{p}}(0) = n_{\mathfrak{n}}(0) = 0,
\end{equation}
together with $n_{\mathfrak{p}^{\ast}}(0) = n_{\mathfrak{n}\ast}(0) = 0$. At the respective thresholds we then numerically initialize the partner densities $n_{\mathfrak{p}^{\ast}}$ and $n_{\mathfrak{n}^{\ast}}$ with a tiny seed ($10^{-10}\ \mathrm{fm}^{-3}$), because otherwise they would remain zero.

\begin{figure}[ht]
	\centering
    \includegraphics[scale=1.0]{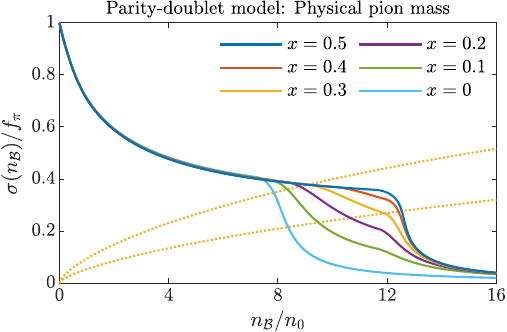}
	\caption{Integrated continuous solution $\sigma(n_{\mathcal{B}})$ for different proton fractions $x$ and for physical pion mass in the parity-doublet model. The dotted lines indicate the thresholds for $x = 0.3$, with the upper line indicating the $\mathfrak{n}^{\ast}$-onset, while the lower line indicating the $\mathfrak{p}^{\ast}$-onset.}
	\label{fig:integration_x_sigma}
\end{figure}

\begin{figure}[ht]
	\centering
    \includegraphics[scale=1.0]{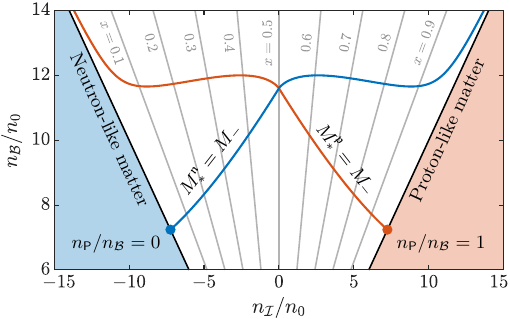}
	\caption{Threshold lines in the plane of the isospin density $n_{\mathcal{I}}$ and the total baryon density $n_{\mathcal{B}}$. The gray lines are lines of constant $x$, and the black lines represent neutron-like matter (left) and proton-like matter (right). The isospin density is constrained by $|n_{\mathcal{I}}| \le n_{\mathcal{B}}$, such that the threshold lines have endpoints on the respective black lines.}
	\label{fig:thresholds_diagram_x}
\end{figure}

Figure~\ref{fig:integration_x_sigma} shows the solution  $\sigma(n_{\mathcal{B}})$, for different proton fractions $x$, as obtained from the continuous integration of Eq.~(\ref{eq:ODE1}) together with Eqs.~(\ref{eq:ODE_np}) and (\ref{eq:ODE_nn}). Depending on the value of $x$, the lines exhibit differences in their behavior, but all of them nearly coincide for small baryon densities: recall that the behavior of $\sigma$  at small $n_{\mathcal{B}}$ is determined by the sigma term, which does not depend on $x$ (see Eq.~(\ref{eq:sigma_term2}) below). This behavior is also readily explained by noticing that below the two thresholds (\ref{eq:thresholds})  the  equations (\ref{eq:ODE_np}) and (\ref{eq:ODE_nn}) reduce to (\ref{eq:ODE_den_simp}). Thus they decouple from the equation that determines $\sigma$, which is then given by Eq.~(\ref{eq:ODE_sigma2}). Moreover, in the presence of a large value of $m_{0}$, Eq.~(\ref{eq:ODE_sigma2}) is well approximated by 
\begin{equation}
    \frac{\rmd \sigma}{\rmd n_{\mathcal{B}}} \simeq -y_{+} \left(\frac{\rmd^{2} U}{\rmd\sigma^{2}} + n_{\mathcal{B}}\frac{\rmd^{2} M_{+}}{\rmd\sigma^{2}}\right)^{-1}, \label{eq:dsigmadn_simp}
\end{equation}
which itself is blind to $x$. This equation was derived in (I), using that $M_{\ast}^{\mathsf{P}} \approx M_{\ast}^{\mathsf{N}} \approx M_{+}$ for small $n_{\mathcal{B}}$, as well as the approximate equality of the scalar density and the baryon density, i.e.\ $n_{\mathrm{s}}^{+} \approx n_{\mathcal{B}}$. Equivalently, the explicit $x$ dependence of $\sigma$ is weak in this regime since, according to Eq.~(\ref{eq:dsigmadx}),
\begin{equation}
    \left.\frac{\rmd \sigma}{\rmd x}\right|_{n_{\mathcal{B}}} \stackrel{M_{\ast}^{\mathfrak{p}} \approx M_{\ast}^{\mathfrak{n}} \approx M_{+}}{\approx} 0 .
\end{equation}
For the choice of $m_{0} = 800\ \mathrm{MeV}$, these approximations are justified up to $n_{\mathcal{B}} \approx 28 \, n_{0}$ for symmetric matter: at this density,  the Fermi momentum becomes of the order of $m_{0}$, and the scalar density $n_{\mathrm{s}}^{+}$  starts to differ significantly from $n_{\mathcal{B}}$. For asymmetric matter, this deviation happens much earlier, e.g.\ for neutron matter at exactly  half  the density quoted above, $n_{\mathcal{B}} \approx 14 \, n_{0}$, but still well beyond the corresponding threshold lines given in Fig.~\ref{fig:thresholds_diagram_x} (the gray lines of constant $x$ in this figure are $n_{\mathcal{B}} = n_{\mathcal{I}}/(2x - 1)$). The  threshold lines in Fig.~\ref{fig:thresholds_diagram_x} help understand the ordering of the various curves in  Fig.~\ref{fig:integration_x_sigma}: as  $x$ goes from $\frac{1}{2}$ to $0$,  the onset of the neutron chiral partners is shifted from $n_{\mathcal{B}} \approx 12 \, n_{0}$ to values below $8 \, n_{0}$. 

The curves for $0 < x < \frac{1}{2}$ in Fig.~\ref{fig:integration_x_sigma} exhibit two kinks in their  shapes, which correspond to the two thresholds (\ref{eq:thresholds}). At each of these kinks,  Eq.~(\ref{eq:ODE_den_simp}) is to be substituted by Eq.~(\ref{eq:ODE_np}) or (\ref{eq:ODE_nn}) in the description through the differential equations. The line representing pure neutron-like matter ($x = 0$) has only one kink, corresponding to the $\mathfrak{n}^{\ast}$-onset. The line representing symmetric matter ($x = \frac{1}{2}$) exhibits also  only one kink, representing in this case the simultaneous onset of the $\mathfrak{p}^{\ast}$ and $\mathfrak{n}^{\ast}$ (with equal densities, $n_{\mathfrak{p}^{\ast}} = n_{\mathfrak{n}^{\ast}}$). In order to illustrate the connection between the kinks and the thresholds, we have plotted in Fig.~\ref{fig:integration_x_sigma} the threshold lines for $x = 0.3$. 

The solutions  $\sigma(n_{\mathcal{B}})$ displayed in Fig.~\ref{fig:integration_x_sigma}, for given proton fractions $x$, do not allow us to determine the order of the  chiral transition. In fact, the variable $x$ that is held constant during the integration of the differential equations is not a good quantity to determine phase coexistence in the case of a first-order transition. For instance, we have seen in Fig.~\ref{fig:liquid_gas_coex_Tzero} that $x$ may vary substantially throughout the liquid-gas transition, except in the case of symmetric matter  $x = \frac{1}{2}$. As mentioned earlier, the conditions for phase coexistence are that the phases share the same physical pressure $P$, as well as the same proton and neutron-chemical potentials $\mu_{\mathsf{P}}$ and $\mu_{\mathsf{N}}$ (which is equivalent to sharing the same baryon and isospin-chemical potentials $\mu_{\mathcal{B}}$ and $\mu_{\mathcal{I}}$). In order to determine the chiral phase transition based on the solutions in Fig.~\ref{fig:integration_x_sigma} and the corresponding solutions for $n_{\mathfrak{p}}$ and $n_{\mathfrak{n}}$ it is convenient to compute the convex envelope of the free energy density, to which we turn to in Sec.~\ref{sec:phasecoex}.

Let us finally discuss the solutions of $\rmd \sigma/\rmd n_{\mathcal{B}}$ in the singlet model, again for a given $x$. The relevant equations  are Eqs.~(\ref{eq:ODE_sigma2}) and (\ref{eq:ODE_den_simp}). As was the case in the doublet model, the small-$n_{\mathcal{B}}$ behavior is dictated by the nucleon sigma term and is independent of $x$. The long plateau observed for the doublet model is of course absent, and the transition towards the chirally symmetric state is smooth for all values of $x$. Contrary to the doublet model, the value of $\sigma$ tends to zero at larger densities the smaller $x$, which we shall explain in the upcoming section about the chiral limit, where the chiral transition becomes second order.
\begin{figure}[ht]
	\centering
    \includegraphics[scale=1.0]{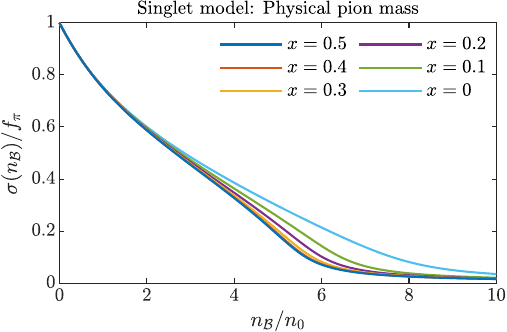}
	\caption{Integrated continuous solution $\sigma(n_{\mathcal{B}})$ for different proton fractions $x$ and for physical pion mass in the singlet model. Analogous plot to Fig.~\ref{fig:integration_x_sigma}.}
	\label{fig:integration_x_sigma_singlet}
\end{figure}

\subsubsection{Neutron matter}

In neutron matter $x = 0$, and the differential equations (\ref{eq:ODE1}) to (\ref{eq:ODE3}) reduce to
\begin{IEEEeqnarray}{rCl}
    \frac{\rmd \sigma}{\rmd n} & = & \frac{1}{m_{\sigma}^{2} M_{\ast}} \! \left[ \left(y_{-} M_{-} - y_{+} M_{+}\right) \!\frac{\rmd n_{\mathfrak{n}}}{\rmd n} - y_{-} M_{-} \right] ,  \qquad \quad \label{eq:dsigmadn_neutron_matter} \\[0.2cm]
    \frac{\rmd n_{\mathfrak{n}}}{\rmd n} & = & \frac{n}{8 \tilde{S}_{\mathsf{N}}} \! \left[ \mathfrak{N}_{0}^{\mathsf{N}} + \frac{y_{-} M_{-} \left(y_{+} M_{+} - y_{-} M_{-}\right)}{m_{\sigma}^{2} M_{\ast}^{2}} \right] , \label{eq:dndnB_neutron_matter}
\end{IEEEeqnarray}
and $\rmd n_{\mathfrak{p}}/\rmd n = 0$ ($n_{\mathsf{P}} = 0$). Here we have alleviated the notation, in the same way as it was done in Sec.~\ref{sec:neutron_matter}. Before the $\mathfrak{n}^{\ast}$-onset is reached, the differential equation for $\sigma$ simply reads
\begin{equation}
    \frac{\rmd\sigma}{\rmd n} = - \frac{y_{+}}{m_{\sigma}^{2}} \frac{M_{+}}{M_{\ast}} .
\end{equation}
This equation has a specific limit for $n \rightarrow 0$,
\begin{equation}
    \left.\frac{\rmd\sigma}{\rmd n}\right|_{n\, =\, 0} = - \left.\frac{y_{+}}{m_{\sigma}^{2}}\right|_{\sigma\, =\, f_{\pi}} = \mathrm{const.} ,
\end{equation}
which we can integrate to
\begin{equation}
    \frac{\sigma_{n}}{f_{\pi}} = 1 - \frac{y_{+} n}{m_{\sigma}^{2} f_{\pi}} \equiv 1 - \frac{\sigma_{N} n}{m_{\pi}^{2} f_{\pi}^{2}},
    \label{eq:sigma_term}
\end{equation}
where $\sigma_{N}$ the pion-nucleon sigma term \cite{Eser:2023oii},
\begin{equation}\label{eq:sigma_term2}
    \sigma_{N} = m_{\pi}^{2} f_{\pi} \frac{y_{+}}{m_{\sigma}^{2}} .
\end{equation}
The sigma term controls  the initial decrease of the isoscalar condensate $\sigma$ with increasing baryon density. Its value is independent of $x$, and from the estimates given in  (I) we have  $\sigma_{N} = 68.6\ \mathrm{MeV}$ in the doublet model, and $\sigma_{N} = 43.7\ \mathrm{MeV}$ in the singlet model.

In the intermediate regime of densities in the doublet model, after the initial decrease (and before the thresholds are crossed), the $\sigma$ field changes only little, as we have seen in Figs.~\ref{fig:phase_diagram_density_PD}, \ref{fig:sigma_neutron_matter}, and \ref{fig:integration_x_sigma}. As stated above, this characteristic behavior can be attributed to the mass $m_{0}$ and the minimum of the fermion mass  at $\sigma=\sigma_{\mathrm{min}}$. Indeed, for $\sigma \approx \mathrm{const.}$, the differential equation (\ref{eq:dsigmadn_simp}) for $\sigma$ takes the form 
\begin{equation}
    \frac{\rmd\sigma}{\rmd n} \simeq - \left(a + b n\right)^{-1}, \qquad
    a, b \approx \mathrm{const.}\ . 
    \label{eq:log_sol}
\end{equation}
This simplified differential equation  describes the slow logarithmic decrease of $\sigma(n)$ that is observed below threshold.

\begin{figure}[ht]
	\centering
    \includegraphics[scale=1.0]{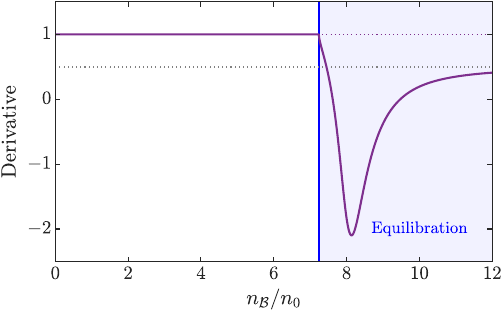}
	\caption{Numerical value of the derivative $\rmd n_\mathfrak{n}/\rmd n_{\mathcal{B}}$ for $x = 0$, as computed by integrating the differential equations (\ref{eq:dsigmadn_neutron_matter}) and (\ref{eq:dndnB_neutron_matter}). The blue vertical line indicates the onset of the $\mathfrak{n}^{\ast}$. The derivative corresponds to Fig.~\ref{fig:densities_neutron_matter}.}
	\label{fig:dn_neutron_matter}
\end{figure}

\begin{figure}[ht]
	\centering
    \includegraphics[scale=1.0]{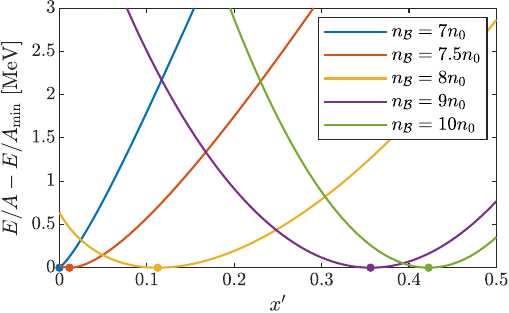}
	\caption{Energy per particle $E/A$ in neutron matter as a function of the parity ratio $x'$, and for various values of the baryon density $n_{\mathcal{B}}$. For each line we subtracted the respective minimum value of $E/A$. The dots indicate the respective locations of the minima (with respect to $x'$).}
	\label{fig:EAmin_neutron_matter}
\end{figure}

Above the $\mathfrak{n}^{\ast}$-threshold, the differential equation for $\sigma$ has to be solved together with the differential equation for $n_{\mathfrak{n}}$, which we rewrite as follows:
\begin{IEEEeqnarray}{rCl}
    \frac{\rmd n_{\mathfrak{n}}}{\rmd n} & = & \frac{n}{8 \tilde{S}_{\mathsf{N}}} \left[\frac{1}{N_{0}^{\mathfrak{n}^{\ast}}} + \tilde{\mathfrak{f}}_{\mathsf{N}}(0)\right] \nonumber\\[0.2cm]
    & \equiv & \left(\frac{1}{N_{0}^{\mathfrak{n}^{\ast}}} + \tilde{f}_{\mathsf{N}}^{\mathfrak{n}^{\ast}} - \tilde{f}_{\mathsf{N}}^{\mathfrak{n}}\right) \bigg\slash \left(\frac{1}{N_{0}^{\mathfrak{n}}} + \frac{1}{N_{0}^{\mathfrak{n}^{\ast}}}\right) , \qquad \label{eq:dndnB_neutron}
\end{IEEEeqnarray}
where we used that
\begin{IEEEeqnarray}{rCl}
    & & \left.\frac{\rmd\sigma}{\rmd n}\right|_{x\, =\, 0} = \frac{\partial\sigma}{\partial n} - \frac{\partial\sigma}{\partial n_{\mathcal{I}}} \nonumber\\[0.2cm]
    & & = \frac{M_{\ast}^{\mathsf{N}}}{y_{+} M_{+} - y_{-} M_{-}} \left(f_{0}^{\mathfrak{n}} - f_{0}^{\mathfrak{n}^{\ast}} + f_{0}^{\prime\mathfrak{n}^{\ast}} - f_{0}^{\prime\mathfrak{n}}\right) , \qquad
\end{IEEEeqnarray}
and the relations (\ref{eq:PNbasis1}) and (\ref{eq:PNbasis2}). In the limit $n_{\mathfrak{n}^{\ast}} \rightarrow 0$, $1/N_{0}^{\mathfrak{n}^{\ast}} \rightarrow \infty$, and the second line of  Eq.~(\ref{eq:dndnB_neutron}) indicates  then that $\rmd n_{\mathfrak{n}}/\rmd n \rightarrow 1$.\footnote{This result can also be seen directly from the first line of Eq.~(\ref{eq:dndnB_neutron}) by noticing that near the $\mathfrak{n}^{\ast}$-onset, $8 \tilde{S}_{\mathsf{N}}/n \sim 1/{N_{0}^{\mathfrak{n}^{\ast}}}$, as can be deduced from Eq.~(\ref{eq:SN2}). This leads to a singular drop of the parity symmetry energy just beyond threshold, which is due to the fact that the parity symmetry energy $\tilde{S}_{\mathsf{N}}$ is evaluated in the vicinity of the $\mathfrak{n}^{\ast}$-onset where the density of state ${N_{0}^{\mathfrak{n}^{\ast}}}$, essentially determined by kinetic energies, vanishes.}
As can be seen in Fig.~\ref{fig:dn_neutron_matter},  as  $n_{\mathfrak{n}^{\ast}}$ increases beyond threshold, the derivative $\rmd n_{\mathfrak{n}}/\rmd n$  rapidly becomes negative, the number of neutrons decreasing in favor of the rapidly increasing population of the $\mathfrak{n}^{\ast}$. In this stage,  the populations adjust themselves in order to minimize the parity symmetry energy.  
In Fig.~\ref{fig:EAmin_neutron_matter} we plot the corresponding energy per particle as a function of the parity fraction $x'$, for different values of the baryon density. For $n = 7\, n_{0}$, thus below the $\mathfrak{n}^{\ast}$-onset, $E/A$ is minimum for vanishing density of the $\mathfrak{n}^{\ast}$ ($x' = 0$), but for $n = 7.5\, n_{0}$ a minimum at $x' > 0$ develops. Further increasing the density then moves the minimum to larger and larger values of $x'$, with the energy per particle steadily growing. For $n = 10\, n_{0}$ the populations of the $\mathfrak{n}$ and $\mathfrak{n}^{\ast}$ are almost equal, with $x'$ already larger than $0.4$.

\begin{figure}[ht]
	\centering
    \includegraphics[scale=1.0]{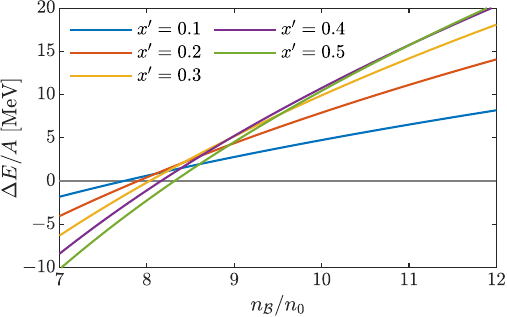}
	\caption{Difference in the energy per particle $\Delta E/A$ as defined in Eq.~(\ref{eq:DeltaEA}), as a function of the baryon density $n_{\mathcal{B}}$, and for various values of the parity ratio $x'$.}
	\label{fig:Delta_EA_neutron_matter}
\end{figure}

To further emphasize the behavior of the energy per particle around threshold (and beyond), we consider the difference
\begin{equation}
    \Delta E/A(n; x') = E/A(n; x' = 0) - E/A(n; x') , \label{eq:DeltaEA}
\end{equation}
which quantifies the change in energy per particle $E/A = \mathcal{E}/n - M_{N}$ (for a constant baryon density), as one changes the parity ratio to values $x' > 0$. This expression obviously relates to the parity symmetry energy, as it was discussed above. Figure~\ref{fig:Delta_EA_neutron_matter} shows the values $\Delta E/A$ for different choices of $x'$, and as a function of increasing $n$, around the onset of the parity partner $\mathfrak{n}^{\ast}$. From Fig.~\ref{fig:EAmin_neutron_matter} we know that at $n = 7\, n_{0}$ the energy per particle is minimum at $x' = 0$, meaning that the $\mathfrak{n}^{\ast}$ are not yet populated. Clearly, any increase of $x'$, keeping $n$ fixed, yields a larger value of $E/A$ and hence a negative value of $\Delta E/A$. Now, when $n$ increases, the minimum in $E/A$ starts to depart from $x' = 0$, and $x' > 0$ becomes energetically favored. $\Delta E/A(n; x')$ eventually turns positive, the earlier the smaller $x'$. As $n$ increases further, the vertical ordering of the lines in Fig.~\ref{fig:Delta_EA_neutron_matter} gets eventually flipped, which shows that the parity-asymmetric state $x' = 0$ becomes more and more unfavored compared to the parity-symmetric state $x' \to \frac{1}{2}$.

\subsection{Determination of phase coexistence in asymmetric matter}
\label{sec:phasecoex}

Following the strategy of Ref.~\cite{Ducoin:2005aa}, we determine now phase coexistence in asymmetric matter near  the chiral transition by computing the convex envelope of the (modified) free energy density $\bar{\mathcal{F}}$,
\begin{IEEEeqnarray}{rCl}
    \bar{\mathcal{F}}(n_{\mathsf{P}}, \mu_{\mathsf{N}}, T) & = & \mathcal{F} - \mu_{\mathsf{N}} n_{\mathsf{N}} = \mathcal{F} - \mu_{\mathsf{N}} (1 - x) n_{\mathcal{B}} \nonumber\\[0.1cm]
    & \equiv & - P + \mu_{\mathsf{P}} n_{\mathsf{P}},
\end{IEEEeqnarray}
which is built as a function of the proton-like density $n_{\mathsf{P}}$, the neutron-chemical potential $\mu_{\mathsf{N}}$, and the temperature $T$. In the formula above, the neutron-like density $n_{\mathsf{N}}$, the proton fraction $x$, and the baryon density $n_{\mathcal{B}}$ (as well as the pressure $P$ and the proton-chemical potential $\mu_{\mathsf{P}}$) are to be considered as functions of $n_{\mathsf{P}}$ and $\mu_{\mathsf{N}}$ (and $T$). By using the function $\bar{\mathcal{F}}(n_{\mathsf{P}}, \mu_{\mathsf{N}}, T)$ defined  above, one can reduce the determination of the phase coexistence  to a one-dimensional problem, i.e.\ at fixed $\mu_{\mathsf{N}}$ and $T$ we may effectively compute the mapping $\bar{\mathcal{F}}(n_{\mathsf{P}})$, and  determine the coexisting points of equal pressure $P$. Note that, at zero temperature, the Helmholtz free energy density $\mathcal{F}$ equals the internal energy density $\mathcal{E}$,
\begin{equation}
    \mathcal{F} = - P + \mu_{\mathsf{P}} n_{\mathsf{P}} + \mu_{\mathsf{N}} n_{\mathsf{N}} \equiv \mathcal{E}.
\end{equation}
For different fixed values of $\mu_{\mathsf{N}}$ we either find an already convex function $\bar{\mathcal{F}}$, hence no phase coexistence, or we find a region in which $\bar{\mathcal{F}}$ becomes concave, thus signaling a first-order phase transition with phase coexistence.

\begin{figure}[ht]
    \centering
    \includegraphics{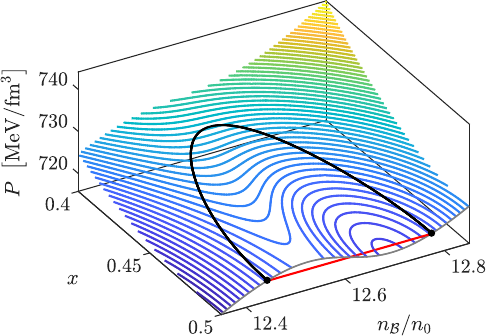}
    \caption{Zero-temperature pressure function around the chiral transition (in the parity-doublet model; physical pion mass). Contours of constant $P$ are given, as a function of the baryon density $n_{\mathcal{B}}$ and the proton fraction $x$. The region inside the black line is unphysical, corresponding to the concave region in the free energy density. For symmetric matter ($x = \frac{1}{2}$) the straight red line demonstrates the Maxwell construction, which connects the two coexisting phases (the black points).}
    \label{fig:pressure_x}
\end{figure}

Figure~\ref{fig:pressure_x} shows the general pressure function (for zero temperature) of the parity-doublet model around the chiral transition, from $x = 0.4$ to the symmetric case ($x = \frac{1}{2}$). This function was obtained by solving the above set of differential equations at different constant values of $x$, and then smoothly interpolating the results. From this function we have to cut out the region that corresponds to the free energy density becoming concave, and the outcome of this procedure is given by the black line. The region that is surrounded by this black line contains unphysical configurations where $P$ shrinks with increasing baryon density. Moreover, the profile of the pressure function at $x = \frac{1}{2}$ demonstrates how the Maxwell construction (red line) connects the two coexisting phases (black dots) in this case (see again the detailed discussion in (I)). Note finally that the chiral transition gap in Fig.~\ref{fig:chiral_transition_nI_nB} coincides with the black cut of Fig.~\ref{fig:chiral_transition_T_den_chiral_limit}, being projected onto the plane of $n_{\mathcal{B}}$ and $x$.

\begin{figure}[ht]
    \centering
    \includegraphics{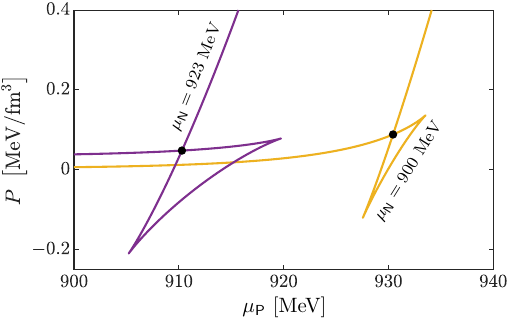}
    \caption{Isotherms at $T = 10\ \mathrm{MeV}$ for different neutron-chemical potentials $\mu_{\mathsf{N}}$ in the parity-doublet model; two examples are shown: $\mu_{\mathsf{N}} = 900\ \mathrm{MeV}$ and $\mu_{\mathsf{N}} = \mu_{0} = M_{N} + E_{\mathrm{bind}} = 923\ \mathrm{MeV}$, as also indicated on the curves. The black dots represent the coexisting gaseous and liquid phases of nuclear matter. The curves are parametrized by the proton density $n_{\mathsf{P}}$, which differs among the coexisting phases.} 
    \label{fig:butterfly}
\end{figure}

Another (equivalent) way of determining phase coexistence, which we also used in practice, in particular at nonzero temperatures, consists in computing isotherms for different constant neutron-chemical potentials $\mu_{\mathsf{N}}$. Self-intersection points of these isotherms correspond to two coexisting phases, with the same values of $P$, $\mu_{\mathsf{N}}$, and $\mu_{\mathsf{P}}$, see the examples plotted in Fig.~\ref{fig:butterfly}. The intersection occurs between two physical branches of the isotherms, where the pressure $P$ increases with increasing chemical potential. 

\subsection{Chiral limit}
\label{sec:chiral_limit}

Further interesting features of the models emerge when one considers the chiral limit of vanishing pion mass, which we do in the last part of this appendix. This limit is achieved by letting $h \rightarrow 0$ in the definition of the mesonic Lagrangian (\ref{eq:mesonL}), while keeping all other parameters unchanged. The initial condition for the differential equation $\rmd\sigma /\rmd n_{\mathcal{B}}$ is then replaced by $f_{\pi} \mapsto \sigma_{\chi}$, where $\sigma_{\chi}$ is the minimum of the bosonic potential in vacuum for $h = 0$. Numerically, we find that $\sigma_{\chi} = 79.8\ \mathrm{MeV}$ in the parity-doublet model, whereas $\sigma_{\chi} = 88.8\ \mathrm{MeV}$ in the singlet model, the latter being much closer to $f_{\pi}$. As we already observed in (I), the comparably large distance between $\sigma_{\chi}$ and $f_{\pi}$ in the doublet model is accompanied by enhanced non linearities in the relations connecting the chiral limit and the point of physical pion mass, which leads for instance to a larger spread in the various estimates of the nucleon sigma term.

\subsubsection{The chiral transition}

\begin{figure}[ht]
    \centering
	\includegraphics[scale=1.0]{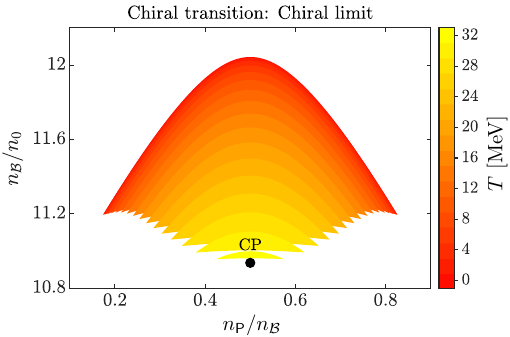}
	\caption{Temperature-dependent coexistence region corresponding to the chiral transition (in the chiral limit), as a function of the proton fraction $x = n_{\mathsf{P}}/n_{\mathcal{B}}$ and the baryon density $n_{\mathcal{B}}$. The black dot shows the tricritical point (``CP'') of the chiral transition for symmetric matter ($x = \frac{1}{2}$) at $(n_{\mathrm{tri}}, T_{\mathrm{tri}}) \approx (1.75\ \mathrm{fm}^{-3}, 33\ \mathrm{MeV})$ (see also the coexistence region in Ref.~\cite{Eser:2023oii}).}
	\label{fig:chiral_transition_T_den_chiral_limit}
\end{figure}

Figure~\ref{fig:chiral_transition_T_den_chiral_limit} is the analog of  Fig.~\ref{fig:chiral_transition_T_den} in the chiral limit. It shows  the coexistence region for the chiral transition as a function of the proton fraction $x$ and the baryon density $n_{\mathcal{B}}$, for different temperatures  up to the temperature of the tricritical point (denoted ``CP'') at $x = \frac{1}{2}$, where the first-order chiral transition turns into a second-order phase transition. The temperature $T_{\mathrm{tri}}$ is much larger than the critical temperature $T_{c}$ in the case of physical pion mass, as it was already determined in (I). 

\begin{figure}[ht]
	\centering
    \includegraphics[scale=1.0]{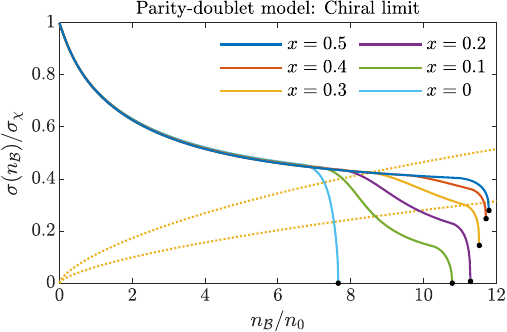}
	\caption{Integrated solution $\sigma(n_{\mathcal{B}})$ for different $x$ and for zero pion mass in the parity-doublet model.  The threshold (dotted) lines shown for $x = 0.3$ are identical to those in  Fig.~\ref{fig:integration_x_sigma}, since they do not depend on the pion mass. The solutions in the chiral-restored phase beyond the singularities (where $\sigma \equiv 0$) are not shown.}
	\label{fig:integration_x_sigma_chiral_limit}
\end{figure}

The corresponding solution of the differential equation for the condensate $\sigma$ at $T = 0$ is shown in Fig.~\ref{fig:integration_x_sigma_chiral_limit}, for different values of $x$. The solution as a function of the total baryon density $n_{\mathcal{B}}$ runs into a divergence at finite $n_{\mathcal{B}}$, which is marked as little black dot, and whose position depends on $x$. For small $x$ it appears just at the density where $\sigma$ reaches zero, while for larger $x$ the derivative $\rmd \sigma/\rmd n_{\mathcal{B}}$ already becomes negatively infinite for positive  $\sigma $. In the first case, the system undergoes a  second-order transition, while the second case implies a gap in $\sigma$ that one may naturally associate with the first-order transition gap (positive jump in density) shown in Fig.~\ref{fig:chiral_transition_T_den_chiral_limit}. Thus, purely based on these solutions of the differential equations and their discontinuities, one  expects a first-order phase transition for $0.2 \lesssim x \lesssim 0.8$ and a second-order transition otherwise, and this is indeed so. In fact, the first-order discontinuity of the physical solution turns out to happen before the divergence in the differential equations is reached (and after the thresholds are crossed).

Above the divergence in the differential equations the system may only exist in the chirally restored phase, where $\sigma = 0$, and thus
\begin{equation}
    M_{+} = M_{-} = m_{0}, \qquad
    y_{+} = \frac{y_{a} - y_{b}}{2} \equiv - y_{-},
\end{equation}
as mentioned earlier, as well as
\begin{equation}
    N_{0}^{\mathfrak{p}} = N_{0}^{\mathfrak{p}^{\ast}}, \qquad
    N_{0}^{\mathfrak{n}} = N_{0}^{\mathfrak{n}^{\ast}} .
\end{equation}
The differential equations then simplify drastically as
\begin{IEEEeqnarray}{rCl}
    \left.\frac{\rmd \sigma}{\rmd n_{\mathcal{B}}} \right|_{x} & = & 0 = \mathrm{const.}, \\[0.2cm]
    \left.\frac{\rmd n_{\mathfrak{p}}}{\rmd n_{\mathcal{B}}} \right|_{x} & = & \frac{x}{2} = \mathrm{const.}, \\[0.2cm]
    \left.\frac{\rmd n_{\mathfrak{n}}}{\rmd n_{\mathcal{B}}} \right|_{x} & = & \frac{1 - x}{2} = \mathrm{const.},
\end{IEEEeqnarray}
which are trivially integrated as
\begin{IEEEeqnarray}{rCl}
    \sigma(n_{\mathcal{B}}) & \equiv & 0, \\[0.2cm]
    n_{\mathfrak{p}}(n_{\mathcal{B}}) & = & \frac{x}{2} n_{\mathcal{B}} \equiv n_{\mathfrak{p}^{\ast}}(n_{\mathcal{B}}), \\[0.2cm]
    n_{\mathfrak{n}}(n_{\mathcal{B}}) & = & \frac{1 - x}{2} n_{\mathcal{B}} \equiv n_{\mathfrak{n}^{\ast}}(n_{\mathcal{B}}).
\end{IEEEeqnarray}
The discontinuity associated to the first-order chiral transition  is determined from the equality of the pressures and the chemical potentials on each side of the transition. The pressure in the broken phase is
\begin{IEEEeqnarray}{rCl}
    P & = & - \sum_{i\,\in\, \lbrace \mathfrak{p},\mathfrak{n},\mathfrak{p}^{\ast},\mathfrak{n}^{\ast} \rbrace} \mathcal{E}_{\mathrm{qp}}^{i}(n_{i}; \sigma) - U(\sigma) \nonumber\\[0.2cm]
    & & +\, M_{\ast}^{\mathsf{P}} n_{\mathsf{P}} + M_{\ast}^{\mathsf{N}} n_{\mathsf{N}} 
    + \frac{1}{2} \left( G_{v} n_{\mathcal{B}}^{2} + G_{w} n_{\mathcal{I}}^{2} \right) , \qquad
\end{IEEEeqnarray}
while in the restored phase, we have
\begin{IEEEeqnarray}{rCl}
    P_{\mathrm{rs}} & = & - \mathcal{E}_{\mathrm{qp}}^{\mathrm{rs}}(x n_{\mathcal{B}}) - \mathcal{E}_{\mathrm{qp}}^{\mathrm{rs}}((1 - x) n_{\mathcal{B}}) - U_{0} \nonumber\\[0.1cm]
    & & + \left[x \sqrt{p_{\mathsf{P}}^{2} + m_{0}^{2}} + (1 - x) \sqrt{p_{\mathsf{N}}^{2} + m_{0}^{2}}\right] n_{\mathcal{B}} \nonumber\\[0.1cm]
    & & +\, \frac{1}{2} \left[G_{v} + (2x - 1)^{2} G_{w}\right] n_{\mathcal{B}}^{2},
\end{IEEEeqnarray}
where
\begin{IEEEeqnarray}{rCl}
    & & \mathcal{E}_{\mathrm{qp}}^{\mathrm{rs}}(n_{\mathsf{P},\mathsf{N}}) = 4 \int_{|p|\, \le\, p_{\mathsf{P},\mathsf{N}}} \frac{\rmd^{3}p}{(2\pi)^{3}} \sqrt{p^{2} + m_{0}^{2}} , \nonumber\\[0.2cm]
    & & = \frac{m_{0}^{4}}{4 \pi^{2}} \left[ \frac{\sqrt{1 + z_{\mathsf{P},\mathsf{N}}^{2}} \left(2 + z_{\mathsf{P},\mathsf{N}}^{2}\right)}{z_{\mathsf{P},\mathsf{N}}^{4}} - \mathrm{csch}^{-1} z_{\mathsf{P},\mathsf{N}} \right], \qquad
\end{IEEEeqnarray}
$z_{\mathsf{P},\mathsf{N}} = m_{0}/p_{\mathsf{P},\mathsf{N}}$, and $U_{0} = U(\sigma = 0)$. The coexistence points are then determined with the method that leads to Fig.~\ref{fig:butterfly}, and one obtains eventually the coexistence region displayed in Fig.~\ref{fig:chiral_transition_T_den_chiral_limit}.

\begin{figure}[ht]
	\centering
    \includegraphics[scale=1.0]{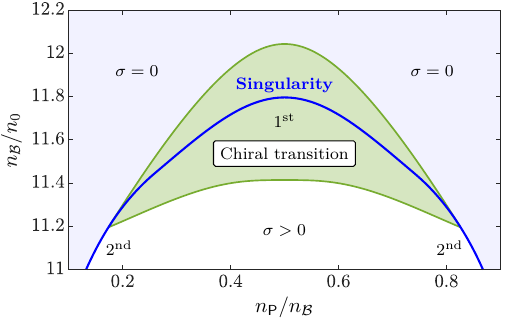}
	\caption{The chiral transition gap at $T = 0$ (in the chiral limit) in the plane of $n_{\mathsf{P}}/n_{\mathcal{B}}$ and the baryon density $n_{\mathcal{B}}$. The green area depicts the positive jump in baryon density the system experiences in the case of a first-order chiral transition. The blue line shows the singularity of the differential equation for $\sigma$; the chiral transition becomes second order starting at the points where the blue line leaves the green area.}
	\label{fig:chiral_transition_x_nB_chiral_limit}
\end{figure}

To better understand the connection between the singularity of the solution of the differential equations and the gap in the $\sigma$ field, we plot in Fig.~\ref{fig:chiral_transition_x_nB_chiral_limit} a blue line which represents the location of the singularity as a function of $x$. Also indicated in the plot is the region (green zone) of the first-order transition limited by two lines, one with $
\sigma=0$ bordering the chirally symmetric phase and the other line with positive $
\sigma$ bordering the chirally broken phase. If the singularity lies inside the green zone, the chiral transition is of first order, and the divergence in the density dependence of the $\sigma$ field (indicated by the black dots in Fig.~\ref{fig:integration_x_sigma_chiral_limit}) occurs at positive $\sigma$. Outside of the green zone, the singularity happens at $\sigma = 0$, indicating a second-order transition. We shall investigate the critical behavior of the doublet model (for neutron matter) at the end of this Appendix.

\begin{figure}[ht]
	\centering
    \includegraphics[scale=1.0]{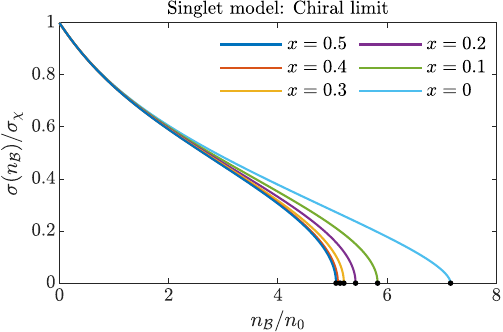}
	\caption{Integrated solution $\sigma(n_{\mathcal{B}})$ for different $x$ and for zero pion mass in the singlet model. Corresponding plot to Fig.~\ref{fig:integration_x_sigma_singlet}; the solutions in the chiral-restored phase ($\sigma \equiv 0$) are not shown.}
	\label{fig:integration_x_sigma_chiral_limit_singlet}
\end{figure}

In the singlet model, see Fig.~\ref{fig:integration_x_sigma_chiral_limit_singlet}, the chiral transition is second order in the chiral limit, for any value of $x$. The corresponding solutions of the differential equation reach zero at the respective critical density, which increases with decreasing $x$ (as observed for physical pion mass, consider again Fig.~\ref{fig:integration_x_sigma_singlet}). This behavior is opposite as compared to the doublet model, where $\sigma$ approaches zero more rapidly the smaller $x$. This finding is easily explained by the fact that the critical Fermi momentum (and also the critical density) in neutron matter in the singlet model is enlarged by a factor $\sqrt{2}$ compared to symmetric matter, which we deduce from the analysis in (I).

\subsubsection{The composition of dense matter in the chiral limit}

\begin{figure}[ht]
	\centering
    \includegraphics[scale=1.0]{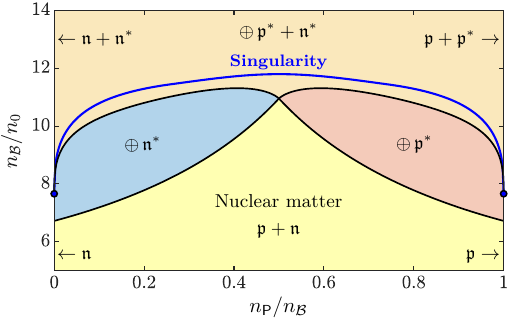}
	\caption{The composition of matter in the chiral limit as a function of the proton fraction $n_{\mathsf{P}}/n_{\mathcal{B}}$ and the total baryon density $n_{\mathcal{B}}$, according to the thresholds (\ref{eq:thresholds}).}
	\label{fig:thresholds_chiral_limit}
\end{figure}

The onset lines of chiral partners in the chiral limit are plotted in Fig.~\ref{fig:thresholds_chiral_limit}, as a function of $x$ and $n_{\mathcal{B}}$. Compared to Fig.~\ref{fig:thresholds_diagram}, which corresponds to the physical pion mass,  the upper parts of the lines now bend downwards as $x \rightarrow 0$ or $x \rightarrow 1$. In fact, these lines  roughly follow the singularity in $\rmd\sigma/\rmd n_{\mathcal{B}}$, which is again marked in the plot as a blue line. The $\mathfrak{p}^{\ast}$-onset meets the singularity line at $x = 0$, and at the critical density $n_{c} \approx 7.66 \, n_{0}$ of neutron matter. The lower parts of the onset lines are also shifted downwards, although their general shape remains unchanged, as compared to Fig.~\ref{fig:thresholds_diagram}.

The point where the $\mathfrak{p}^{\ast}$-onset hits the $n_{\mathcal{B}}$-axis (at $x = 0$) on the plot of Fig.~\ref{fig:thresholds_chiral_limit} is marked as a blue-filled black dot (and likewise for the $\mathfrak{n}^{\ast}$-onset at $x = 1$). This type of dot shall indicate that while we do find a singularity in $\rmd\sigma/\rmd n_{\mathcal{B}}$ at $x = 0$ (neutron matter), i.e.\ the second-order chiral transition, there is no onset of the $\mathfrak{p}^{\ast}$, even though $\sigma = 0$ at the transition point (implying that $M_{+} = M_{-}$). The reason for this is that the threshold condition for the $\mathfrak{p}^{\ast}$ does not hold at $x = 0$ (and the same being true for the $\mathfrak{n}^{\ast}$-threshold at $x = 1$). This property of the composition diagram is also indicated in the upper left (``$\mathfrak{n} + \mathfrak{n}^{\ast}$'') and upper right (``$\mathfrak{p} + \mathfrak{p}^{\ast}$'') corners, stating that dense matter beyond $n \approx 7\, n_{0}$ at $x = 0$ ($x = 1$) exclusively consists of neutrons (protons) and their respective chiral partners.

\subsubsection{Critical behavior in neutron matter and the isoscalar mass}


In order to close this appendix, we eventually consider the critical behavior in neutron matter in the parity-doublet model. This constitutes an interesting application of the formalism of the differential equations, as it yields direct access to the entire development of the $\sigma$ condensate, and in particular to the critical behavior, as well as to the critical mode, whose mass vanishes at the second-order transition. As we have seen, the chiral transition in neutron matter is indeed second order in the chiral limit, cf.\ again Fig.~\ref{fig:integration_x_sigma_chiral_limit}, and we are thus able to describe the critical behavior in terms of a rather simple differential equation. 

\begin{figure}[ht]
	\centering
    \includegraphics[scale=1.0]{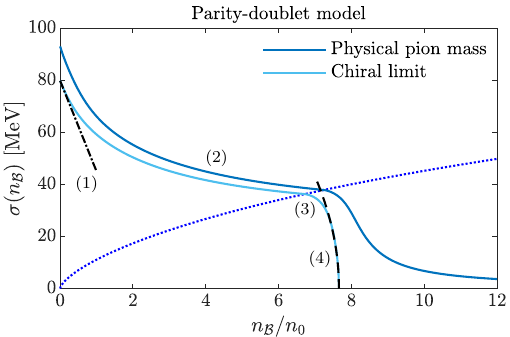}
	\caption{Decomposition of the density dependence of the isoscalar condensate in the parity-doublet model into different stages. The blue dotted line shows the threshold for the onset of the $\mathfrak{n}^{\ast}$, which is independent of the pion mass. The labels (1) to (4) are explained in the main text.}
	\label{fig:integration_x_sigma_decompose_PD}
\end{figure}

In the parity-doublet model, we may now identify four different stages of the solution of the gap equation (\ref{eq:gap}): (1) the initial decrease of $\sigma$ due to the pion-nucleon sigma term in the chiral limit, analogously to Eq.~(\ref{eq:sigma_term}), i.e.
\begin{equation}
    \frac{\sigma_{n_{\mathcal{B}}}}{\sigma_{\chi}} = 1 - \frac{\sigma_{N}^{\chi} n_{\mathcal{B}}}{m_{\pi}^{2} \sigma_{\chi}^{2}},
\end{equation}
with 
\begin{equation}
    \sigma_{N}^{\chi} = m_{\pi}^{2} \sigma_{\chi} \frac{y_{+}(\sigma_{\chi})}{m_{\sigma}^{2}(\sigma_{\chi})} \approx 43.1\ \mathrm{MeV}.
\end{equation}
This linear decrease is shown as the black dash-dotted line in Fig.~\ref{fig:integration_x_sigma_decompose_PD}. (2) After the initial linear behavior it follows that the $\sigma$ condensate decreases more slowly with increasing $n_{\mathcal{B}}$ in a log-like manner, according to the discussion around Eq.~(\ref{eq:log_sol}). (3) The solution of the gap equation then crosses the threshold of the onset of the chiral partner $\mathfrak{n}^{\ast}$, $M_{\ast}^{\mathfrak{n}} = M_{-}$, which leads again to a substantial speedup in its decrease. Let us also note here again that the blue dotted threshold line in Fig.~\ref{fig:integration_x_sigma_decompose_PD} is the same for both cases of zero pion mass and physical pion mass, as it does not depend on any bosonic parameters of the model, and it is only the initial condition of the differential equations that distinguishes between different pion masses. (4) The solution in the chiral limit follows a critical square-root behavior, which is shown as the black dashed line, and which we shall determine next. For physical pion mass the second-order chiral transition is smeared out to a smooth crossover.

In order to determine the critical behavior in the vicinity of the chiral transition, where $\sigma \rightarrow 0$, we may approximate the sigma potential as a quartic polynomial \cite{Eser:2023oii},
\begin{equation}
    U(\sigma) \simeq U_{0} - \frac{r}{2} \sigma^{2} + \frac{u}{4} \sigma^{4} .
\end{equation}
The gap equation for small $\sigma$ then reads
\begin{equation}
    - \tilde{r} \sigma + \mathsf{c} \sigma^{2} + \tilde{u} \sigma^{3} = 0,  \label{eq:gap_neutron_crit}
\end{equation}
with the (renormalised) parameters $\tilde{r}$, $\mathsf{c}$, and $\tilde{u}$, containing corrections from the scalar densities. These parameters are straightforwardly calculated in the same way as we computed the critical behavior for symmetric matter in our previous publication (I) (thus we omit the corresponding analytic expressions here).
To obtain this simplified form of the gap equation we eliminated the Fermi momentum $p_{+}$ of the neutrons by exploiting the equality (\ref{eq:constr2}) of the effective Landau masses above the threshold,
\begin{IEEEeqnarray}{rCl}
    & & p_{+} \simeq \frac{m_{0}}{z_{-}} - z_{-} (y_{a} - y_{b}) \sigma - z_{-}^{3} \frac{(y_{a} - y_{b})^{2}}{2 m_{0}} \sigma^{2} \nonumber\\[0.1cm]
    & & - \frac{z_{-}}{2 m_{0}^{2}} (y_{a} - y_{b}) \left[ z_{-}^{4} (y_{a} - y_{b})^{2} 
    + \frac{1}{4} (y_{a} + y_{b})^{2} \right] \sigma^{3} , \qquad\  \label{eq:ms_correction}
\end{IEEEeqnarray}
where $z_{-} = m_{0}/p_{-}$ (and $p_{-}$ the Fermi momentum of the $\mathfrak{n}^{\ast}$). This elimination is crucial in order to obtain the correct expression for the gap equation for small $\sigma$, and essentially results in the term $\mathsf{c}\sigma^{2}$. Leaving aside the trivial solution $\sigma = 0$, the relevant solution of Eq.~(\ref{eq:gap_neutron_crit}) at which $\sigma \rightarrow 0$ is given by $z_{-} \approx 1.54$. This value corresponds to a critical Fermi momentum of $p_{c} \approx 518.5\ \mathrm{MeV}$, and yields a critical baryon density of
\begin{equation}
    n_{c} = \frac{2 p_{c}^{3}}{3 \pi^{2}} \approx 7.66\, n_{0} .
\end{equation}
The critical behavior of the $\sigma$ field prior to the transition density $n_{c}$ can be inferred from the differential equations (\ref{eq:dsigmadn_neutron_matter}), (\ref{eq:dndnB_neutron_matter}), and (\ref{eq:dndnB_neutron}). Combining these equations we find
\begin{equation}
    \frac{\rmd \sigma}{\rmd n} = - \frac{1}{\tilde{m}_{\sigma}^{2} M_{\ast}} \frac{y_{\pm} M_{\pm} N_{0}^{\pm}}{N_{0}} , 
    \label{eq:dsdn_total}
\end{equation}
where we used again the summation convention, and
\begin{equation}
    \tilde{m}_{\sigma}^{2} = m_{\sigma}^{2} - \frac{\left(y_{-} M_{-} - y_{+} M_{+}\right)^{2}}{M_{\ast}^{2}} \left(\frac{1}{N_{0}^{+}} + \frac{1}{N_{0}^{-}}\right)^{-1} .
\end{equation}
For densities close to the critical density the second factor in Eq.~(\ref{eq:dsdn_total}) is linear in $\sigma$,
\begin{IEEEeqnarray}{rCl}
    \frac{y_{\pm} M_{\pm} N_{0}^{\pm}}{N_{0}} & \simeq & \frac{1}{2} \left[\left(y_{a}^{2} + y_{b}^{2}\right) - \frac{1}{2} \left(y_{a} - y_{b}\right)^{2} z_{c}^{2}\right] \sigma \nonumber\\[0.2cm]
    & \equiv & \mathcal{C}_{\mathrm{num}} \sigma ,
\end{IEEEeqnarray}
while the denominator of the first factor evaluates as
\begin{widetext}
\begin{IEEEeqnarray}{rCl}
    \tilde{m}_{\sigma}^{2} M_{\ast} & \simeq & m_{0} \biggg( 2 u \frac{\sqrt{1 + z_{c}^{2}}}{z_{c}} + \frac{1}{24 \pi^{2}} \Bigg\lbrace \Big[ 3 \left(y_{a} + y_{b}\right)^{4} 
    + 7 \left(y_{a} - y_{b}\right)^{4} + 12 \left(y_{a} + y_{b}\right)^{2}  \left(y_{a} - y_{b}\right)^{2} \Big] \frac{1}{z_{c}} \nonumber\\[0.2cm] 
    & & \qquad - \left(y_{a} - y_{b}\right)^{4} z_{c} \left(1 - 4 z_{c}^{2}\right) 
    - 24 \left(y_{a}^{4} + y_{b}^{4}\right) \frac{\sqrt{1 + z_{c}^{2}}}{z_{c}} \ln \frac{\sqrt{1 + z_{c}^{2}} + 1}{z_{c}} \Bigg\rbrace \biggg) \sigma^{2} \nonumber\\[0.2cm]
    & \equiv & \mathcal{C}_{\mathrm{denom}} \sigma^{2} ,
\end{IEEEeqnarray}
\end{widetext}
where $z_{c} = m_{0}/p_{c}$, and we used again the gap equation, evaluated at $z_{c}$. Furthermore, we exploited the approximation
\begin{IEEEeqnarray}{rCl}
    \tilde{m}_{\sigma}^{2} & \simeq & \bigg( 2 \tilde{u} - \frac{(y_{a} - y_{b})^{2}}{8 \pi^{2}} \frac{z_{c}^{2}}{(1 + z_{c}^{2})^{3/2}} \bigg\lbrace 2 \left(y_{a}^{2} + y_{b}^{2}\right) \quad \nonumber\\[0.2cm]
    & & + \left[ 2 \left(y_{a} + y_{b}\right)^{2} + 3 \left(y_{a} - y_{b}\right)^{2} \right] z_{c}^{2} \bigg\rbrace \bigg) \sigma^{2} .
    \label{eq:ms_tilde_small_sigma}
\end{IEEEeqnarray}
This finally leads to
\begin{equation}
    \frac{\rmd \sigma}{\rmd \Delta n} = - \frac{\rmd \sigma}{\rmd n} \simeq \frac{\mathcal{C}_{\mathrm{num}}}{\mathcal{C}_{\mathrm{denom}}} \frac{1}{\sigma} ,
\end{equation}
with $\Delta n = n_{c} - n$, and which we integrate as
\begin{equation}
    \sigma \simeq \sqrt{\frac{2 \mathcal{C_{\mathrm{num}}}}{\mathcal{C_{\mathrm{denom}}}}} \sqrt{\Delta n} \approx 53\ \mathrm{MeV} \sqrt{\frac{n_{c} - n}{n_{0}}} , \quad
\end{equation}
corresponding to the black dashed line in Fig.~\ref{fig:integration_x_sigma_decompose_PD}.

\begin{figure}[ht]
	\centering
    \includegraphics[scale=1.0]{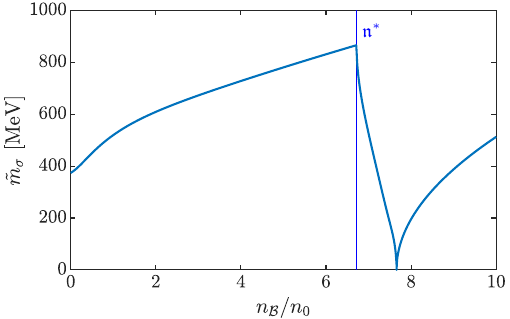}
	\caption{Isoscalar mass $\tilde{m}_{\sigma}$ in neutron matter in the parity-doublet model, as a function of the baryon density at zero temperature. The mass becomes zero at the critical density of the second-order chiral transition, $n_{c} \approx 7.66\, n_{0}$.}
	\label{fig:isoscalar_mass}
\end{figure}

Another important aspect of the critical behavior at the second-order chiral transition is that the isoscalar (chiral) mode becomes massless. From the differential equations and the divergence at the critical density we may also obtain direct information about the vanishing of the corresponding isoscalar mass (here at $T = 0$). 
The mass $\tilde{m}_{\sigma}$ indeed vanishes once $\sigma = 0$, see again Eq.~(\ref{eq:ms_tilde_small_sigma}), and it is responsible for the divergence in $\rmd \sigma/ \rmd n$. We plot the mass $\tilde{m}_{\sigma}$ in Fig.~\ref{fig:isoscalar_mass}, as a function of the baryon density $n$, where we immediately see that it becomes zero at the critical density $n_{c} \approx 7.66\, n_{0}$, as expected. Before this happens, it monotonously increases from its vacuum value (in the chiral limit) of $\tilde{m}_{\sigma} \equiv m_{\sigma} \approx 374\ \mathrm{MeV}$, and then suddenly shrinks drastically once it passed the $\mathfrak{n}^{\ast}$-onset.



\bibliography{references}

\end{document}